\documentclass[12pt,a4paper,twoside]{article}
\usepackage{makeidx}
\usepackage{latexsym,amsmath,amssymb,amsfonts,amsthm}

\newtheorem{proposition}{Proposition}[section]
\newtheorem{definition}{Definition}[section]
\newtheorem{corollary}{Corollary}[section]
\newtheorem{theorem}{Theorem}[section]
\newtheorem{remark}{Remark}[section]
\def\nabla{\bigtriangledown}
\newcommand{ \R} {\mbox{\rm I$\!$R}}

\makeindex \textheight 23.0cm
\begin{document}

\title{ Generalized Finsler Geometry in
Einstein, \\ String and Metric--Affine Gravity}
\author{Sergiu I. Vacaru \thanks{%
e-mails:\ vacaru@fisica.ist.utl.pt, ~~ sergiu$_{-}$vacaru@yahoo.com,\ } \\
{\small \textit{Centro Multidisciplinar de Astrofisica - CENTRA,
Departamento de Fisica,}}\\
{\small \textit{Instituto Superior Tecnico, Av. Rovisco Pais 1, Lisboa,
1049-001, Portugal}} }
\date{October 14, 2004 }
\maketitle

\begin{abstract}
We develop the method of anholonomic frames with associated
nonlinear connection (in brief, N--connection) structure and show
explicitly how geometries with local anisotropy (various type of
Finsler--Lagrange--Cartan--Hamilton geometry) can be modeled in
the metric--affine spaces. There are formulated the criteria when
such generalized Finsler metrics are effectively induced in the
Einstein, teleparallel, Riemann--Cartan and metric--affine
gravity. We argue that every generic off--diagonal metric (which
can not be diagonalized by coordinate transforms) is related to
specific N--connection configurations. We elaborate the concept
of generalized Finsler--affine geometry for spaces provided with
arbitrary N--connection, metric and linear connection structures
and characterized by gravitational field strengths, i. e. by
nontrivial N--connection curvature, Riemannian curvature, torsion
and nonmetricity. We apply a irreducible decomposition techniques
(in our case with additional N--connection splitting) and study
the dynamics of metric--affine gravity fields generating Finsler
like configurations. The classification of basic eleven classes
of metric--affine spaces with generic local anisotropy is
presented.

\vskip5pt

Pacs:\ 04.50.+h, 02.40.-k,

\end{abstract}

\tableofcontents


\section{Introduction}

Brane worlds and related string and gauge theories define the paradigm of
modern physics and have generated enormous interest in higher--dimensional
spacetimes amongst particle and astrophysics theorists (see recent advances
in Refs. \cite{branes,sgr,gauge} and an outline of the gauge idea and
gravity in\ Refs. \cite{mag,ggrav}). The \ unification scheme in the
framework of string/ brane theory indicates that the classical (psedo)
Riemannian description is not valid on all scales of interactions. It turns
out that low--energy dilaton and axi--dilaton interactions are tractable in
terms of non--Riemannian mathematical structures possessing in particular
anholonomic (super) frame [equivalently, (super) vielbein] fields \cite%
{cartan}, noncommutative geometry \cite{ncg}, quantum group structures \cite%
{majid} all containing, in general, nontrivial torsion and nonmetricity
fields. For instance, in the closest alternatives to general relativity
theory, the teleparallel gravity models \cite{tgm}, the spacetime is of
Witzenbock type with trivial curvature but nontrivial torsion. The frame or
coframe filed (tetrad, vierbein, in four dimensions, 4D) is the basic
dynamical variable treated as the gauge potential corresponding to the group
of local translations.

Nowadays, it was established a standard point of view that a number of low
energy (super) string and particle physics interactions, at least the
nongravitational ones, are described by (super) gauge potentials interpreted
as linear connections in suitable (super) bundle spaces. The formal identity
between the geometry of fiber bundles \cite{gfb} is recognized since the
works \cite{gfgfb} (see a recent discussing in connection to a unified
description in of interactions in terms of composite fiber bundles in Ref. %
\cite{tres}).

The geometry of fiber bundles and the moving frame method originating from
the E.\ Cartan works \cite{cartan} constitute a modern approach to Finsler
geometry and generalizations (also suggested by E. Cartan \cite{carf} but
finally elaborated in R.\ Miron and M. Anastasiei works \cite{ma}), see some
earlier and recent developments in Refs. \cite{fg,rund,as,asanovinv,bog}. \
Various type of geometries with local anisotropy (Finsler, Lagrange,
Hamilton, Cartan and their generalizations, according to the terminology
proposed in \cite{ma}), are modeled on (co) vector / tangent bundles and
their higher order generalizations \cite{mat1,mhss} with different
applications in Lagrange and Hamilton mechanics or in generalized Finsler
gravity. Such constructions were defined in low energy limits of (super)
string theory \ and supergravity \cite{vst,vsup} and generalized for spinor
bundles \cite{vsp} and affine-- de Sitter frame bundles \cite{vd} provided
with nonlinear connection (in brief, N--connection) structure of first and
higher order anisotropy.

The gauge and moving frame geometric background is also presented in the
metric--affine gravity (MAG) \cite{mag}. The geometry of this theory is very
general being described by the two--forms of curvature and of torsion and
the one--form of nonmetricity treated respectively as the gravitational
field strengths for the linear connection, coframe and metric. The kinematic
scheme of MAG is well understood at present time as well certain dynamical
aspects of the vacuum configurations when the theory can be reduced to an
effective Einstein--Proca model with nontrivial torsion and nonmetricity %
\cite{oveh,obet2,hm,ghlms}. There were constructed a number of exact
solutions in MAG connecting the theory to modern string gravity and another
extra dimension generalizations \cite{esolmag,cwmag,sbhmag}. Nevertheless,
one very important aspect has not been yet considered. As a gauge theory,
the MAG can be expressed with respect to arbitrary frames and/or coframes.
So, if we introduce frames with associated N--connection structure, the MAG
should incorporate models with generic local anisotropy (Finsler like ones
and their generalizations) which are distinguished by certain prescriptions
for anholonomic frame transforms, N--connection coefficients and metric and
linear connection structures adapted to such anholonomic configurations.
Roughly speaking, the MAG contains the bulk of known generalized Finsler
geometries which can be modeled on metric--affine spaces by defining
splittings on subspaces like on (co) vector/ tangent bundles and considering
certain anholonomically constrained moving frame dynamics and associated
N--connection geometry.

Such metric--affine spaces with local anisotropy are enabled with generic
off--diagonal metrics which can not be diagonalized by any coordinate
transforms. The off--diagonal coefficients can be mapped into the components
of a specific class of anholonomic frames, defining also the coefficients of
the N--connection structure.\ It is possible to redefine equivalently all
geometrical values like tensors, spinors and connections with respect to
such N--adapted anholonomic bases. If the N--connection, metric and linear
connections are chosen for an explicit type of Finsler geometry, a such
geometric structure is modeled on a metric--affine space (we claim that a
Finsler--affine geometry is constructed). The point is to find explicitly by
what type of frames and connections a locally anisotropic structure can be
modeled by exact solutions in the framework of MAG. Such constructions can
be performed in the Einstein--Proca sector of the MAG gravity and they can
be defined even in general relativity theory (see the partners of this paper
with field equations and exact solutions in MAG modeling Finsler like
metrics and generalizations \cite{exsolmag}).

Within the framework of moving frame method \cite{cartan}, we investigated
in a series of works \cite{v1,v2,ncgg,vncggf} the conditions when various
type of metrics with noncommutative symmetry and/or local anisotropy can be
effectively modeled by anholonomic frames on (pseudo) Riemannian and
Riemann--Cartan spaces \cite{rcg}. We constructed explicit classes of such
exact solutions in general relativity theory and extra dimension gravity
models. They are parametrized by generic off--diagonal metrics which can not
diagonalized by any coordinate transforms but only by anholonomic frame
transforms. The new classes of solutions describe static black ellipsoid
objects, locally anistoropic configurations with toroidal and/or ellipsoidal
symmetries, wormholes/ flux tubes and Taub-NUT metrics with polarizied
constants and various warped spinor--soliton--dilaton configurations. For
certain conditions, some classes of such solutions preserve the four
dimensional (4D) local Lorentz symmetry but, in general, they are with
violated Lorentz symmetry in the bulk.

Our ongoing effort is to model different classes of geometries following a
general approach to the geometry of (co) vector/tangent bundles and
affine--de Sitter frame bundles \cite{vd} and superbundles \cite{vsup} and
or anisotropic spinor spaces \cite{vsp} provided with N--connection
stuctures. The basic geometric objects \ on such spaces are defined by
proper classes of anholonomic frames and associated N--connections and
correspondingly adapted metric and linear connections.\ There are examples
when certain Finsler like configurations are modeled even by some exact
solutions in Einstein or Einstein--Cartan gravity and, inversely (the
outgoing effort), by using the almost Hermitian formulation \cite%
{ma,mhss,vsp} of Lagrange/Hamilton and Finsler/Cartan geometry, we can
consider Einstein and gauge gravity models defined on tangen/cotangent and
vector/covector bundles. Recently, there were also obtained some explicit
results demonstrating that the anholonomic frames geometry has a natural
connection to noncommutative geometry in string/M--theory and noncommutative
gauge models of gravity \cite{ncgg,vncggf} (on existing approaches to
noncommutative geometry and gravity we cite Refs. \cite{ncg}).

We consider torsion fields induced by anholonomic vielbein transforms when
the theory can be extended to a gauge \cite{ggrav}, metric--affine \cite{mag}%
,  a more particular Riemann--Cartan case \cite{rcg}, or to string gravity
with $B$--field \cite{sgr}. We are also interested to define the conditions
when an exact solution possesses hidden noncommutative symmetries, induced
torsion and/or locally anisotropic configurations constructed, for instance,
in the framework of the Einstein theory. This direction of investigation
develops the results obtained in Refs. \cite{v2} and should be distinguished
from our previous works on the geometry of Clifford and spinor structures in
generalized Finsler and Lagrange--Hamilton spacetimes \cite{vsp}. Here we
emphasize that the works \cite{v1,v2,ncgg,vncggf,vsp} were elaborated
following general methods of the geometry of anholonomic frames with
associated N--connections in vector (super) bundles \cite{ma,mhss,vsup}. The
concept of N--connection was proposed in Finsler geometry \cite%
{fg,as,asanovinv,bog,rund,carf}. As a set of coefficients it was firstly
present the E. Cartan's monograph\cite{carf} and then was elaborated in a
more explicit form by A. Kawaguchi \cite{kaw}. It was proven that the
N--connection structures can be defined also on (pseudo) Riemannian spaces
and certain methods work effectively in constructing exact solutions in
Einstein gravity \cite{vsp,v1,v2}.

In order to avoid possible terminology ambiguities, we note that for us the
definition of N--connection is that proposed in global form by W. Barthel in
1963 \cite{barthel} when a N--connection is defined as an exact sequence
related to a corresponding Whitney sum of the vertical and horizontal
subbundles, for instance, \ in a tangent vector bundle. \footnote{%
Instead of a vector bundle we can consider a tangent bundle, or
cotangent/covector ones, or even general manifolds of necessary smooth class
with adapted definitions of global sums of horizonal and vertical subspaces.
The geometry of N--connections is investigated in details in Refs. \cite%
{vilms,ma,vsup,vsp,vd} for various type of spaces.} This concept is
different from that accepted in Ref. \cite{tm} were the term 'nonlinear
connection' is used for tetrads as N--connections which do not transform
inhomogeneously under local frame rotations. That approach invokes nonlinear
realizations of the local spacetime group (see also an early model of gauge
gravity with nonlinear gauge group realizations \cite{ts} and its extensions
to Finsler like \cite{vd} or noncommutative gauge gravity theories \cite%
{ncgg}).

In summary, the aim of the present work is to develop a unified scheme of
anholonomic frames with associated N--connection structure for a large
number of gauge and gravity models (in general, with locally isotropic and
anisotropic interactions and various torsion and nonmetricity contributions)
and effective generalized Finsler--Weyl--Riemann--Cartan geometries derived
from MAG. We elaborate a detailed classification of such spaces with
nontrivial N--connection geometry. The unified scheme and classification
were inspired by a number of exact solutions parametrized by generic
off--diagonal metrics and anholonomic frames in Einstein, Einstein--Cartan
and string gravity. The resulting formalism admits inclusion of locally
anisotropic spinor interactions and extensions to noncommutative geometry
and string/brane gravity \cite{vst,vsup,v1,v2,ncgg,vncggf}. Thus, the
geometry of metric--affine spaces enabled with an additional N--connection
structure is sufficient not only to model the bulk of physically important
non--Riemannian geometries on (pseudo) Riemannian spaces but also states the
conditions when effective spaces with generic anisotropy can be derived as
exact solutions of gravitational and matter field equations. In the present
work we pay attention  to the geometrical (pre--dynamical) aspects of the
generalized Finsler--affine gravity which constitute a theoretical
background for constructing a number of exact solutions in MAG in the
partner papers \cite{exsolmag}.

The article is organized as follows. We begin, in Sec. 2, with a review of
the main concepts from the metric--affine geometry and the geometry of
anholonomic frames with associated N--connections. We introduce the basic
definitions and formulate and prove the main theorems for the N--connection,
linear connection and metric structures on metric--affine spaces and derive
the formulas for torsion and curvature distinguished by N--connections.
Next, in Sec. 3, we state the main properties of the linear and nonlinear
connections modeling Finsler spaces and their generalizations and consider
how the N--connection structure can be derived from a generic off--diagonal
metric in a metric--affine space. Section 4 is devoted to the definition and
investigation of generalized Finsler--affine spaces. We illustrate how by
corresponding parametrizations of the off--diagonal metrics, anholonomic
frames, N--connections and distinguished connections every type of
generalized Finsler--Lagrange--Cartan--Hamilton geometry can be modeled in
the metric--affine gravity or any its restrictions to the Einstein--Cartan
and general relativity theory. In Sec. 5, we conclude the results and point
out how the synthesis of the Einstein, MAG and generalized Finsler gravity
models can be realized and connected to the modern string gravity. In
Appendix we elaborate a detailed classification of eleven classes of spaces
with generic local anisotropy (i. e. possessing nontrivial N--connection
structure) and various types of curvature, torsion and nonmetricity
distinguished by N--connections.

Our basic notations and conventions combine those from Refs. \cite%
{mag,ma,v1,v2} and contain an interference of traditions from MAG and
generalized Finsler geometry. The spacetime is modeled as a manifold $V^{n+m}
$ of necessary smoothly class of dimension $n+m.$ The Greek indices $\alpha
,\beta ,...$ can split into subclasses like $\alpha =\left( i,a\right) ,$ $%
\beta =\left( j,b\right) ...$ where the Latin indices from the middle of the
alphabet, $i,j,k,...$ run values $1,2,...n$ and the Latin indices from the
beginning of the alphabet, $\ a,b,c,...$ run values $n+1,n+2,$ ..., $n+m.$
We follow the Penrose convention on abstract indices \cite{pen} and use
underlined indices like $\underline{\alpha }=\left( \underline{i},\underline{%
a}\right) ,$ for decompositions with respect to coordinate frames. The
notations for connections $\Gamma _{\ \beta \gamma }^{\alpha },$ metrics $%
g_{\alpha \beta }$ and frames $e_{\alpha }$ and coframes $\vartheta ^{\beta
},$ or other geometrical and physical objects, are the standard ones from
MAG if a nonlinear connection (N--connection) structure is not emphasized on
the spacetime. If a N--connection and corresponding anholonomic frame
structure are prescribed, we use ''boldfaced'' symbols with possible
splitting of the objects and indices like $\mathbf{V}^{n+m},$ $\mathbf{%
\Gamma }_{\ \beta \gamma }^{\alpha }=\left(
L_{jk}^{i},L_{bk}^{a},C_{jc}^{i},C_{bc}^{a}\right) ,$\textbf{\ }$\mathbf{g}%
_{\alpha \beta }=\left( g_{ij},h_{ab}\right) ,$ $\mathbf{e}_{\alpha }=\left(
e_{i},e_{a}\right) ,$ ...being distinguished by N--connection (in brief, we
use the terms d--objects, d--tensor, d--connection in order to say that they
are for a metric--affine space modeling a generalized Finsler, or another
type, anholonomic frame geometry). The symbol ''\ $\doteqdot $'' will be
used is some formulas which state that the relation is introduced ''by
definition'' and the end of proofs will be stated by symbol $\blacksquare .$

\section{Metric--Affine Spaces and Nonlinear Connections}

We outline the geometry of anholonomic frames and associated nonlinear
connections (in brief, N--connections) in metric--affine spaces which in
this work are necessary smooth class manifolds, or (co) vector/ tangent
bundles provided \ with, in general, independent nonlinear and linear
connections and metrics, and correspondingly derived strengths like
N--connection curvature, Riemannian curvature, torsion and nonmetricity. The
geometric formalism will be applied in the next Sections where we shall
prove that every class of (pseudo) Riemannian, Kaluza--Klein,
Einstein--Cartan, metric--affine and \ generalized Lagrange--Finsler and
Hamilton--Cartan spaces is characterized by corresponding N--connection,
metric and linear connection structures.

\subsection{Linear connections, metrics and anholonomic frames}

\label{standres}We briefly review the standard results on linear connections
and metrics (and related formulas for torsions, curvatures, Ricci and
Einstein tensors and Bianchi identities) defined with respect to arbitrary
anholonomic bases in order to fix a necessary reference which will be
compared with generalized Finsler--affine structures we are going to propose
in the next sections for spaces provided with N--connection. The results are
outlined in a form with conventional splitting into horizontal and vertical
subspaces and subindices. We follow the Ref. \cite{stw} but we use Greek
indices and denote a covariant derivative by $D$ preserving the symbol $%
\bigtriangledown $ for the Levi--Civita (metric and torsionless) connection.
Similar formulas can be found, for instance. in Ref. \cite{mtw}.

Let $V^{n+m}$ be a $\left( {n+m}\right) $--dimensional underlying manifold
of necessary smooth class and denote by $TV^{n+m}$ the corresponding tangent
bundle. The local coordinates on $V^{n+m},u=\{u^{\underline{\alpha }}=\left(
x^{\underline{i}},y^{\underline{a}}\right) \}$ conventionally split into two
respective subgroups of ''horizontal'' coordinates (in brief,
h--coordinates), $x=(x^{\underline{i}}),$ and ''vertical'' coordinates
(v--coordinates), $y=\left( y^{\underline{a}}\right) ,$ with respective
indices running the values $\underline{i},\underline{j},...=1,2,...,n$ and $%
\underline{a},\underline{b},...=n+1,n+2,...,n+m.$ The splitting of
coordinates is treated as a formal labeling if any fiber and/or the
N--connection structures are not defined. Such a splitting of abstract
coordinates $u^{\alpha }=(x^{i},y^{a})$ may be considered, for instance, for
a general (pseudo) Riemannian manifold with $x^{i}$ being some 'holonomic''
variables (unconstrained) and $y^{a}$ being ''anholonomic'' variables
(subjected to some constraints), or in order to parametrize locally a vector
bundle $\left( E,\mu ,F,M\right) $ defined by an injective surjection~$\mu
:E\rightarrow M$ from the total space $E$ to the base space $M$ of dimension
$\dim M=n,$ with $F$ being the typical vector space of dimension $\dim F=m.$
For our purposes, we consider that both $M$ and $F$ can be, in general,
provided with metric structures of arbitrary signatures. On vector bundles,
the values $\ x=(x^{i})$ are coordinates on the base and $y=(y^{a})$ are
coordinates in the fiber. If $\dim M=\dim F,$ the vector bundle $E$
transforms into the tangent bundle $TM.$ The same conventional coordinate
notation $u^{\alpha }=(x^{i},y^{a}\rightarrow p_{a})$ can be used for a dual
vector bundle $\left( E,\mu ,F^{\ast },M\right) $ with the typical fiber $%
F^{\ast }$ being a covector space (of 1-forms) dual to $F,$ where $p_{a}$
are local (dual) coordinates. For simplicity, we shall label $y^{a}$ as
general coordinates even for dual spaces if this will not result in
ambiguities. In general, our geometric constructions will be elaborated for
a manifold $V^{n+m}$ (a general metric--affine spaces) with some additional
geometric structures and fibrations to be stated or modeled latter (for
generalized Finsler geometries) on spacetimes under consideration.

At each point $p\in V^{n+m},$ there are defined basis vectors (local frames,
vielbeins) $e_{\alpha }=A_{\alpha }^{\ \underline{\alpha }}(u)\partial _{%
\underline{\alpha }}\in TV^{n+m},$ with $\partial _{\underline{\alpha }%
}=\partial /\partial u^{\underline{\alpha }}$ being tangent vectors to the
local coordinate lines $u^{\underline{\alpha }}=u^{\underline{\alpha }}(\tau
)$ with parameter $\tau .$ In every point $p,$ there is also a dual basis $%
\vartheta ^{\beta }=A_{\ \underline{\beta }}^{\beta }(u)du^{\underline{\beta
}}$ with $du^{\underline{\beta }}$ considered as coordinate one forms. The
duality conditions can be written in abstract form by using the interior
product $\rfloor ,$ $\ e_{\alpha }\rfloor \vartheta ^{\beta }=\delta
_{\alpha }^{\beta },$ or in coordinate form $A_{\alpha }^{\ \underline{%
\alpha }}A_{\ \underline{\alpha }}^{\beta }=\delta _{\alpha }^{\beta },$
where the Einstein rule of summation on index $\underline{\alpha }$ is
considered, $\delta _{\alpha }^{\beta }$ is the Kronecker symbol. The ''not
underlined'' indices $\alpha ,\beta ,...,$ or $i,j,...$ and $a,b,...$ are
treated as abstract labels (as suggested by R. Penrose). We shall underline
the coordinate indices only in the cases when it will be necessary to
distinguish them from the abstract ones.

Any vector and 1--form fields, for instance, $X$ and, respectively, $%
\widetilde{Y}$ on $V^{n+m}$ are decomposed in h-- and v--irreducible
components,
\begin{equation*}
X=X^{\alpha }e_{\alpha }=X^{i}e_{i}+X^{a}e_{a}=X^{\underline{\alpha }%
}\partial _{\underline{\alpha }}=X^{\underline{i}}\partial _{\underline{i}%
}+X^{\underline{a}}\partial _{\underline{a}}
\end{equation*}%
and
\begin{equation*}
\widetilde{Y}=\widetilde{Y}_{\alpha }\vartheta ^{\alpha }=\widetilde{Y}%
_{i}\vartheta ^{i}+\widetilde{Y}_{a}\vartheta ^{a}=\widetilde{Y}_{\underline{%
\alpha }}du^{\underline{\alpha }}=\widetilde{Y}_{\underline{i}}dx^{%
\underline{i}}+\widetilde{Y}_{\underline{a}}dy^{\underline{a}}.
\end{equation*}%
We shall omit labels like ''$\widetilde{}"$ for forms if this will not
result in ambiguities.

\begin{definition}
\label{deflcon}A linear (affine) connection $D$ on $V^{n+m}$ is a linear map
(operator) sending every pair of smooth vector fields $\left( X,Y\right) $
to a vector field $D_{X}Y$ such that
\begin{equation*}
D_{X}\left( sY+Z\right) =sD_{X}Y+D_{X}Z
\end{equation*}%
for any scalar $s=const$ and for any scalar function $f\left( u^{\alpha
}\right) ,$
\begin{equation*}
D_{X}\left( fY\right) =fD_{X}Y+\left( Xf\right) Y\mbox{ and
}D_{X}f=Xf.
\end{equation*}
\end{definition}

$D_{X}Y$ is called the covariant derivative of $Y$ with respect to $X$ (this
is not a tensor). But we can always define a tensor $DY:$ $X\rightarrow
D_{X}Y.$ The value $DY$ is a $\left( 1,1\right) $ tensor field and called
the covariant derivative of $Y.$

With respect to a local basis $e_{\alpha },$ we can define the scalars $%
\Gamma _{\ \beta \gamma }^{\alpha },$ called the components of the linear
connection $D,$ such that
\begin{equation*}
D_{\alpha }e_{\beta }=\Gamma _{\ \beta \alpha }^{\gamma }e_{\gamma }%
\mbox{
and }D_{\alpha }\vartheta ^{\beta }=-\Gamma _{\ \gamma \alpha }^{\beta
}\vartheta ^{\gamma }
\end{equation*}%
were, by definition, $D_{\alpha }\doteqdot D_{e_{\alpha }}$ and because $%
e_{\beta }\vartheta ^{\beta }=const.$

We can decompose
\begin{equation}
D_{X}Y=\left( D_{X}Y\right) ^{\beta }e_{\beta }=\left[ e_{\alpha }(Y^{\beta
})+\Gamma _{\ \gamma \alpha }^{\beta }\vartheta ^{\gamma }\right] e_{\beta
}\doteqdot Y_{\ ;\alpha }^{\beta }X^{\alpha }  \label{covderrul}
\end{equation}%
where $Y_{\ ;\alpha }^{\beta }$ are the components of the tensor $DY.$

It is a trivial proof that any change of basis (vielbein transform), $%
e_{\alpha ^{\prime }}=B_{\alpha ^{\prime }}^{\ \alpha }e_{\alpha },$ with
inverse $B_{\ \alpha }^{\alpha ^{\prime }},$ results in a corresponding
(nontensor) rule of transformation of the components of the linear
connection,%
\begin{equation}
\Gamma _{\ \beta ^{\prime }\gamma ^{\prime }}^{\alpha ^{\prime }}=B_{\
\alpha }^{\alpha ^{\prime }}\left[ B_{\beta ^{\prime }}^{\ \beta }B_{\gamma
^{\prime }}^{\ \gamma }\Gamma _{\ \beta \gamma }^{\alpha }+B_{\gamma
^{\prime }}^{\ \gamma }e_{\gamma }\left( B_{\beta ^{\prime }}^{\ \alpha
}\right) \right] .  \label{lcontr}
\end{equation}

\begin{definition}
A local basis $e_{\beta }$ is anhlonomic (nonholonomic) if there are
satisfied the conditions%
\begin{equation}
e_{\alpha }e_{\beta }-e_{\beta }e_{\alpha }=w_{\alpha \beta }^{\gamma
}e_{\gamma }  \label{anh}
\end{equation}%
for certain nontrivial anholonomy coefficients $w_{\alpha \beta }^{\gamma
}=w_{\alpha \beta }^{\gamma }(u^{\tau }).$ A such basis is holonomic if $%
w_{\alpha \beta }^{\gamma }\doteqdot 0.$
\end{definition}

For instance, any coordinate basis $\partial _{\alpha }$ is holonomic. Any
holonomic basis can be transformed into a coordinate one by certain
coordinate transforms.

\begin{definition}
\label{deftors}The torsion tensor is a tensor field $\mathcal{T}$ defined by
\begin{equation}
\mathcal{T}\left( X,Y\right) =D_{X}Y-D_{Y}X-[X,Y],  \label{torsa}
\end{equation}%
where $[X,Y]=XY-YX,$ for any smooth vector fields $X$ and $Y.$
\end{definition}

The components $T_{\ \alpha \beta }^{\gamma }$ of a torsion $\mathcal{T}$
with respect to a basis $e_{\alpha }$ are computed by introducing $%
X=e_{\alpha }$ and $Y=e_{\beta }$ in (\ref{torsa}),%
\begin{equation*}
\mathcal{T}\ \left( e_{\alpha },e_{\beta }\right) =D_{\alpha }e_{\beta
}-D_{\beta }e_{\alpha }-[e_{\alpha },e_{\beta }]=T_{\ \alpha \beta }^{\gamma
}e_{\gamma }
\end{equation*}%
where
\begin{equation}
T_{\ \alpha \beta }^{\gamma }=\Gamma _{\ \beta \alpha }^{\gamma }-\Gamma _{\
\alpha \beta }^{\gamma }-w_{\alpha \beta }^{\gamma }.  \label{torsac}
\end{equation}%
We note that with respect to anholonomic frames the coefficients of
anholonomy $w_{\alpha \beta }^{\gamma }$ are contained in the formula for
the torsion coefficients (so any anholonomy induces a specific torsion).

\begin{definition}
\label{defcurv}The Riemann curvature tensor $\mathcal{R}$\ is defined as a
tensor field
\begin{equation}
\mathcal{R}\left( X,Y\right) Z=D_{Y}D_{X}Z-D_{X}D_{Y}Z+D_{[X,Y]}Z.
\label{curva}
\end{equation}
\end{definition}

We can compute the components $R_{\ \beta \gamma \tau }^{\alpha }$of
curvature $\mathcal{R}$, with respect to a basis $e_{\alpha }$ are computed
by introducing $X=e_{\gamma },Y=e_{\tau }$, $Z=e_{\beta }$ in (\ref{curva}).
One obtains
\begin{equation*}
\mathcal{R}\left( e_{\gamma },e_{\tau }\right) e_{\beta }=R_{\ \beta \gamma
\tau }^{\alpha }e_{\alpha }
\end{equation*}%
where
\begin{equation}
R_{\ \beta \gamma \tau }^{\alpha }=e_{\tau }\left( \Gamma _{\ \beta \gamma
}^{\alpha }\right) -e_{\gamma }\left( \Gamma _{\ \beta \tau }^{\alpha
}\right) +\Gamma _{\ \beta \gamma }^{\nu }\Gamma _{\ \nu \tau }^{\alpha
}-\Gamma _{\ \beta \tau }^{\nu }\Gamma _{\ \nu \gamma }^{\alpha }+w_{\gamma
\tau }^{\nu }\Gamma _{\ \beta \nu }^{\alpha }.  \label{curvc}
\end{equation}%
We emphasize that the anholonomy and vielbein coefficients are contained in
the formula for the curvature components (\ref{curva}). With respect to
coordinate frames, $e_{\tau }=\partial _{\tau },$ with $w_{\gamma \tau
}^{\nu }=0,$ we have the usual coordinate formula.

\begin{definition}
The Ricci tensor $\mathcal{R}i$ is a tensor field obtained by contracting
the Riemann tensor,%
\begin{equation}
R_{\ \beta \tau }=R_{\ \beta \tau \alpha }^{\alpha }.  \label{rt}
\end{equation}
\end{definition}

We note that for a general affine (linear) connection the Ricci tensor is
not symmetric $R_{\ \beta \tau }\doteqdot R_{\ \tau \beta }.$

\begin{definition}
\label{defm}A metric tensor is a $\left( 0,2\right) $ symmetric tensor field
\begin{equation*}
\mathit{g}=g_{\alpha \beta }(u^{\gamma })\vartheta ^{\alpha }\otimes
\vartheta ^{\beta }
\end{equation*}%
defining the quadratic (length) linear element,%
\begin{equation*}
ds^{2}=g_{\alpha \beta }(u^{\gamma })\vartheta ^{\alpha }\vartheta ^{\beta
}=g_{\underline{\alpha }\underline{\beta }}(u^{\underline{\gamma }})du^{%
\underline{\alpha }}du^{\underline{\beta }}.
\end{equation*}
\end{definition}

For physical applications, we consider spaces with local Minkowski
singnature, when locally, in a point \ $u_{0}^{\underline{\gamma }},$ the
diagonalized metric is $g_{\underline{\alpha }\underline{\beta }}(u_{0}^{%
\underline{\gamma }})=\eta _{\underline{\alpha }\underline{\beta }}=\left(
1,-1,-1,...\right) $ or, for our further convenience, we shall use metrics
with the local diagonal ansatz being defined by any permutation of this
order.

\begin{theorem}
\label{tmetricity}If a manifold $V^{n+m}$ is enabled with a metric structure
$\mathit{g,}$ then there is a unique torsionless connection, the
Levi--Civita connection $D=\bigtriangledown ,$ satisfying the metricity
condition
\begin{equation}
\bigtriangledown \mathit{g}=0.  \label{lcmc}
\end{equation}
\end{theorem}

The proof, as an explicit construction, is given in Ref. \cite{stw}. Here we
present the formulas for the components $\Gamma _{\bigtriangledown \ \beta
\tau }^{\alpha }$ of the connection $\bigtriangledown ,$ computed with
respect to a basis $e_{\tau },$%
\begin{eqnarray}
\Gamma _{\bigtriangledown \ \alpha \beta \gamma } &=&g\left( e_{\alpha
},\bigtriangledown _{\gamma }e_{\beta }\right) =g_{\alpha \tau }\Gamma
_{\bigtriangledown \ \alpha \beta }^{\tau }  \label{lccoef} \\
&=&\frac{1}{2}\left[ e_{\beta }\left( g_{\alpha \gamma }\right) +e_{\gamma
}\left( g_{\beta \alpha }\right) -e_{\alpha }\left( g_{\gamma \beta }\right)
+w_{\ \gamma \beta }^{\tau }g_{\alpha \tau }+w_{\ \alpha \gamma }^{\tau
}g_{\beta \tau }-w_{\ \beta \gamma }^{\tau }g_{\alpha \tau }\right] .  \notag
\end{eqnarray}%
By straightforward calculations , we can check that%
\begin{equation*}
\bigtriangledown _{\alpha }g_{\beta \gamma }=e_{\alpha }\left( g_{\beta
_{\gamma }}\right) -\Gamma _{\bigtriangledown \ \beta \alpha }^{\tau
}g_{\tau _{\gamma }}-\Gamma _{\bigtriangledown \ \gamma \alpha }^{\tau
}g_{\beta \tau }\equiv 0
\end{equation*}%
and, using the formula (\ref{torsac}),
\begin{equation*}
T_{\bigtriangledown \ \alpha \beta }^{\gamma }=\Gamma _{\bigtriangledown \
\beta \alpha }^{\gamma }-\Gamma _{\ \bigtriangledown \alpha \beta }^{\gamma
}-w_{\alpha \beta }^{\gamma }\equiv 0.
\end{equation*}%
We emphasize that the vielbein and anholonomy coefficients are contained in
the formulas for the components of the Levi--Civita connection $\Gamma
_{\bigtriangledown \ \alpha \beta }^{\tau }$ (\ref{lccoef}) given with
respect to an anholonomic basis $e_{\alpha }.$ The torsion of this
connection, by definition, vanishes with respect to all bases, anholonomic
or holonomic ones. With respect to a coordinate base $\partial _{\alpha }$,
the components $\Gamma _{\bigtriangledown \ \alpha \beta \gamma }$ (\ref%
{lccoef}) transforms into the so--called 1-st type Christoffel symbols%
\begin{equation}
\Gamma _{\alpha \beta \gamma }^{\bigtriangledown }=\Gamma _{\alpha \beta
\gamma }^{\{\}}=\{\alpha \beta \gamma \}=\frac{1}{2}\left( \partial _{\beta
}g_{\alpha \gamma }+\partial _{\gamma }g_{\beta \alpha }-\partial _{\alpha
}g_{\gamma \beta }\right) .  \label{christ}
\end{equation}

If a space $V^{n+m}$ posses a metric tensor, we can use $g_{\alpha \beta }$
and the inverse values $g^{\alpha \beta }$ for lowering and upping indices
as well to contract tensor objects.

\begin{definition}
\label{defrset}{\ } \newline
a) The Ricci scalar $R$ is defined
\begin{equation*}
R\doteqdot g^{\alpha \beta }R_{\alpha \beta },
\end{equation*}%
where $R_{\alpha \beta }$ is the Ricci tensor (\ref{rt}).\newline
b) The Einstein tensor $\mathcal{G}$ has the coefficients
\begin{equation*}
G_{\alpha \beta }\doteqdot R_{\alpha \beta }-\frac{1}{2}Rg_{\alpha \beta },
\end{equation*}%
with respect to any anholonomic or anholonomic frame $e_{\alpha }.$\
\end{definition}

We note that $G_{\alpha \beta }$ and $R_{\alpha \beta }$ are symmetric only
for the Levi--Civita connection $\bigtriangledown $ and that $%
\bigtriangledown _{\alpha }G^{\alpha \beta }=0.$

It should be emphasized that for any general affine connection $D$ and
metric $\mathit{g}$ structures the metric compatibility conditions (\ref%
{lcmc}) are not satisfied.

\begin{definition}
\label{dnmf}The nonmetricity field
\begin{equation*}
\ \mathcal{Q}=Q_{\alpha \beta }\ \vartheta ^{\alpha }\otimes \vartheta
^{\beta }
\end{equation*}%
on a space $V^{n+m}$ is defined by a tensor field with the coefficients
\begin{equation}
Q_{\gamma \alpha \beta }\doteqdot -D_{\gamma }g_{\alpha \beta }  \label{nmfa}
\end{equation}%
where the covariant derivative $D$ is defined by a linear connection $\ $%
1--form $\Gamma _{\ \alpha }^{\gamma }=\Gamma _{\ \alpha \beta }^{\gamma
}\vartheta ^{\beta }.$
\end{definition}

In result, we can generalize the concept of (pseudo) Riemann space [defined
only by a locally (pseudo) Euclidean metric inducing the Levi--Civita
connection with vanishing torsion] and Riemann--Cartan space [defined by any
independent metric and linear connection with nontrivial torsion but with
vanishing nonmetricity] (see details in Refs. \cite{mag,rcg}):

\begin{definition}
\label{defmas} A metric--affine space is a manifold of necessary smooth
class provided with independent linear connection and metric structures. In
general, such spaces posses nontrivial curvature, torsion and nonmetricity
(called strength fields).
\end{definition}

We can extend the geometric formalism in order to include into consideration
the Finsler spaces and their generalizations. This is possible by
introducing an additional fundamental geometric object called the
N--connection.

\subsection{Anholonomic frames and associated N--connections}

Let us define the concept of nonlinear connection on a manifold $V^{n+m}.$
\footnote{%
\bigskip see Refs. \cite{barthel,ma} for original results and constructions
on vector and tangent bundles.
\par
{}} We denote by $\pi ^{T}:TV^{n+m}\rightarrow TV^{n}$ $\ $the differential
of the map $\pi :V^{n+m}\rightarrow V^{n}$ defined as a fiber--preserving
morphism of the tangent bundle $\left( TV^{n+m},\tau _{E},V^{n}\right) $ to $%
V^{n+m}$ and of tangent bundle $\left( TV^{n},\tau ,V^{n}\right) .$ The
kernel of the morphism $\pi ^{T}$ is a vector subbundle of the vector bundle
$\left( TV^{n+m},\tau _{E},V^{n+m}\right) .$ This kernel is denoted $\left(
vV^{n+m},\tau _{V},V^{n+m}\right) $ and called the vertical subbundle over $%
V^{n+m}.$ We denote the inclusion mapping \ by $i:vV^{n+m}\rightarrow
TV^{n+m}$ when the local coordinates of a point $u\in V^{n+m}$ are written $%
u^{\alpha }=\left( x^{i},y^{a}\right) ,$ where the values of indices are $%
i,j,k,...=1,2,...,n$ and $a,b,c,...=n+1,n+2,...,n+m.$

A vector $X_{u}\in TV^{n+m},$ tangent in the point $u\in V^{n+m},$ is
locally represented as $\ (x,y,$ $X,\widetilde{X})$ $=$ $\left(
x^{i},y^{a},X^{i},X^{a}\right) ,$where $\left( X^{i}\right) \in $$\R$$^{n}$
and $\left( X^{a}\right) \in $$\R$$^{m}$ are defined by the equality $%
X_{u}=X^{i}\partial _{i}+X^{a}\partial _{a}$ [$\partial _{\alpha }=\left(
\partial _{i},\partial _{a}\right) $ are usual partial derivatives on
respective coordinates $x^{i}$ and $y^{a}$]. For instance, $\pi ^{T}\left(
x,y,X,\widetilde{X}\right) =\left( x,X\right) $ and the submanifold $%
vV^{n+m} $ contains elements of type $\left( x,y,0,\widetilde{X}\right) $
and the local fibers of the vertical subbundle are isomorphic to $\R$$^{m}.$
Having $\pi ^{T}\left( \partial _{a}\right) =0,$ one comes out that $%
\partial _{a}$ is a local basis of the vertical distribution $u\rightarrow
v_{u}V^{n+m}$ on $V^{n+m},$ which is an integrable distribution.

\begin{definition}
\label{dnlc}A nonlinear connection (N--connection) $\mathbf{N}$ in a space $%
\left( V^{n+m},\pi ,V^{n}\right) $ is defined by the splitting on the left
of the exact sequence
\begin{equation}
0\rightarrow vV^{n+m}\rightarrow TV^{n+m}/vV^{n+m}\rightarrow 0,
\label{eseq}
\end{equation}%
i. e. a morphism of manifolds $N:TV^{n+m}\rightarrow vV^{n+m}$ such that $%
C\circ i$ is the identity on $vV^{n+m}.$
\end{definition}

The kernel of the morphism $\mathbf{N}$ \ is a subbundle of $\left(
TV^{n+m},\tau _{E},V^{n+m}\right) ,$ it is called the horizontal subspace
(being a subbundle for vector bundle constructions) and denoted by $\left(
hV^{n+m},\tau _{H},V^{n+m}\right) .$ Every tangent bundle $(TV^{n+m},$ $\tau
_{E},$ $V^{n+m})$ provided with a N--connection structure is a Whitney sum
of the vertical and horizontal subspaces (in brief, h- and v-- subspaces),
i. e.
\begin{equation}
TV^{n+m}=hV^{n+m}\oplus vV^{n+m}.  \label{wihit}
\end{equation}%
It is proven that for every vector bundle $\left( V^{n+m},\pi ,V^{n}\right) $
over a compact manifold $V^{n}$ there exists a nonlinear connection \cite{ma}
(the proof is similar if the bundle structure is modeled on a manifold).%
\footnote{%
We note that the exact sequence (\ref{eseq}) defines the N--connection in a
global coordinate free form. In a similar form, the N--connection can be
defined for covector bundles or, as particular cases for (co) tangent
bundles. Generalizations for superspaces and noncommutative spaces are
considered respectively in Refs. \cite{vsup} and \cite{ncgg,vncggf}.}

A N--connection $\mathbf{N}$ is defined locally by a set of coefficients $%
N_{i}^{a}(u^{\alpha })$ $=$ $N_{i}^{a}(x^{j},y^{b})$ transforming as
\begin{equation}
N_{i^{\prime }}^{a^{\prime }}\frac{\partial x^{i^{\prime }}}{\partial x^{i}}%
=M_{a}^{a^{\prime }}N_{i}^{a}-\frac{\partial M_{a}^{a^{\prime }}}{\partial
x^{i}}y^{a}  \label{ncontr}
\end{equation}%
under coordinate transforms on the space $\left( V^{n+m},\mu ,M\right) $
when $x^{i^{\prime }}=x^{i^{\prime }}\left( x^{i}\right) $ and $y^{a^{\prime
}}=M_{a}^{a^{\prime }}(x)y^{a}.$ The well known class of linear connections
consists a particular parametization of its coefficients $N_{i}^{a}$ to be
linear on variables $y^{b},$
\begin{equation*}
N_{i}^{a}(x^{j},y^{b})=\Gamma _{bi}^{a}(x^{j})y^{b}.
\end{equation*}

A N--connection structure can be associated to a prescribed ansatz of
vielbein transforms%
\begin{eqnarray}
A_{\alpha }^{\ \underline{\alpha }}(u) &=&\mathbf{e}_{\alpha }^{\ \underline{%
\alpha }}=\left[
\begin{array}{cc}
e_{i}^{\ \underline{i}}(u) & N_{i}^{b}(u)e_{b}^{\ \underline{a}}(u) \\
0 & e_{a}^{\ \underline{a}}(u)%
\end{array}%
\right] ,  \label{vt1} \\
A_{\ \underline{\beta }}^{\beta }(u) &=&\mathbf{e}_{\ \underline{\beta }%
}^{\beta }=\left[
\begin{array}{cc}
e_{\ \underline{i}}^{i\ }(u) & -N_{k}^{b}(u)e_{\ \underline{i}}^{k\ }(u) \\
0 & e_{\ \underline{a}}^{a\ }(u)%
\end{array}%
\right] ,  \label{vt2}
\end{eqnarray}%
in particular case $e_{i}^{\ \underline{i}}=\delta _{i}^{\underline{i}}$ and
$e_{a}^{\ \underline{a}}=\delta _{a}^{\underline{a}}$ with $\delta _{i}^{%
\underline{i}}$ and $\delta _{a}^{\underline{a}}$ being the Kronecker
symbols, defining a global splitting of $\mathbf{V}^{n+m}$ into
''horizontal'' and ''vertical'' subspaces with the N--vielbein structure%
\begin{equation*}
\mathbf{e}_{\alpha }=\mathbf{e}_{\alpha }^{\ \underline{\alpha }}\partial _{%
\underline{\alpha }}\mbox{ and }\mathbf{\vartheta }_{\ }^{\beta }=\mathbf{e}%
_{\ \underline{\beta }}^{\beta }du^{\underline{\beta }}.
\end{equation*}%
In this work, we adopt the convention that for the spaces provided with
N--connection structure the geometrical objects can be denoted by
''boldfaced'' symbols if it would be necessary to dinstinguish such objects
from similar ones for spaces without N--connection. The results from
subsection \ref{standres} can be redefined in order to be compatible with
the N--connection structure and rewritten in terms of ''boldfaced'' values.

A N--connection $\mathbf{N}$ in a space $\mathbf{V}^{n+m}$ is parametrized,
with respect to a local coordinate base,
\begin{equation}
\partial _{\alpha }=(\partial _{i},\partial _{a})\equiv \frac{\partial }{%
\partial u^{\alpha }}=\left( \frac{\partial }{\partial x^{i}},\frac{\partial
}{\partial y^{a}}\right) ,  \label{pder}
\end{equation}%
and dual base (cobase),
\begin{equation}
d^{\alpha }=(d^{i},d^{a})\equiv du^{\alpha }=(dx^{i},dy^{a}),  \label{pdif}
\end{equation}%
by its components $N_{i}^{a}(u)=N_{i}^{a}(x,y),$
\begin{equation*}
\mathbf{N}=N_{i}^{a}(u)d^{i}\otimes \partial _{a}.
\end{equation*}%
It is characterized by the N--connection curvature $\mathbf{\Omega }%
=\{\Omega _{ij}^{a}\}$ as a Nijenhuis tensor field $N_{v}\left( X,Y\right) $
associated to $\mathbf{N\ },$
\begin{equation*}
\mathbf{\Omega }=N_{v}=\left[ vX,vY\right] +v\left[ X,Y\right] -v\left[ vX,Y%
\right] -v\left[ X,vY\right] ,
\end{equation*}%
for $X,Y\in \mathcal{X}\left( V^{n+m}\right) $ \cite{vilms} and $\left[ ,%
\right] $ denoting commutators. In local form one has%
\begin{equation*}
\mathbf{\Omega }=\frac{1}{2}\Omega _{ij}^{a}d^{i}\wedge d^{j}\otimes
\partial _{a},
\end{equation*}%
\begin{equation}
\Omega _{ij}^{a}=\delta _{\lbrack j}N_{i]}^{a}=\frac{\partial N_{i}^{a}}{%
\partial x^{j}}-\frac{\partial N_{j}^{a}}{\partial x^{i}}+N_{i}^{b}\frac{%
\partial N_{j}^{a}}{\partial y^{b}}-N_{j}^{b}\frac{\partial N_{i}^{a}}{%
\partial y^{b}}.  \label{ncurv}
\end{equation}%
The 'N--elongated' operators $\delta _{j}$ from (\ref{ncurv}) are defined
from a certain vielbein configuration induced by the N--connection, the
N--elongated partial derivatives (in brief, N--derivatives)
\begin{equation}
\mathbf{e}_{\alpha }\doteqdot \delta _{\alpha }=\left( \delta _{i},\partial
_{a}\right) \equiv \frac{\delta }{\delta u^{\alpha }}=\left( \frac{\delta }{%
\delta x^{i}}=\partial _{i}-N_{i}^{a}\left( u\right) \partial _{a},\frac{%
\partial }{\partial y^{a}}\right)  \label{dder}
\end{equation}%
and the N--elongated differentials (in brief, N--differentials)
\begin{equation}
\mathbf{\vartheta }_{\ }^{\beta }\doteqdot \delta \ ^{\beta }=\left(
d^{i},\delta ^{a}\right) \equiv \delta u^{\alpha }=\left( \delta
x^{i}=dx^{i},\delta y^{a}=dy^{a}+N_{i}^{a}\left( u\right) dx^{i}\right)
\label{ddif}
\end{equation}%
called also, respectively, the N--frame and N--coframe. \footnote{%
We shall use both type of denotations $\mathbf{e}_{\alpha }\doteqdot \delta
_{\alpha }$ and $\mathbf{\vartheta }_{\ }^{\beta }\doteqdot \delta \
^{\alpha }$ in order to preserve a connection to denotations from Refs. \cite%
{ma,vsup,vsp,v1,v2,vd,vncggf}. The 'boldfaced' symbols $\mathbf{e}_{\alpha }$
and $\mathbf{\vartheta }_{\ }^{\beta }$ are written in order to emphasize
that they define N--adapted vielbeins and the symbols $\delta _{\alpha }$
and $\delta \ ^{\beta }$ will be used for the N--elongated partial
derivatives and, respectively, differentials.
\par
{}}

The N--coframe (\ref{ddif}) is anholonomic because there are satisfied the
anholonomy relations (\ref{anh}),
\begin{equation}
\left[ \delta _{\alpha },\delta _{\beta }\right] =\delta _{\alpha }\delta
_{\beta }-\delta _{\beta }\delta _{\alpha }=\mathbf{w}_{\ \alpha \beta
}^{\gamma }\left( u\right) \delta _{\gamma }  \label{anhr}
\end{equation}%
for which the anholonomy coefficients $\mathbf{w}_{\beta \gamma }^{\alpha
}\left( u\right) $ are computed to have certain nontrivial values
\begin{equation}
\mathbf{w}_{~ji}^{a}=-\mathbf{w}_{~ij}^{a}=\Omega _{ij}^{a},\ \mathbf{w}%
_{~ia}^{b}=-\mathbf{w}_{~ai}^{b}=\partial _{a}N_{i}^{b}.  \label{anhc}
\end{equation}

We emphasize that the N--connection formalism is a natural one for
investigating physical systems with mixed sets of holonomic--anholonomic
variables. The imposed anholonomic constraints (anisotropies) are
characterized by the coefficients of N--connection which defines a global
splitting of the components of geometrical objects with respect to some
'horizontal' (holonomic) and 'vertical' (anisotropic) directions. In brief,
we shall use respectively the terms h- and/or v--components, h- and/or
v--indices, and h- and/or v--subspaces

A N--connection structure on $\mathbf{V}^{n+m}$ defines the algebra of
tensorial distinguished \ (by N--connection structure) fields $dT\left( T%
\mathbf{V}^{n+m}\right) $ (d--fields, d--tensors, d--objects, if to follow
the terminology from \cite{ma}) on $\mathbf{V}^{n+m}$ introduced as the
tensor algebra $\mathcal{T}=\{\mathcal{T}_{qs}^{pr}\}$ of the distinguished
tangent bundle $\mathcal{V}_{(d)},$ $p_{d}:\ h\mathbf{V}^{n+m}\oplus v%
\mathbf{V}^{n+m}\rightarrow \mathbf{V}^{n+m}.$ An element $\mathbf{t}\in
\mathcal{T}_{qs}^{pr},$ a d--tensor field of type $\left(
\begin{array}{cc}
p & r \\
q & s%
\end{array}%
\right) ,$ can be written in local form as%
\begin{equation*}
\mathbf{t}=t_{j_{1}...j_{q}b_{1}...b_{r}}^{i_{1}...i_{p}a_{1}...a_{r}}\left(
u\right) \delta _{i_{1}}\otimes ...\otimes \delta _{i_{p}}\otimes \partial
_{a_{1}}\otimes ...\otimes \partial _{a_{r}}\otimes d^{j_{1}}\otimes
...\otimes d^{j_{q}}\otimes \delta ^{b_{1}}...\otimes \delta ^{b_{r}}.
\end{equation*}

There are used the denotations $\mathcal{X}\left( \mathcal{V}_{(d)}\right) $
(or $\mathcal{X}(\mathbf{V}^{n+m}{),\wedge }^{p}\left( \mathcal{V}%
_{(d)}\right) $ (or ${\wedge }^{p}\left( \mathbf{V}^{n+m}\right) $ and $%
\mathcal{F}\left( \mathcal{V}_{(d)}\right) $ (or $\mathcal{F}$ $\left(
\mathbf{V}^{n+m}\right) $) for the module of d--vector fields on $\mathcal{V}%
_{(d)}$ (or $\mathbf{V}^{n+m}$ ), the exterior algebra of p--forms on $%
\mathcal{V}_{(d)}$ (or $\mathbf{V}^{n+m})$ and the set of real functions on $%
\mathcal{V}_{(d)}$ (or $\mathbf{V}^{n+m}).$

\subsection{Distinguished linear connection and metric structures}

\label{dlcms} The d--objects on $\mathcal{V}_{(d)}$ are introduced in a
coordinate free form as geometric objects adapted to the N--connection
structure. In coordinate form, we can characterize such objects (linear
connections, metrics or any tensor field) by certain group and coordinate
transforms adapted to the N-connection structure on $\mathbf{V}^{n+m},$ i.
e. to the global space splitting (\ref{wihit}) into h- and v--subspaces.

\subsubsection{d--connections}

We analyze the general properties of a class of linear connections being
adapted to the N--connection structure (called d--connections).

\begin{definition}
\label{defdcon}A d--connection $\mathbf{D}$ on $\mathcal{V}_{(d)}$ is
defined as a linear connection $D,$ see Definition \ref{deflcon}, on $%
\mathcal{V}_{(d)}$ conserving under a parallelism the global decomposition
of $T\mathbf{V}^{n+m}$ (\ref{wihit}) into the horizontal subbundle, $h%
\mathbf{V}^{n+m},$ and vertical subbundle, $v\mathbf{V}^{n+m},$ of $\mathcal{%
V}_{(d)}.$
\end{definition}

A N-connection induces decompositions of d--tensor indices into sums of
horizontal and vertical parts, for example, for every d--vector $\mathbf{X}%
\in \mathcal{X}\left( \mathcal{V}_{(d)}\right) $ and 1-form $\widetilde{%
\mathbf{X}}\in \Lambda ^{1}\left( \mathcal{V}_{(d)}\right) $ we have
respectively
\begin{equation*}
X=hX+vX\ \mbox{and \quad
}\widetilde{X}=h\widetilde{X}+v\widetilde{X}.
\end{equation*}%
For simplicity, we shall not use boldface symbols for d--vectors and
d--forms if this will not result in ambiguities. In consequence, we can
associate to every d--covariant derivation $\mathbf{D}_{X}=X\rfloor \mathbf{D%
}$ two new operators of h- and v--covariant derivations, $\mathbf{D}%
_{X}=D_{X}^{[h]}+D_{X}^{[v]},$ defined respectively
\begin{equation*}
D_{X}^{[h]}Y=\mathbf{D}_{hX}Y\quad \mbox{ and \quad }D_{X}^{[v]}Y=\mathbf{D}%
_{vX}Y,
\end{equation*}%
for which the following conditions hold:%
\begin{eqnarray}
\mathbf{D}_{X}Y &=&D_{X}^{[h]}Y+D_{X}^{[v]}Y,  \label{hvder} \\
D_{X}^{[h]}f &=&(hX\mathbf{)}f\mbox{ \quad and\quad
}D_{X}^{[v]}f=(vX)f,\quad  \notag
\end{eqnarray}%
for any $X,Y\in \mathcal{X}\left( E\right) ,f\in \mathcal{F}\left(
V^{n+m}\right) .$

The N--adapted components $\mathbf{\Gamma }_{\beta \gamma }^{\alpha }$ of a
d-connection $\mathbf{D}_{\alpha }=(\delta _{\alpha }\rfloor \mathbf{D})$
are defined by the equations%
\begin{equation*}
\mathbf{D}_{\alpha }\delta _{\beta }=\mathbf{\Gamma }_{\ \alpha \beta
}^{\gamma }\delta _{\gamma },
\end{equation*}%
from which one immediately follows
\begin{equation}
\mathbf{\Gamma }_{\ \alpha \beta }^{\gamma }\left( u\right) =\left( \mathbf{D%
}_{\alpha }\delta _{\beta }\right) \rfloor \delta ^{\gamma }.  \label{dcon1}
\end{equation}%
The operations of h- and v-covariant derivations, $D_{k}^{[h]}=%
\{L_{jk}^{i},L_{bk\;}^{a}\}$ and $D_{c}^{[v]}=\{C_{jk}^{i},C_{bc}^{a}\}$
(see (\ref{hvder})) are introduced as corresponding h- and
v--parametrizations of (\ref{dcon1}),%
\begin{eqnarray}
L_{jk}^{i} &=&\left( \mathbf{D}_{k}\delta _{j}\right) \rfloor d^{i},\quad
L_{bk}^{a}=\left( \mathbf{D}_{k}\partial _{b}\right) \rfloor \delta ^{a}
\label{hcov} \\
C_{jc}^{i} &=&\left( \mathbf{D}_{c}\delta _{j}\right) \rfloor d^{i},\quad
C_{bc}^{a}=\left( \mathbf{D}_{c}\partial _{b}\right) \rfloor \delta ^{a}.
\label{vcov}
\end{eqnarray}%
A set of h--components \ (\ref{hcov}) and v--components (\ref{vcov}),
distinguished in the form $\mathbf{\Gamma }_{\ \alpha \beta }^{\gamma }$ $%
=(L_{jk}^{i},$ $L_{bk}^{a},$ $C_{jc}^{i},C_{bc}^{a}),$ completely defines
the local action of a d--connection $\mathbf{D}$ in $\mathbf{V}^{n+m}.$ For
instance, having taken a d--tensor field of type $\left(
\begin{array}{cc}
1 & 1 \\
1 & 1%
\end{array}%
\right) ,$ $\mathbf{t}=t_{jb}^{ia}\delta _{i}\otimes \partial _{a}\otimes
\partial ^{j}\otimes \delta ^{b},$ and a d--vector $\mathbf{X}=X^{i}\delta
_{i}+X^{a}\partial _{a}$ we can write%
\begin{equation*}
\mathbf{D}_{X}\mathbf{t=}D_{X}^{[h]}\mathbf{t+}D_{X}^{[v]}\mathbf{t=}\left(
X^{k}t_{jb|k}^{ia}+X^{c}t_{jb\perp c}^{ia}\right) \delta _{i}\otimes
\partial _{a}\otimes d^{j}\otimes \delta ^{b},
\end{equation*}%
where the h--covariant derivative is
\begin{equation*}
t_{jb|k}^{ia}=\frac{\delta t_{jb}^{ia}}{\delta x^{k}}%
+L_{hk}^{i}t_{jb}^{ha}+L_{ck}^{a}t_{jb}^{ic}-L_{jk}^{h}t_{hb}^{ia}-L_{bk}^{c}t_{jc}^{ia}
\end{equation*}%
and the v--covariant derivative is
\begin{equation*}
t_{jb\perp c}^{ia}=\frac{\partial t_{jb}^{ia}}{\partial y^{c}}%
+C_{hc}^{i}t_{jb}^{ha}+C_{dc}^{a}t_{jb}^{id}-C_{jc}^{h}t_{hb}^{ia}-C_{bc}^{d}t_{jd}^{ia}.
\end{equation*}%
For a scalar function $f\in \mathcal{F}\left( V^{n+m}\right) $ we have
\begin{equation*}
D_{k}^{[h]}=\frac{\delta f}{\delta x^{k}}=\frac{\partial f}{\partial x^{k}}%
-N_{k}^{a}\frac{\partial f}{\partial y^{a}}\mbox{ and }D_{c}^{[v]}f=\frac{%
\partial f}{\partial y^{c}}.
\end{equation*}%
We note that these formulas are written in abstract index form and specify
for d--connections the covariant derivation rule (\ref{covderrul}).

\subsubsection{Metric structures and d--metrics}

We introduce arbitrary metric structures on a space $\mathbf{V}^{n+m}$ and
consider the possibility to adapt them to N--connection structures.

\begin{definition}
A metric structure $\mathbf{g}$ on a space $\mathbf{V}^{n+m}$ is defined as
a symmetric covariant tensor field of type $\left( 0,2\right) ,$ $g_{\alpha
\beta ,}$ being nondegenerate and of constant signature on $\mathbf{V}%
^{n+m}. $
\end{definition}

This Definition is completely similar to Definition \ref{defm} but in our
case it is adapted to the N--connection structure. A N--connection $\mathbf{%
N=}\{N_{\underline{i}}^{\underline{b}}\left( u\right) \}$ and a metric
structure%
\begin{equation}
\mathbf{g}=g_{\underline{\alpha }\underline{\beta }}du^{\underline{\alpha }%
}\otimes du^{\underline{\beta }}  \label{mstr}
\end{equation}%
on $\mathbf{V}^{n+m}$ are mutually compatible if there are satisfied the
conditions
\begin{equation}
\mathbf{g}\left( \delta _{\underline{i}},\partial _{\underline{a}}\right) =0,%
\mbox{ or equivalently,
}g_{\underline{i}\underline{a}}\left( u\right) -N_{\underline{i}}^{%
\underline{b}}\left( u\right) h_{\underline{a}\underline{b}}\left( u\right)
=0,  \label{comp1}
\end{equation}%
where $h_{\underline{a}\underline{b}}\doteqdot \mathbf{g}\left( \partial _{%
\underline{a}},\partial _{\underline{b}}\right) $ and $g_{\underline{i}%
\underline{a}}\doteqdot \mathbf{g}\left( \partial _{\underline{i}},\partial
_{\underline{a}}\right) \,$ resulting in
\begin{equation}
N_{i}^{b}\left( u\right) =h^{ab}\left( u\right) g_{ia}\left( u\right)
\label{nconstr}
\end{equation}%
(the matrix $h^{ab}$ is inverse to $h_{ab};$ for simplicity, we do not
underly \ the indices in the last formula). In consequence, we obtain a
h--v--decomposition of metric (in brief, d--metric)%
\begin{equation}
\mathbf{g}(X,Y)\mathbf{=}h\mathbf{g}(X,Y)+v\mathbf{g}(X,Y),  \label{block1}
\end{equation}%
where the d-tensor $h\mathbf{g}(X,Y)=\mathbf{g}(hX,hY)$ is of type $\left(
\begin{array}{cc}
0 & 0 \\
2 & 0%
\end{array}%
\right) $ and the d-tensor \newline
$\ v\mathbf{g}(X,Y)\mathbf{=h}(vX,vY)$ is of type $\left(
\begin{array}{cc}
0 & 0 \\
0 & 2%
\end{array}%
\right) .$ With respect to a N--coframe (\ref{ddif}), the d--metric (\ref%
{block1}) is written
\begin{equation}
\mathbf{g}=\mathbf{g}_{\alpha \beta }\left( u\right) \delta ^{\alpha
}\otimes \delta ^{\beta }=g_{ij}\left( u\right) d^{i}\otimes
d^{j}+h_{ab}\left( u\right) \delta ^{a}\otimes \delta ^{b},  \label{block2}
\end{equation}%
where $g_{ij}\doteqdot \mathbf{g}\left( \delta _{i},\delta _{j}\right) .$
The d--metric (\ref{block2}) can be equivalently written in
''off--diagonal'' form if the basis of dual vectors consists from the
coordinate differentials (\ref{pdif}),
\begin{equation}
\underline{g}_{\alpha \beta }=\left[
\begin{array}{cc}
g_{ij}+N_{i}^{a}N_{j}^{b}h_{ab} & N_{j}^{e}h_{ae} \\
N_{i}^{e}h_{be} & h_{ab}%
\end{array}%
\right] .  \label{ansatz}
\end{equation}%
It is easy to check that one holds the relations%
\begin{equation*}
\mathbf{g}_{\alpha \beta }=\mathbf{e}_{\alpha }^{\ \underline{\alpha }}%
\mathbf{e}_{\beta }^{\ \underline{\beta }}\underline{g}_{\underline{\alpha }%
\underline{\beta }}
\end{equation*}%
or, inversely,
\begin{equation*}
\underline{g}_{\underline{\alpha }\underline{\beta }}=\mathbf{e}_{\
\underline{\alpha }}^{\alpha }\mathbf{e}_{\ \underline{\beta }}^{\beta }%
\mathbf{g}_{\alpha \beta }
\end{equation*}%
as it is stated by respective vielbein transforms (\ref{vt1}) and (\ref{vt2}%
).

\begin{remark}
\label{rgod}A metric, for instance, parametrized in the form (\ref{ansatz})\
is generic off--diagonal if it can not be diagonalized by any coordinate
transforms. If the anholonomy coefficients (\ref{anhc}) vanish for a such
parametrization, we can define certain coordinate transforms to diagonalize
both the off--diagonal form (\ref{ansatz}) and the equivalent d--metric (\ref%
{block2}).
\end{remark}

\begin{definition}
The nonmetricity d--field
\begin{equation*}
\ \mathcal{Q}=\mathbf{Q}_{\alpha \beta }\mathbf{\vartheta }^{\alpha }\otimes
\mathbf{\vartheta }^{\beta }=\mathbf{Q}_{\alpha \beta }\delta \ ^{\alpha
}\otimes \delta ^{\beta }
\end{equation*}%
on a space $\mathbf{V}^{n+m}$ provided with N--connection structure is
defined by a d--tensor field with the coefficients
\begin{equation}
\mathbf{Q}_{\alpha \beta }\doteqdot -\mathbf{Dg}_{\alpha \beta }  \label{nmf}
\end{equation}%
where the covariant derivative $\mathbf{D}$ is for a d--connection $\mathbf{%
\Gamma }_{\ \alpha }^{\gamma }=\mathbf{\Gamma }_{\ \alpha \beta }^{\gamma }%
\mathbf{\vartheta }^{\beta },$ see (\ref{dcon1}) with the respective
splitting $\mathbf{\Gamma }_{\alpha \beta }^{\gamma }=\left(
L_{jk}^{i},L_{bk}^{a},C_{jc}^{i},C_{bc}^{a}\right) ,$ as to be adapted to
the N--connection structure.
\end{definition}

This definition is similar to that given for metric--affine spaces (see
definition \ref{dnmf}) and Refs. \cite{mag}, but in our case the
N--connection establishes some 'preferred' N--adapted local frames (\ref%
{dder}) and (\ref{ddif}) splitting all geometric objects into irreducible h-
and v--components. A linear connection $D_{X}$ is compatible\textbf{\ } with
a d--metric $\mathbf{g}$ if%
\begin{equation}
D_{X}\mathbf{g}=0,  \label{mc}
\end{equation}%
$\forall X\mathbf{\in }\mathcal{X}\left( V^{n+m}\right) ,$ i. e. if $%
Q_{\alpha \beta }\equiv 0.$ In a space provided with N--connection
structure, the metricity condition (\ref{mc}) may split into a set of
compatibility conditions on h- and v-- subspaces. We should consider
separately which of the conditions
\begin{equation}
D^{[h]}(h\mathbf{g)}=0,D^{[v]}(h\mathbf{g)}=0,D^{[h]}(v\mathbf{g)}%
=0,D^{[v]}(v\mathbf{g)}=0  \label{mca}
\end{equation}%
are satisfied, or not, for a given d--connection $\mathbf{\Gamma }_{\ \alpha
\beta }^{\gamma }.$ For instance, if $D^{[v]}(h\mathbf{g)}=0$ and $D^{[h]}(v%
\mathbf{g)}=0,$ but, in general, $D^{[h]}(h\mathbf{g)}\neq 0$ and $D^{[v]}(v%
\mathbf{g)}\neq 0$ we can consider a nonmetricity d--field (d--nonmetricity)
$\mathbf{Q}_{\alpha \beta }=\mathbf{Q}_{\gamma \alpha \beta }\vartheta
^{\gamma }$ with irreducible h--v--components (with respect to the
N--connection decompositions), $\ \mathbf{Q}_{\gamma \alpha \beta }=\left(
Q_{ijk},Q_{abc}\right) .$

By acting on forms with the covariant derivative $D,$ in a metric--affine
space, we can also define another very important geometric objects (the
'gravitational field potentials', see \cite{mag}):

\begin{equation}
\mbox{ torsion }\ \mathcal{T}^{\alpha }\doteqdot D\vartheta ^{\alpha
}=d\vartheta ^{\alpha }+\Gamma _{\ \beta }^{\gamma }\wedge \vartheta ^{\beta
},\mbox{ see Definition \ref{deftors}}  \label{dta}
\end{equation}%
and
\begin{equation}
\mbox{ curvature }\ \mathcal{R}_{\ \beta }^{\alpha }\doteqdot D\Gamma _{\
\beta }^{\alpha }=d\Gamma _{\ \beta }^{\alpha }-\Gamma _{\ \beta }^{\gamma
}\wedge \Gamma _{\ \gamma }^{\alpha },\mbox{ see Definition \ref{defcurv}}.
\label{dra}
\end{equation}

The Bianchi identities are%
\begin{equation}
DQ_{\alpha \beta }\equiv \mathcal{R}_{\alpha \beta }+\mathcal{R}_{\beta
\alpha },\ D\mathcal{T}^{\alpha }\equiv \mathcal{R}_{\gamma }^{\ \alpha
}\wedge \vartheta ^{\gamma }\mbox{ and }D\mathcal{R}_{\gamma }^{\ \alpha
}\equiv 0,  \label{bi}
\end{equation}%
where we stress the fact that $Q_{\alpha \beta },T^{\alpha }$ and $R_{\beta
\alpha }$ are called also the strength fields of a metric--affine theory.

For spaces provided with N--connections, we write the corresponding formulas
by using ''boldfaced'' symbols and change the usual differential $d$ $\ $%
into N-adapted operator $\delta .$%
\begin{equation}
\ \mathbf{T}^{\alpha }\doteqdot \mathbf{D\vartheta }^{\alpha }=\delta
\mathbf{\vartheta }^{\alpha }+\mathbf{\Gamma }_{\ \beta }^{\gamma }\wedge
\mathbf{\vartheta }^{\beta }  \label{dt}
\end{equation}%
and
\begin{equation}
\ \mathbf{R}_{\ \beta }^{\alpha }\doteqdot \mathbf{D\Gamma }_{\ \beta
}^{\alpha }=\delta \mathbf{\Gamma }_{\ \beta }^{\alpha }-\mathbf{\Gamma }_{\
\beta }^{\gamma }\wedge \mathbf{\Gamma }_{\ \ \gamma }^{\alpha }  \label{dc}
\end{equation}%
where the Bianchi identities written in 'boldfaced' symbols split into h-
and v--irreducible decompositions induced by the N--connection. \footnote{%
see similar details in Ref. \cite{ma} for the case of vector/tangent bundles
provided with mutually compatible N--connection, d--connection and d--metric
structure.
\par
{}} We shall examine and compute the general form of torsion and curvature
d--tensors in spaces provided with N--connection structure in section \ref%
{torscurv}.

We note that the bulk of works on Finsler geometry and generalizations \cite%
{fg,ma,mhss,rund,as,bog,carf,vsup,vsp,vncggf} consider very general linear
connection and metric fields being addapted to the N--connection structure.
In another turn, the researches on metric--affine gravity \cite{mag,rcg}
concern generalizations to nonmetricity but not N--connections. In this
work, we elaborate a unified moving frame geometric approach to both Finlser
like and metric--affine geometries.

\subsection{ Torsions and curvatures of d--connections}

\label{torscurv}We define and calculate the irreducible components of
torsion and curvature in a space $\mathbf{V}^{n+m}$ provided with additional
N--connection structure (these could be any metric--affine spaces \cite{mag}%
, or their particular, like Riemann--Cartan \cite{rcg}, cases with vanishing
nonmetricity and/or torsion, or any (co) vector / tangent bundles like in
Finsler geometry and generalizations).

\subsubsection{d--torsions and N--connections}

We give a definition being equivalent to (\ref{dt}) but in d--operator form
(the Definition \ref{deftors} was for the spaces not possessing
N--connection structure):

\begin{definition}
The torsion $\mathbf{T}$ of a d--connection $\mathbf{D}=\left(
D^{[h]},D^{[v]}\right) \mathbf{\ }$ in space $\mathbf{V}^{n+m}$ is defined
as an operator (d--tensor field) adapted to the N--connection structure
\begin{equation}
\mathbf{T}\left( X,Y\right) =\mathbf{D}_{X}Y\mathbf{-D}_{Y}X\mathbf{\ -}%
\left[ X,Y\right] \mathbf{.}  \label{torsion}
\end{equation}
\end{definition}

One holds the following h- and v--decompositions%
\begin{equation}
\mathbf{T}\left( X,Y\right) \mathbf{=T}\left( hX,hY\right) \mathbf{+T}\left(
hX,vY\right) \mathbf{+T}\left( vX,hY\right) \mathbf{+T}\left( vX,vY\right)
\mathbf{.}  \label{hvtorsion}
\end{equation}%
We consider the projections: $h\mathbf{T}\left( X,Y\right) \mathbf{,}v%
\mathbf{T}\left( hX,hY\right) \mathbf{,}h\mathbf{T}\left( hX,hY\right)
\mathbf{,...}$ and say that, for instance, $\mathbf{\ }h\mathbf{T}\left(
hX,hY\right) $ is the h(hh)-torsion of $D$ , $vT\left( hX,hY\right) \mathbf{%
\ }$ is the v(hh)-torsion of $\mathbf{D}$ and so on.

The torsion (\ref{torsion}) is locally determined by five d--tensor fields,
d--torsions (irreducible N--adapted h--v--decompositions) defined as%
\begin{eqnarray}
T_{jk}^{i} &=&h\mathbf{T}\left( \delta _{k},\delta _{j}\right) \rfloor
d^{i},\quad T_{jk}^{a}=v\mathbf{T}\left( \delta _{k},\delta _{j}\right)
\rfloor \delta ^{a},\quad P_{jb}^{i}=h\mathbf{T}\left( \partial _{b},\delta
_{j}\right) \rfloor d^{i},  \notag \\
\quad P_{jb}^{a} &=&v\mathbf{T}\left( \partial _{b},\delta _{j}\right)
\rfloor \delta ^{a},\quad \ S_{bc}^{a}=v\mathbf{T}\left( \partial
_{c},\partial _{b}\right) \rfloor \delta ^{a}.  \notag
\end{eqnarray}%
Using the formulas (\ref{dder}), (\ref{ddif}), and (\ref{ncurv}), we can
calculate the h--v--components of torsion (\ref{hvtorsion}) for a
d--connection, i. e. we can prove \footnote{%
see also the original proof for vector bundles in \cite{ma}}

\begin{theorem}
\label{tdtors}The torsion $\mathbf{T}_{.\beta \gamma }^{\alpha
}=(T_{.jk}^{i},T_{ja}^{i},T_{.ij}^{a},T_{.bi}^{a},T_{.bc}^{a})$ of a
d--connection\newline
$\mathbf{\Gamma }_{\alpha \beta }^{\gamma }=\left(
L_{jk}^{i},L_{bk}^{a},C_{jc}^{i},C_{bc}^{a}\right) $(\ref{dcon1}) has
irreducible h- v--components (d--torsions)
\begin{eqnarray}
T_{.jk}^{i} &=&-T_{kj}^{i}=L_{jk}^{i}-L_{kj}^{i},\quad
T_{ja}^{i}=-T_{aj}^{i}=C_{.ja}^{i},\ T_{.ji}^{a}=-T_{.ij}^{a}=\frac{\delta
N_{i}^{a}}{\delta x^{j}}-\frac{\delta N_{j}^{a}}{\delta x^{i}}=\Omega
_{.ji}^{a},  \notag \\
T_{.bi}^{a} &=&-T_{.ib}^{a}=P_{.bi}^{a}=\frac{\partial N_{i}^{a}}{\partial
y^{b}}-L_{.bj}^{a},\
T_{.bc}^{a}=-T_{.cb}^{a}=S_{.bc}^{a}=C_{bc}^{a}-C_{cb}^{a}.\   \label{dtorsb}
\end{eqnarray}
\end{theorem}

We note that on (pseudo) Riemanian spacetimes the d--torsions can be induced
by the N--connection coefficients and reflect an anholonomic frame
structures. Such objects vanishes when we transfer our considerations with
respect to holonomic bases for a trivial N--connection and zero ''vertical''
dimension.

\subsubsection{d--curvatures and N--connections}

In operator form, the curvature (\ref{dc}) is stated from the

\begin{definition}
The curvature $\mathbf{R}$ of a d--connection $\mathbf{D}=\left(
D^{[h]},D^{[v]}\right) \mathbf{\ }$ in space $\mathbf{V}^{n+m}$ is defined
as an operator (d--tensor field) adapted to the N--connection structure
\begin{equation}
\mathbf{R}\left( X,Y\right) Z=\mathbf{D}_{X}\mathbf{D}_{Y}Z-\mathbf{D}_{Y}%
\mathbf{D}_{X}Z-\mathbf{D}_{[X,Y]}Z\mathbf{.}  \label{curvaturea}
\end{equation}
\end{definition}

This Definition is similar to the Definition \ref{defcurv} being a
generalization for the spaces provided with N--connection. One holds certain
properties for the h- and v--decompositions of curvature:%
\begin{equation}
v\mathbf{R}\left( X,Y\right) hZ=0,\ h\mathbf{R}\left( X,Y\right) vZ\mathbf{=}%
0,\ \mathbf{R}\left( X,Y\right) Z=h\mathbf{R}\left( X,Y\right) hZ\mathbf{+}v%
\mathbf{R}\left( X,Y\right) vZ\mathbf{.}  \notag
\end{equation}%
From (\ref{curvaturea}) and the equation $\mathbf{R}\left( X,Y\right)
\mathbf{=-R}\left( Y,X\right) ,$ we get that the curvature of a
d-con\-necti\-on $\mathbf{D}$ in $\mathbf{V}^{n+m}$ is completely determined
by the following six d--tensor fields (d--curvatures):%
\begin{eqnarray}
R_{\ hjk}^{i} &=&d^{i}\rfloor \mathbf{R}\left( \delta _{k},\delta
_{j}\right) \delta _{h},~R_{\ bjk}^{a}=\delta ^{a}\rfloor \mathbf{R}\left(
\delta _{k},\delta _{j}\right) \partial _{b},  \label{curvaturehv} \\
P_{\ jkc}^{i} &=&d^{i}\rfloor \mathbf{R}\left( \partial _{c},\partial
_{k}\right) \delta _{j},~P_{\ bkc}^{a}=\delta ^{a}\rfloor \mathbf{R}\left(
\partial _{c},\partial _{k}\right) \partial _{b},  \notag \\
S_{\ jbc}^{i} &=&d^{i}\rfloor \mathbf{R}\left( \partial _{c},\partial
_{b}\right) \delta _{j},~S_{\ bcd}^{a}=\delta ^{a}\rfloor \mathbf{R}\left(
\partial _{d},\partial _{c}\right) \partial _{b}.  \notag
\end{eqnarray}%
By a direct computation, using (\ref{dder}), (\ref{ddif}), (\ref{hcov}), (%
\ref{vcov}) and (\ref{curvaturehv}), we prove

\begin{theorem}
The curvature $\mathbf{R}_{.\beta \gamma \tau }^{\alpha }=(R_{\
hjk}^{i},R_{\ bjk}^{a},P_{\ jka}^{i},P_{\ bka}^{c},S_{\ jbc}^{i},S_{\
bcd}^{a})$ of a d--con\-nec\-ti\-on $\mathbf{\Gamma }_{\alpha \beta
}^{\gamma }=\left( L_{jk}^{i},L_{bk}^{a},C_{jc}^{i},C_{bc}^{a}\right) $(\ref%
{dcon1}) has irreducible h- v--components (d--curvatures)
\begin{eqnarray}
R_{\ hjk}^{i} &=&\frac{\delta L_{.hj}^{i}}{\delta x^{k}}-\frac{\delta
L_{.hk}^{i}}{\delta x^{j}}%
+L_{.hj}^{m}L_{mk}^{i}-L_{.hk}^{m}L_{mj}^{i}-C_{.ha}^{i}\Omega _{.jk}^{a},
\label{dcurv} \\
R_{\ bjk}^{a} &=&\frac{\delta L_{.bj}^{a}}{\delta x^{k}}-\frac{\delta
L_{.bk}^{a}}{\delta x^{j}}%
+L_{.bj}^{c}L_{.ck}^{a}-L_{.bk}^{c}L_{.cj}^{a}-C_{.bc}^{a}\ \Omega
_{.jk}^{c},  \notag \\
P_{\ jka}^{i} &=&\frac{\partial L_{.jk}^{i}}{\partial y^{k}}-\left( \frac{%
\partial C_{.ja}^{i}}{\partial x^{k}}%
+L_{.lk}^{i}C_{.ja}^{l}-L_{.jk}^{l}C_{.la}^{i}-L_{.ak}^{c}C_{.jc}^{i}\right)
+C_{.jb}^{i}P_{.ka}^{b},  \notag \\
P_{\ bka}^{c} &=&\frac{\partial L_{.bk}^{c}}{\partial y^{a}}-\left( \frac{%
\partial C_{.ba}^{c}}{\partial x^{k}}+L_{.dk}^{c%
\,}C_{.ba}^{d}-L_{.bk}^{d}C_{.da}^{c}-L_{.ak}^{d}C_{.bd}^{c}\right)
+C_{.bd}^{c}P_{.ka}^{d},  \notag \\
S_{\ jbc}^{i} &=&\frac{\partial C_{.jb}^{i}}{\partial y^{c}}-\frac{\partial
C_{.jc}^{i}}{\partial y^{b}}+C_{.jb}^{h}C_{.hc}^{i}-C_{.jc}^{h}C_{hb}^{i},
\notag \\
S_{\ bcd}^{a} &=&\frac{\partial C_{.bc}^{a}}{\partial y^{d}}-\frac{\partial
C_{.bd}^{a}}{\partial y^{c}}+C_{.bc}^{e}C_{.ed}^{a}-C_{.bd}^{e}C_{.ec}^{a}.
\notag
\end{eqnarray}
\end{theorem}

The components of the Ricci d-tensor
\begin{equation*}
\mathbf{R}_{\alpha \beta }=\mathbf{R}_{\ \alpha \beta \tau }^{\tau }
\end{equation*}%
with respect to a locally adapted frame (\ref{dder}) has four irreducible h-
v--components, $\mathbf{R}_{\alpha \beta }=\{R_{ij},R_{ia},R_{ai},S_{ab}\},$
where%
\begin{eqnarray}
R_{ij} &=&R_{\ ijk}^{k},\quad R_{ia}=-\ ^{2}P_{ia}=-P_{\ ika}^{k},
\label{dricci} \\
R_{ai} &=&\ ^{1}P_{ai}=P_{\ aib}^{b},\quad S_{ab}=S_{\ abc}^{c}.  \notag
\end{eqnarray}%
We point out that because, in general, $^{1}P_{ai}\neq ~^{2}P_{ia}$ the
Ricci d--tensor is non symmetric.

Having defined a d--metric of type (\ref{block2}) in $\mathbf{V}^{n+m},$ we
can introduce the scalar curvature of a d--connection $\mathbf{D,}$
\begin{equation}
{\overleftarrow{\mathbf{R}}}=\mathbf{g}^{\alpha \beta }\mathbf{R}_{\alpha
\beta }=R+S,  \label{dscal}
\end{equation}%
where $R=g^{ij}R_{ij}$ and $S=h^{ab}S_{ab}$ and define the distinguished
form of the Einstein tensor (the Einstein d--tensor), see Definition \ref%
{defrset},%
\begin{equation}
\mathbf{G}_{\alpha \beta }\doteqdot \mathbf{R}_{\alpha \beta }-\frac{1}{2}%
\mathbf{g}_{\alpha \beta }{\overleftarrow{\mathbf{R}}.}  \label{deinst}
\end{equation}

The Ricci and Bianchi identities (\ref{bi}) of d--connections are formulated
in h- v- irreducible forms on vector bundle \cite{ma}. The same formulas
hold for arbitrary metric compatible d--connections on $\mathbf{V}^{n+m}$
(for simplicity, we omit such details in this work).

\section{Some Classes of Linear and Nonlinear Connections}

\label{lconnections}The geometry of d--connections in a space $\mathbf{V}%
^{n+m}$ provided with N--connection structure is very reach (works \cite{ma}
and \cite{vsup} contain results on generalized Finsler spaces and
superspaces). If a triple of fundamental geometric objects $(N_{i}^{a}\left(
u\right) ,\mathbf{\Gamma }_{\ \beta \gamma }^{\alpha }\left( u\right) ,$ $%
\mathbf{g}_{\alpha \beta }\left( u\right) )$ is fixed on $\mathbf{V}^{n+m}{,}
$ in general, with respect to N--adapted frames, a multiconnection structure
is defined (with different rules of covariant derivation). In this Section,
we analyze a set of linear connections and associated covariant derivations
being very important for investigating spacetimes provided with anholonomic
frame structure and generic off--diagonal metrics.

\subsection{The Levi--Civita connection and N--connections}

The Levi--Civita connection $\bigtriangledown =\{\mathbf{\Gamma }%
_{\bigtriangledown \beta \gamma }^{\tau }\}$ with coefficients
\begin{equation}
\mathbf{\Gamma }_{\alpha \beta \gamma }^{\bigtriangledown }=g\left( \mathbf{e%
}_{\alpha },\bigtriangledown _{\gamma }\mathbf{e}_{\beta }\right) =\mathbf{g}%
_{\alpha \tau }\mathbf{\Gamma }_{\bigtriangledown \beta \gamma }^{\tau },\,
\label{lccon1}
\end{equation}%
is torsionless,
\begin{equation*}
\mathbf{T}_{\bigtriangledown }^{\alpha }\doteqdot \bigtriangledown \mathbf{%
\vartheta }^{\alpha }=d\mathbf{\vartheta }^{\alpha }+\mathbf{\Gamma }%
_{\bigtriangledown \beta \gamma }^{\tau }\wedge \mathbf{\vartheta }^{\beta
}=0,
\end{equation*}%
and metric compatible, $\bigtriangledown \mathbf{g}=0,$ see see Definition %
\ref{deflcon}. The formula (\ref{lccon1}) states that the operator $%
\bigtriangledown $ can be defined on spaces provided with N--connection
structure (we use 'boldfaced' symbols) but this connection is not adapted to
the N--connection splitting \ (\ref{wihit}). It is defined as a linear
connection but not as a d--connection, see Definition \ref{defdcon}. The
Levi--Civita connection is usually considered on (pseudo) Riemannian spaces
but it can be also introduced, for instance, in (co) vector/tangent bundles
both with respect to coordinate and anholonomic frames \cite{ma,v1,v2}. One
holds a Theorem similar to the Theorem \ref{tmetricity},

\begin{theorem}
If a space $\mathbf{V}^{n+m}$ is provided with both N--connection $\mathbf{N}
$\ and d--metric $\mathbf{g}$ structures, there is a unique linear symmetric
and torsionless connection $\mathbf{\bigtriangledown },$ being metric
compatible such that $\bigtriangledown _{\gamma }\mathbf{g}_{\alpha \beta
}=0 $ $\ $for $\mathbf{g}_{\alpha \beta }=\left( g_{ij},h_{ab}\right) ,$ see
(\ref{block2}), with the coefficients
\begin{equation*}
\mathbf{\Gamma }_{\alpha \beta \gamma }^{\bigtriangledown }=\mathbf{g}\left(
\delta _{\alpha },\bigtriangledown _{\gamma }\delta _{\beta }\right) =%
\mathbf{g}_{\alpha \tau }\mathbf{\Gamma }_{\bigtriangledown \beta \gamma
}^{\tau },\,
\end{equation*}%
computed as
\begin{equation}
\mathbf{\Gamma }_{\alpha \beta \gamma }^{\bigtriangledown }=\frac{1}{2}\left[
\delta _{\beta }\mathbf{g}_{\alpha \gamma }+\delta _{\gamma }\mathbf{g}%
_{\beta \alpha }-\delta _{\alpha }\mathbf{g}_{\gamma \beta }+\mathbf{g}%
_{\alpha \tau }\mathbf{w}_{\gamma \beta }^{\tau }+\mathbf{g}_{\beta \tau }%
\mathbf{w}_{\alpha \gamma }^{\tau }-\mathbf{g}_{\gamma \tau }\mathbf{w}%
_{\beta \alpha }^{\tau }\right]  \label{lcsym}
\end{equation}%
with respect to N--frames $\mathbf{e}_{\beta }\doteqdot \delta _{\beta }$ (%
\ref{dder}) and N--coframes $\mathbf{\vartheta }_{\ }^{\alpha }\doteqdot
\delta ^{\alpha }$ (\ref{ddif}).
\end{theorem}

The proof is that from Theorem \ref{tmetricity}, see also Refs. \cite%
{stw,mtw}, with $e_{\beta }\rightarrow \mathbf{e}_{\beta }$ and $\vartheta
^{\beta }\rightarrow \mathbf{\vartheta }^{\beta }$ substituted directly in
formula (\ref{lccoef}).

With respect to coordinate frames $\partial _{\beta }$ (\ref{pder}) and $%
du^{\alpha }$ (\ref{pdif}), the metric (\ref{block2}) transforms
equivalently into (\ref{mstr}) with coefficients (\ref{ansatz})\ and the
coefficients of (\ref{lcsym}) transform into the usual Christoffel symbols (%
\ref{christ}). We emphasize that we shall use the coefficients just in the
form (\ref{lcsym}) in order to compare the properties of different classes
of connections given with respect to N--adapted frames. The coordinate form (%
\ref{christ}) is not ''N--adapted'', being less convenient for geometric
constructions on spaces with anholonomic frames and associated N--connection
structure.

We can introduce the 1-form formalism and express
\begin{equation*}
\mathbf{\Gamma }_{\ \gamma \alpha }^{\bigtriangledown }=\mathbf{\Gamma }%
_{\gamma \alpha \beta }^{\bigtriangledown }\mathbf{\vartheta }^{\beta }
\end{equation*}%
where
\begin{equation}
\mathbf{\Gamma }_{\ \gamma \alpha }^{\bigtriangledown }=\frac{1}{2}\left[
\mathbf{e}_{\gamma }\rfloor \ \delta \mathbf{\vartheta }_{\alpha }-\mathbf{e}%
_{\alpha }\rfloor \ \delta \mathbf{\vartheta }_{\gamma }-\left( \mathbf{e}%
_{\gamma }\rfloor \ \mathbf{e}_{\alpha }\rfloor \ \delta \mathbf{\vartheta }%
_{\beta }\right) \wedge \mathbf{\vartheta }^{\beta }\right] ,
\label{christa}
\end{equation}%
contains h- v-components, $\mathbf{\Gamma }_{\bigtriangledown \alpha \beta
}^{\gamma }=\left( L_{\bigtriangledown jk}^{i},L_{\bigtriangledown
bk}^{a},C_{\bigtriangledown jc}^{i},C_{\bigtriangledown bc}^{a}\right) ,$
defined similarly to (\ref{hcov}) and (\ref{vcov}) but using the operator $%
\bigtriangledown ,$
\begin{equation*}
L_{\bigtriangledown jk}^{i}=\left( \bigtriangledown _{k}\delta _{j}\right)
\rfloor d^{i},\quad L_{\bigtriangledown bk}^{a}=\left( \bigtriangledown
_{k}\partial _{b}\right) \rfloor \delta ^{a},\ C_{\bigtriangledown
jc}^{i}=\left( \bigtriangledown _{c}\delta _{j}\right) \rfloor d^{i},\quad
C_{\bigtriangledown bc}^{a}=\left( \bigtriangledown _{c}\partial _{b}\right)
\rfloor \delta ^{a}.
\end{equation*}%
In explicit form, the components $L_{\bigtriangledown
jk}^{i},L_{\bigtriangledown bk}^{a},C_{\bigtriangledown jc}^{i}$ and $%
C_{\bigtriangledown bc}^{a}$ are defined by formula (\ref{christa}) if we
consider N--frame $\mathbf{e}_{\gamma }=\left( \delta _{i}=\partial
_{i}-N_{i}^{a}\partial _{a},\partial _{a}\right) $ and N--coframe $\mathbf{%
\vartheta }^{\beta }=(dx^{i},\delta y^{a}=dy^{a}+ N_{i}^{a}dx^{i})$ and a
d--metric $\mathbf{g}=\left( g_{ij,}h_{ab}\right) .$ In these formulas, we
write $\delta \mathbf{\vartheta }_{\alpha }$ instead of absolute
differentials $d\vartheta _{\alpha }$ from Refs. \cite{mag,rcg} because the
N--connection is considered. The coefficients (\ref{christa}) transforms
into the usual Levi--Civita (or Christoffel) ones for arbitrary anholonomic
frames $e_{\gamma }$ and $\vartheta ^{\beta }$ and for a metric
\begin{equation*}
g=g_{\alpha \beta }\vartheta ^{\alpha }\otimes \vartheta ^{\beta }
\end{equation*}%
if $\mathbf{e}_{\gamma }\rightarrow e_{\gamma },$ $\mathbf{\vartheta }%
^{\beta }\rightarrow \vartheta ^{\beta }$ and $\delta \mathbf{\vartheta }%
_{\beta }\rightarrow d\vartheta _{\beta }.$

Finally, we note that if the N--connection structure is not trivial, we can
define arbitrary vielbein transforms starting from $\mathbf{e}_{\gamma }$
and $\mathbf{\vartheta }^{\beta },$ i. e. $e_{\alpha }^{[N]}=A_{\alpha }^{\
\alpha ^{\prime }}(u)\mathbf{e}_{\alpha ^{\prime }}$ and $\vartheta
_{\lbrack N]}^{\beta }=A_{\ \beta ^{\prime }}^{\beta }(u)\mathbf{\vartheta }%
^{\beta ^{\prime }}$ (we put the label $[N]$ in order to emphasize that such
object were defined by vielbein transforms starting from certain N--adapted
frames). This way we develop a general anholonomic frame formalism adapted
to the prescribed N--connection structure. If we consider geometric objects
with respect to coordinate frames $\mathbf{e}_{\alpha ^{\prime }}\rightarrow
\partial _{\underline{\alpha }}=\partial /\partial u^{\underline{\alpha }}$
and coframes $\mathbf{\vartheta }^{\beta ^{\prime }}\rightarrow du^{%
\underline{\beta }},$ the N--connection strucuture is 'hidden' in the
off--diagonal metric coefficients (\ref{ansatz}) and performed geometric
constructions, in general, are not N--adapted.

\subsection{The canonical d--connection and the Levi--Civita connection}

The Levi--Civita connection $\bigtriangledown $ is constructed only from the
metric coefficients, being torsionless and satisfying the metricity
conditions $\bigtriangledown _{\alpha }g_{\beta \gamma }=0$. Because the
Levi--Civita connection is not adapted to the N--connection structure, we
can not state its coefficients in an irreducible form for the h-- and
v--subspaces. We need a type of d--connection which would be similar to the
Levi--Civita connection but satisfy certain metricity conditions adapted to
the N--connection.

\begin{proposition}
There are metric d--connections $\mathbf{D=}\left( D^{[h]},D^{[v]}\right) $
in a space \ $\mathbf{V}^{n+m},$ see (\ref{hvder}), satisfying the metricity
conditions if and only if
\begin{equation}
D_{k}^{[h]}g_{ij}=0,\ D_{a}^{[v]}g_{ij}=0,\ D_{k}^{[h]}h_{ab}=0,\
D_{a}^{[h]}h_{ab}=0.  \label{mcas}
\end{equation}
\end{proposition}

The general proof of existence of such metric d--connections on vector
(super) bundles is given in Ref. \cite{ma}. Here we note that the equations (%
\ref{mcas}) on $\mathbf{V}^{n+m}$ are just the conditions (\ref{mca}). In
our case the existence may be proved by constructing an explicit example:

\begin{definition}
The canonical d--connection $\widehat{\mathbf{D}}$\ \ $\mathbf{=}\left(
\widehat{D}^{[h]},\widehat{D}^{[v]}\right) ,$ equivalently $\widehat{\mathbf{%
\Gamma }}_{\ \alpha }^{\gamma }=\widehat{\mathbf{\Gamma }}_{\ \alpha \beta
}^{\gamma }\mathbf{\vartheta }^{\beta },$\ is defined by the h--
v--irreducible components $\widehat{\mathbf{\Gamma }}_{\ \alpha \beta
}^{\gamma }=\left( \widehat{L}_{jk}^{i},\widehat{L}_{bk}^{a},\widehat{C}%
_{jc}^{i},\widehat{C}_{bc}^{a}\right) ,$%
\begin{eqnarray}
\widehat{L}_{jk}^{i} &=&\frac{1}{2}g^{ir}\left( \frac{\delta g_{jk}}{\delta
x^{k}}+\frac{\delta g_{kr}}{\delta x^{j}}-\frac{\delta g_{jk}}{\delta x^{r}}%
\right) ,  \label{candcon} \\
\widehat{L}_{bk}^{a} &=&\frac{\partial N_{k}^{a}}{\partial y^{b}}+\frac{1}{2}%
h^{ac}\left( \frac{\delta h_{bc}}{\delta x^{k}}-\frac{\partial N_{k}^{d}}{%
\partial y^{b}}h_{dc}-\frac{\partial N_{k}^{d}}{\partial y^{c}}h_{db}\right)
,  \notag \\
\widehat{C}_{jc}^{i} &=&\frac{1}{2}g^{ik}\frac{\partial g_{jk}}{\partial
y^{c}},  \notag \\
\widehat{C}_{bc}^{a} &=&\frac{1}{2}h^{ad}\left( \frac{\partial h_{bd}}{%
\partial y^{c}}+\frac{\partial h_{cd}}{\partial y^{b}}-\frac{\partial h_{bc}%
}{\partial y^{d}}\right) .  \notag
\end{eqnarray}%
satisfying the torsionless conditions for the h--subspace and v--subspace,
respectively, $\widehat{T}_{jk}^{i}=\widehat{T}_{bc}^{a}=0.$
\end{definition}

By straightforward calculations with (\ref{candcon}) we can verify that the
conditions (\ref{mcas}) are satisfied and that the d--torsions are subjected
to the conditions $\widehat{T}_{jk}^{i}=\widehat{T}_{bc}^{a}=0$ (see section %
\ref{torscurv})). We emphasize that the canonical d--torsion posses
nonvanishing torsion components,%
\begin{equation*}
\widehat{T}_{.ji}^{a}=-\widehat{T}_{.ij}^{a}=\frac{\delta N_{i}^{a}}{\delta
x^{j}}-\frac{\delta N_{j}^{a}}{\delta x^{i}}=\Omega _{.ji}^{a},~\widehat{T}%
_{ja}^{i}=-\widehat{T}_{aj}^{i}=\widehat{C}_{.ja}^{i},~\widehat{T}%
_{.bi}^{a}=-\widehat{T}_{.ib}^{a}=\widehat{P}_{.bi}^{a}=\frac{\partial
N_{i}^{a}}{\partial y^{b}}-\widehat{L}_{.bj}^{a}
\end{equation*}%
induced by $\widehat{L}_{bk}^{a},$ $\widehat{C}_{jc}^{i}$ and N--connection
coefficients $N_{i}^{a}$ and their partial derivatives $\partial
N_{i}^{a}/\partial y^{b}$ (as is to be computed by introducing (\ref{candcon}%
) in formulas (\ref{dtorsb})). This is an anholonmic frame effect.

\begin{proposition}
\label{lccdc}The components of the Levi--Civita connection $\mathbf{\Gamma }%
_{\bigtriangledown \beta \gamma }^{\tau }$ and the irreducible components of
the canonical d--connection \ $\widehat{\mathbf{\Gamma }}_{\ \beta \gamma
}^{\tau }$\ are related by formulas%
\begin{equation}
\mathbf{\Gamma }_{\bigtriangledown \beta \gamma }^{\tau }=\left( \widehat{L}%
_{jk}^{i},\widehat{L}_{bk}^{a}-\frac{\partial N_{k}^{a}}{\partial y^{b}},%
\widehat{C}_{jc}^{i}+\frac{1}{2}g^{ik}\Omega _{jk}^{a}h_{ca},\widehat{C}%
_{bc}^{a}\right) ,  \label{lcsyma}
\end{equation}%
where $\Omega _{jk}^{a}$\ \ is the N--connection curvature\ (\ref{ncurv}).
\end{proposition}

The proof follows from an explicit decomposition of N--adapted frame (\ref%
{dder}) and N--adapted coframe (\ref{ddif}) in (\ref{lcsym}) (equivalently,
in (\ref{christa})) and regroupation of the components as to distinguish the
h- and v-- irreducible values (\ref{candcon}) for $\mathbf{g}_{\alpha \beta
}=\left( g_{ij},h_{ab}\right) .$

We conclude from (\ref{lcsyma}) that, in a trivial case, the Levi--Civita
and the canonical d--connection are given by the same h-- v-- components $%
\left( \widehat{L}_{jk}^{i},\widehat{L}_{bk}^{a},\widehat{C}_{jc}^{i},%
\widehat{C}_{bc}^{a}\right) $ if $\Omega _{jk}^{a}=0,\,$\ and $\partial
N_{k}^{a}/\partial y^{b}=0.$ This results in zero anholonomy coefficients (%
\ref{anhc}) when the anholonomic N--basis is reduced to a holonomic one. It
should be also noted that even in this case some components of the
anholonomically induced by d--connection torsion $\widehat{\mathbf{T}}%
_{\beta \gamma }^{\alpha }$ could be nonzero (see formulas (\ref{dtorsions})
just for $\widehat{\mathbf{\Gamma }}_{\ \beta \gamma }^{\tau }).$ For
instance, one holds the

\begin{corollary}
\label{ctlc}The d--tensor components
\begin{equation}
\widehat{T}_{.bi}^{a}=-\widehat{T}_{.ib}^{a}=\widehat{P}_{.bi}^{a}=\frac{%
\partial N_{i}^{a}}{\partial y^{b}}-\widehat{L}_{.bj}^{a}  \label{indtors}
\end{equation}%
for a canonical d--connection (\ref{candcon}) can be nonzero even $\partial
N_{k}^{a}/\partial y^{b}=0$ and $\Omega _{jk}^{a}=0$ and a trivial equality
of the components of the canonical d--connection and of the Levi--Civita
connection, $\mathbf{\Gamma }_{\bigtriangledown \beta \gamma }^{\tau }=$ $%
\widehat{\mathbf{\Gamma }}_{\ \beta \gamma }^{\tau }$ holds just with
respect to coordinate frames.
\end{corollary}

This quite surprising fact follows from the anholonomic character of the
N--connection structure. If a N--connection is defined, there are imposed
specific types of constraints on the frame structure. This is important for
definition of d--connections (being adapted to the N--connection structure)
but not for the Levi--Civita connection which is not a d--connection. Even
such linear connections have the same components with respect to a
N--adapted (co) frame, they are very different geometrical objects because
they are subjected to different rules of transformation with respect to
frame and coordinate transforms. The d--connections' transforms are adapted
to those for the N--connection (\ref{ncontr}) but the Levi--Civita
connection is subjected to general rules of linear connection transforms (%
\ref{lcontr}).\footnote{%
The Corollary \ref{ctlc} is important for constructing various classes of
exact solutions with generic off--diagonal metrics in Einstein gravity, its
higher dimension and/or different gauge, Einstein-\--Cartan and
metric--affine generalizations. Certain type of ansatz were proven to result
in completely integrable gravitational field equations just for the
canonical d--connection (but not for the Levi--Civita one), see details in
Refs. \cite{v1,v2,vd,vncggf}. The induced d--torsion (\ref{indtors}) is
contained in the Ricci d--tensor $R_{ai}=\ ^{1}P_{ai}=P_{a.ib}^{.b},$ see (%
\ref{dricci}), i. e. in the Einstein d--tensor constructed for the canonical
d--connection. If a class of solutions were obtained for a d--connection, we
can select those subclasses which satisfy the \ condition $\mathbf{\Gamma }%
_{\bigtriangledown \beta \gamma }^{\tau }=$ $\widehat{\mathbf{\Gamma }}_{\
\beta \gamma }^{\tau }$ with respect to a frame of reference. In this case
the nontrivial d--torsion $\widehat{T}_{.bi}^{a}$ (\ref{indtors}) can be
treated as an object constructed from some ''pieces'' of a generic
off--diagonal metric and related to certain components of the N--adapted
anholonomic frames.
\par
{}
\par
{}}

\begin{proposition}
\label{ptorsvlc}A canonical d--connection $\widehat{\mathbf{\Gamma }}_{\
\beta \gamma }^{\tau }$ defined by a N--connection $N_{i}^{a}$ and d--metric
$\mathbf{g}_{\alpha \beta }=\left[ g_{ij},h_{ab}\right] $ has zero
d--torsions \ (\ref{dtorsions}) if an only if there are satisfied the
conditions $\Omega _{jk}^{a}=0,\widehat{C}_{jc}^{i}=0$ and $\widehat{L}%
_{.bj}^{a}=\partial N_{i}^{a}/\partial y^{b},$ i. e. $\widehat{\mathbf{%
\Gamma }}_{\ \beta \gamma }^{\tau }=\left( \widehat{L}_{jk}^{i},\widehat{L}%
_{bk}^{a}=\partial N_{i}^{a}/\partial y^{b},0,\widehat{C}_{bc}^{a}\right) $
which is equivalent to
\begin{eqnarray}
g^{ik}\frac{\partial g_{jk}}{\partial y^{c}} &=&0,\   \label{cond01} \\
\frac{\delta h_{bc}}{\delta x^{k}}-\frac{\partial N_{k}^{d}}{\partial y^{b}}%
h_{dc}-\frac{\partial N_{k}^{d}}{\partial y^{c}}h_{db} &=&0,  \label{cond02}
\\
\frac{\partial N_{i}^{a}}{\partial x^{j}}-\frac{\partial N_{j}^{a}}{\partial
x^{i}}+N_{i}^{b}\frac{\partial N_{j}^{a}}{\partial y^{b}}-N_{j}^{b}\frac{%
\partial N_{i}^{a}}{\partial y^{b}} &=&0.  \label{cond03}
\end{eqnarray}%
The Levi--Civita connection defined by the same N--connection and d--metric
structure with respect to N--adapted (co) frames has the components $^{[0]}%
\mathbf{\Gamma }_{\bigtriangledown \beta \gamma }^{\tau }=$ $\widehat{%
\mathbf{\Gamma }}_{\ \beta \gamma }^{\tau }=\left( \widehat{L}_{jk}^{i},0,0,%
\widehat{C}_{bc}^{a}\right) .$
\end{proposition}

\textbf{Proof:} The relations (\ref{cond01})--(\ref{cond03}) follows from
the condition of vanishing of d--torsion coefficients (\ref{dtorsions}) when
the coefficients of the canonical d--connection and the Levi--Civita
connection are computed respectively following formulas (\ref{candcon}) and (%
\ref{lcsyma})

We note a specific separation of variables in the equations (\ref{cond01})--(%
\ref{cond03}). For instance, the equation (\ref{cond01}) is satisfied by any
$g_{ij}=g_{ij}\left( x^{k}\right) .$ We can search a subclass of
N--connections with $N_{j}^{a}=\delta _{j}N^{a}$, i. e. of 1--forms on the
h--subspace, $\widetilde{N}^{a}=\delta _{j}N^{a}dx^{i}$ which are closed on
this subspace,
\begin{equation*}
\delta \widetilde{N}^{a}=\frac{1}{2}\left( \frac{\partial N_{i}^{a}}{%
\partial x^{j}}-\frac{\partial N_{j}^{a}}{\partial x^{i}}+N_{i}^{b}\frac{%
\partial N_{j}^{a}}{\partial y^{b}}-N_{j}^{b}\frac{\partial N_{i}^{a}}{%
\partial y^{b}}\right) dx^{i}\wedge dx^{j}=0,
\end{equation*}%
satisfying the (\ref{cond03}). Having defined such $N_{i}^{a}$ and computing
the values $\partial _{c}N_{i}^{a},$ we may try to solve (\ref{cond02})
rewritten as a system of first order partial differential equations
\begin{equation*}
\frac{\partial h_{bc}}{\partial x^{k}}=N_{k}^{e}\frac{\partial h_{bc}}{%
\partial y^{e}}+\partial _{b}N_{k}^{d}\ h_{dc}+\partial _{c}N_{k}^{d}h_{db}
\end{equation*}%
with known coefficients.$\blacksquare $

We can also associate the nontrivial values of $\widehat{\mathbf{T}}_{\beta
\gamma }^{\tau }$ (in particular cases, of $\widehat{T}_{.bi}^{a})$ to be
related to any algebraic equations in the Einstein--Cartan theory or
dynamical equations for torsion like in string or supergravity models. But
in this case we shall prescribe a specific class of anholonomically
constrained dynamics for the N--adapted frames.

Finally, we note that if a (pseudo) Riemannian space is provided with a
generic off--diagonal metric structure (see Remark \ref{rgod}) we can
consider alternatively to the Levi--Civita connection an infinite number of
metric d--connections, details in the section \ref{srgarcg}. Such
d--connections have nontrivial d--torsions $\mathbf{T}_{\beta \gamma }^{\tau
}$ induced by anholonomic frames and constructed from off--diagonal metric
terms and h- and v--components of d--metrics.

\subsection{The set of metric d--connections}

Let us define the set of all possible metric d--connections, satisfying the
conditions (\ref{mcas}) and being constructed only form $g_{ij},h_{ab}$ and $%
N_{i}^{a}$ and their partial derivatives. Such d--connections satisfy
certain conditions$\ $for d--torsions that$\ T_{~jk}^{i}=T_{~bc}^{a}=0$ and
can be generated by two procedures of deformation of the connection
\begin{eqnarray*}
\widehat{\mathbf{\Gamma }}_{\ \alpha \beta }^{\gamma } &\rightarrow &\ ^{[K]}%
\mathbf{\Gamma }_{\ \alpha \beta }^{\gamma }=\mathbf{\Gamma }_{\ \alpha
\beta }^{\gamma }+\ ^{[K]}\mathbf{Z}_{\ \alpha \beta }^{\gamma }%
\mbox{\
(Kawaguchi's metrization \cite{kaw}) }, \\
\mbox{ or } &\rightarrow &^{[M]}\mathbf{\Gamma }_{\ \alpha \beta }^{\gamma }=%
\widehat{\mathbf{\Gamma }}_{\ \alpha \beta }^{\gamma }+\ ^{[M]}\mathbf{Z}_{\
\alpha \beta }^{\gamma }\mbox{\ (Miron's
connections \cite{ma} )}.
\end{eqnarray*}

\begin{theorem}
\ \label{kmp}Every deformation d--tensor (equivalently, distorsion, or
deflection) \
\begin{eqnarray*}
^{\lbrack K]}\mathbf{Z}_{\ \alpha \beta }^{\gamma } &=&\{\ ^{[K]}Z_{\
jk}^{i}=\frac{1}{2}g^{im}D_{j}^{[h]}g_{mk},\ ^{[K]}Z_{\ bk}^{a}=\frac{1}{2}%
h^{ac}D_{k}^{[h]}h_{cb},\  \\
&&\ ^{[K]}Z_{\ ja}^{i}=\frac{1}{2}g^{im}D_{a}^{[v]}g_{mj},\ ^{[K]}Z_{\
bc}^{a}=\frac{1}{2}h^{ad}D_{c}^{[v]}h_{db}\}
\end{eqnarray*}%
\ transforms a d--connection $\mathbf{\Gamma }_{\ \alpha \beta }^{\gamma
}=\left( L_{jk}^{i},L_{bk}^{a},C_{jc}^{i},C_{bc}^{a}\right) $\ (\ref{dcon1})
into a metric d--connection%
\begin{equation*}
\ ^{[K]}\mathbf{\Gamma }_{\ \alpha \beta }^{\gamma }=\left( L_{jk}^{i}+\
^{[K]}Z_{\ jk}^{i},L_{bk}^{a}+\ ^{[K]}Z_{\ bk}^{a},C_{jc}^{i}+\ ^{[K]}Z_{\
ja}^{i},C_{bc}^{a}+\ ^{[K]}Z_{\ bc}^{a}\right) .
\end{equation*}%
\
\end{theorem}

The proof consists from a straightforward verification which demonstrate
that the conditions (\ref{mcas}) are satisfied on $\mathbf{V}^{n+m}$ for $\
^{[K]}\mathbf{D=\{^{[K]}\mathbf{\Gamma }_{\alpha \beta }^{\gamma }\}}$ and $%
\mathbf{g}_{\alpha \beta }=\left( g_{ij},h_{ab}\right) .$ We note that the
Kawaguchi's metrization procedure contains additional covariant derivations
of the d--metric coefficients, defined by arbitrary d--connection, not only
N--adapted derivatives of the d--metric and N--connection coefficients as in
the case of the canonical d--connection.

\begin{theorem}
\label{mconnections}For a fixed d--metric structure \ (\ref{block2}),\ $%
\mathbf{g}_{\alpha \beta }=\left( g_{ij},h_{ab}\right) ,$ on a space $%
\mathbf{V}^{n+m},$ the set of metric d--connections \ \ $^{[M]}\mathbf{%
\Gamma }_{\ \alpha \beta }^{\gamma }=\widehat{\mathbf{\Gamma }}_{\ \alpha
\beta }^{\gamma }+\ ^{[M]}\mathbf{Z}_{\ \alpha \beta }^{\gamma }$\ \ \ is
defined by the deformation d--tensor \
\begin{eqnarray*}
^{\lbrack M]}\mathbf{Z}_{\alpha \beta }^{\gamma } &=&\{\ ^{[M]}Z_{\
jk}^{i}=\ ^{[-]}O_{km}^{li}Y_{lj}^{m},\ ^{[M]}Z_{\ bk}^{a}=\
^{[-]}O_{bd}^{ea}Y_{ej}^{m},\  \\
&&\ ^{[M]}Z_{\ ja}^{i}=\ ^{[+]}O_{jk}^{mi}Y_{mc}^{k},\ ^{[M]}Z_{\ bc}^{a}=\
^{[+]}O_{bd}^{ea}Y_{ec}^{d}\}
\end{eqnarray*}%
where the so--called Obata operators are defined
\begin{equation*}
\ ^{[\pm ]}O_{km}^{li}=\frac{1}{2}\left( \delta _{k}^{l}\delta _{m}^{i}\pm
g_{km}g^{li}\right) \mbox{ and }\ ^{[\pm ]}O_{bd}^{ea}=\frac{1}{2}\left(
\delta _{b}^{e}\delta _{d}^{a}\pm h_{bd}h^{ea}\right)
\end{equation*}%
and \ $Y_{lj}^{m},$\ $Y_{ej}^{m},Y_{mc}^{k},$\ $Y_{ec}^{d}$ are arbitrary
d--tensor fields.
\end{theorem}

The proof consists from a direct verification of the fact that the
conditions (\ref{mcas}) are satisfied on $\mathbf{V}^{n+m}$ for $\ ^{[M]}%
\mathbf{D=\{^{[M]}\mathbf{\Gamma }_{\ \alpha \beta }^{\gamma }\}.}$ We note
that the relation (\ref{lcsyma}) \ between the Levi--Civita and the
canonical d--connection is a particular case of $^{[M]}\mathbf{Z}_{\alpha
\beta }^{\gamma },$ when $Y_{lj}^{m},$\ $Y_{ej}^{m}$ and $Y_{ec}^{d}$ are
zero, but $Y_{mc}^{k}$ is taken to have $\ ^{[+]}O_{jk}^{mi}Y_{mc}^{k}=\frac{%
1}{2}g^{ik}\Omega _{jk}^{a}h_{ca}.$

There is a very important consequence of the Theorems \ref{kmp} and \ref%
{mconnections}: For a generic off--diagonal metric structure (\ref{ansatz})
we can derive a N--connection structure $N_{i}^{a}$ with a d--metric $%
\mathbf{g}_{\alpha \beta }=\left( g_{ij},h_{ab}\right) $ (\ref{block2}). So,
we may consider an infinite number of d--connections $\{\mathbf{D\},}$ all
constructed from the coefficients of the off--diagonal metrics, satisfying
the metricity conditions $\mathbf{D}_{\gamma }\mathbf{g}_{\alpha \beta }=0$
and having partial vanishing torsions, $T_{jk}^{i}=T_{bc}^{a}=0.$ The
covariant calculi associated to the set $\{\mathbf{D\}}$ are adapted to the
N--connection splitting and alternative to the covariant calculus defined by
the Levi--Civita connection $\bigtriangledown ,$ which is not adapted to the
N--connection.

\subsection{Nonmetricity in Finsler Geometry}

Usually, the N--connection, d--connection and d--metric in generalized
Finsler spaces satisfy certain metric compatibility conditions \cite%
{ma,fg,rund,as}. Nevertheless, there were considered some classes of
d--connections (for instance, related to the Berwald d--connection) with
nontrivial components of the nonmetricity d--tensor. Let us consider some
such examples modeled on metric--affine spaces.

\subsubsection{The Berwald d--connection}

\label{berwdcon}A d--connection of Berwald type (see, for instance, Ref. %
\cite{ma} on such configurations in Finsler and Lagrange geometry), $\ ^{[B]}%
\mathbf{\Gamma }_{\ \alpha }^{\gamma }=\ ^{[B]}\widehat{\mathbf{\Gamma }}_{\
\alpha \beta }^{\gamma }\mathbf{\vartheta }^{\beta },$ is defined by h- and
v--irreducible components
\begin{equation}
\ \ ^{[B]}\mathbf{\Gamma }_{\alpha \beta }^{\gamma }=\left( \widehat{L}_{\
jk}^{i},\frac{\partial N_{k}^{a}}{\partial y^{b}},0,\widehat{C}_{\
bc}^{a}\right) ,  \label{berw}
\end{equation}%
with $\widehat{L}_{\ jk}^{i}$ and $\widehat{C}_{\ bc}^{a}$ taken as in (\ref%
{candcon}), satisfying only partial metricity compatibility conditions for a
d--metric \ (\ref{block2}),\ $\mathbf{g}_{\alpha \beta }=\left(
g_{ij},h_{ab}\right) $ on space $\mathbf{V}^{n+m}$
\begin{equation*}
\ ^{[B]}D_{k}^{[h]}g_{ij}=0\mbox{ and }\ ^{[B]}D_{c}^{[v]}h_{ab}=0.
\end{equation*}%
This is an example of d--connections which may possess nontrivial
nonmetricity components, $\ ^{[B]}\mathbf{Q}_{\alpha \beta \gamma }=\left(
^{[B]}Q_{cij},\ ^{[B]}Q_{iab}\right) $ with
\begin{equation}
\ ^{[B]}Q_{cij}=\ ^{[B]}D_{c}^{[v]}g_{ij}\mbox{ and }\ ^{[B]}Q_{iab}=\
^{[B]}D_{i}^{[h]}h_{ab}.  \label{berwnm}
\end{equation}%
So, the Berwald d--connection defines a metric--affine space $\mathbf{V}%
^{n+m}$ with N--connection structure.

If $\widehat{L}_{\ jk}^{i}=0$ and $\widehat{C}_{\ bc}^{a}=0,$ we obtain a
Berwald type connection $\ $%
\begin{equation*}
^{\lbrack N]}\mathbf{\Gamma }_{\alpha \beta }^{\gamma }=\left( 0,\frac{%
\partial N_{k}^{a}}{\partial y^{b}},0,0\right)
\end{equation*}%
induced by the N--connection structures. It defines a vertical covariant
derivation $^{[N]}D_{c}^{[v]}$ acting in the v--subspace of $\mathbf{V}%
^{n+m},$ with the coefficients being partial derivatives on v--coordinates $%
y^{a}$ of the N--connection coefficients $N_{i}^{a}$ \cite{vilms}.

We can generalize the Berwald connection (\ref{berw}) to contain any fixed
values of d--torsions $T_{.jk}^{i}$ and $T_{.bc}^{a}$ from the h-
v--decomposition (\ref{dtorsions}). We can check by a straightforward
calculations that the d--connection%
\begin{equation}
^{\lbrack B\tau ]}\mathbf{\Gamma }_{\ \alpha \beta }^{\gamma }=\left(
\widehat{L}_{\ jk}^{i}+\tau _{\ jk}^{i},\frac{\partial N_{k}^{a}}{\partial
y^{b}},0,\widehat{C}_{\ bc}^{a}+\tau _{\ bc}^{a}\right)  \label{bct}
\end{equation}%
with
\begin{eqnarray}
\tau _{\ jk}^{i} &=&\frac{1}{2}g^{il}\left(
g_{kh}T_{.lj}^{h}+g_{jh}T_{.lk}^{h}-g_{lh}T_{\ jk}^{h}\right)
\label{tauformulas} \\
\tau _{\ bc}^{a} &=&\frac{1}{2}h^{ad}\left( h_{bf}T_{\ dc}^{f}+h_{cf}T_{\
db}^{f}-h_{df}T_{\ bc}^{f}\right)  \notag
\end{eqnarray}%
results in $^{[B\tau ]}\mathbf{T}_{jk}^{i}=T_{.jk}^{i}$ and $^{[B\tau ]}%
\mathbf{T}_{bc}^{a}=T_{.bc}^{a}.$ The d--connection (\ref{bct}) has certain
nonvanishing irreducible nonmetricity components $^{[B\tau ]}\mathbf{Q}%
_{\alpha \beta \gamma }=\left( ^{[B\tau ]}Q_{cij},\ ^{[B\tau
]}Q_{iab}\right) .$

In general, by using the Kawaguchi metrization procedure (see Theorem \ref%
{kmp}) we can also construct metric d--connections with prescribed values of
d--torsions $T_{.jk}^{i}$ and $T_{.bc}^{a},$ or to express, for instance,
the Levi--Civita connection via coefficients of an arbitrary metric
d--connection (see details, for vector bundles, in \cite{ma}).

Similarly to formulas (\ref{acn}), (\ref{accn}) and (\ref{distan}), we can
express a general affine Berwald d--connection $\ ^{[B\tau ]}\mathbf{D,}$ i.
e. $\ ^{[B\tau ]}\mathbf{\Gamma }_{\ \ \alpha }^{\gamma }=\ ^{[B\tau ]}%
\mathbf{\Gamma }_{\ \alpha \beta }^{\gamma }\mathbf{\vartheta }^{\beta },$
via its deformations from the Levi--Civita connection $\mathbf{\Gamma }%
_{\bigtriangledown \ \beta }^{\alpha },$
\begin{equation}
\ ^{[B\tau ]}\mathbf{\Gamma }_{\ \beta }^{\alpha }=\mathbf{\Gamma }%
_{\bigtriangledown \ \beta }^{\alpha }+\ ^{[B\tau ]}\mathbf{Z}_{\ \ \beta
}^{\alpha },  \label{accnb}
\end{equation}%
$\mathbf{\Gamma }_{\bigtriangledown \ \beta }^{\alpha }$ being expressed as (%
\ref{christa}) (equivalently, defined by (\ref{lcsym})) and
\begin{eqnarray}
\ ^{[B\tau ]}\mathbf{Z}_{\alpha \beta } &=&\mathbf{e}_{\beta }\rfloor \
^{[B\tau ]}\mathbf{T}_{\alpha }-\mathbf{e}_{\alpha }\rfloor \ ^{[B\tau ]}%
\mathbf{T}_{\beta }+\frac{1}{2}\left( \mathbf{e}_{\alpha }\rfloor \mathbf{e}%
_{\beta }\rfloor \ ^{[B\tau ]}\mathbf{T}_{\gamma }\right) \mathbf{\vartheta }%
^{\gamma }  \label{distanb} \\
&&+\left( \mathbf{e}_{\alpha }\rfloor \ ^{[B\tau ]}\mathbf{Q}_{\beta \gamma
}\right) \mathbf{\vartheta }^{\gamma }-\left( \mathbf{e}_{\beta }\rfloor \
^{[B\tau ]}\mathbf{Q}_{\alpha \gamma }\right) \mathbf{\vartheta }^{\gamma }+%
\frac{1}{2}\ ^{[B\tau ]}\mathbf{Q}_{\alpha \beta }.  \notag
\end{eqnarray}%
defined with prescribed d--torsions $^{[B\tau ]}\mathbf{T}%
_{jk}^{i}=T_{.jk}^{i}$ and $^{[B\tau ]}\mathbf{T}_{bc}^{a}=T_{.bc}^{a}.$
This Berwald d--connection can define a particular subclass of
metric--affine connections being adapted to the N--connection structure and
with prescribed values of d--torsions.

\subsubsection{The canonical/ Berwald metric--affine d--connections}

\label{berwdcona}If the deformations of d--metrics in formulas (\ref{accn}) $%
\ $and (\ref{accnb}) are considered not with respect to the Levi--Civita
connection $\mathbf{\Gamma }_{\bigtriangledown \ \beta }^{\alpha }$ but with
respect to the canonical d--connection $\widehat{\mathbf{\Gamma }}_{\ \alpha
\beta }^{\gamma }$ with h- v--irreducible coefficients (\ref{candcon}), we
can construct a set of canonical metric--affine d--connections. Such
metric--affine d--connections $\mathbf{\Gamma }_{\ \ \alpha }^{\gamma }=\
\mathbf{\Gamma }_{\ \alpha \beta }^{\gamma }\mathbf{\vartheta }^{\beta }$
are defined via deformations
\begin{equation}
\ \mathbf{\Gamma }_{\ \beta }^{\alpha }=\widehat{\mathbf{\Gamma }}_{\ \beta
}^{\alpha }+\ \widehat{\mathbf{Z}}_{\ \ \beta }^{\alpha },  \label{mafdc}
\end{equation}%
$\widehat{\mathbf{\Gamma }}_{\ \beta }^{\alpha }$ being the canonical
d--connection (\ref{dcon1}) and
\begin{eqnarray}
\ \widehat{\mathbf{Z}}_{\alpha \beta } &=&\mathbf{e}_{\beta }\rfloor \
\mathbf{T}_{\alpha }-\mathbf{e}_{\alpha }\rfloor \ \mathbf{T}_{\beta }+\frac{%
1}{2}\left( \mathbf{e}_{\alpha }\rfloor \mathbf{e}_{\beta }\rfloor \ \mathbf{%
T}_{\gamma }\right) \mathbf{\vartheta }^{\gamma }  \label{dmafdc} \\
&&+\left( \mathbf{e}_{\alpha }\rfloor \ ^{[B\tau ]}\mathbf{Q}_{\beta \gamma
}\right) \mathbf{\vartheta }^{\gamma }-\left( \mathbf{e}_{\beta }\rfloor \
\mathbf{Q}_{\alpha \gamma }\right) \mathbf{\vartheta }^{\gamma }+\frac{1}{2}%
\ ^{[B\tau ]}\mathbf{Q}_{\alpha \beta }  \notag
\end{eqnarray}%
where $\mathbf{T}_{\alpha }$ and $\mathbf{Q}_{\alpha \beta }$ are arbitrary
torsion and nonmetricity structures.

A metric--affine d--connection $\mathbf{\Gamma }_{\ \ \alpha }^{\gamma }$
can be also considered as a deformation from the Berwald connection $%
^{[B\tau ]}\mathbf{\Gamma }_{\ \alpha \beta }^{\gamma }$
\begin{equation}
\ \mathbf{\Gamma }_{\ \beta }^{\alpha }=\ ^{[B\tau ]}\mathbf{\Gamma }_{\
\alpha \beta }^{\gamma }+\ ^{[B\tau ]}\ \widehat{\mathbf{Z}}_{\ \ \beta
}^{\alpha },  \label{mafbc}
\end{equation}%
$\ ^{[B\tau ]}\mathbf{\Gamma }_{\ \alpha \beta }^{\gamma }$ being the
Berwald d--connection (\ref{bct}) and
\begin{eqnarray}
\ ^{[B\tau ]}\ \widehat{\mathbf{Z}}_{\ \ \beta }^{\alpha } &=&\mathbf{e}%
_{\beta }\rfloor \ \mathbf{T}_{\alpha }-\mathbf{e}_{\alpha }\rfloor \
\mathbf{T}_{\beta }+\frac{1}{2}\left( \mathbf{e}_{\alpha }\rfloor \mathbf{e}%
_{\beta }\rfloor \ \mathbf{T}_{\gamma }\right) \mathbf{\vartheta }^{\gamma }
\label{dmafbc} \\
&&+\left( \mathbf{e}_{\alpha }\rfloor \ ^{[B\tau ]}\mathbf{Q}_{\beta \gamma
}\right) \mathbf{\vartheta }^{\gamma }-\left( \mathbf{e}_{\beta }\rfloor \
\mathbf{Q}_{\alpha \gamma }\right) \mathbf{\vartheta }^{\gamma }+\frac{1}{2}%
\ ^{[B\tau ]}\mathbf{Q}_{\alpha \beta }  \notag
\end{eqnarray}

The h- and v--splitting of formulas can be computed by introducing N--frames
$\mathbf{e}_{\gamma }=\left( \delta _{i}=\partial _{i}-N_{i}^{a}\partial
_{a},\partial _{a}\right) $ and N--coframes $\mathbf{\vartheta }^{\beta
}=\left( dx^{i},\delta y^{a}=dy^{a}+N_{i}^{a}dx^{i}\right) $ and d--metric $%
\mathbf{g}=\left( g_{ij,}h_{ab}\right) $ into (\ref{christa}), (\ref{accnb})
and (\ref{distanb}) for the general Berwald d--connections. In a similar
form we can compute splitting by introducing the N--frames and d--metric
into \ (\ref{dcon1}), (\ref{mafdc}) and (\ref{dmafdc}) \ for the metric
affine canonic d--connections and, respectively, into (\ref{bct}), (\ref%
{mafbc}) and (\ref{dmafbc}) for the metric--affine Berwald d--connections.
For the corresponding classes of d--connections, we can compute the torsion
and curvature tensors by introducing respective connections (\ref{christa}),
(\ref{accn}), (\ref{candcon}), (\ref{berw}), (\ref{bct}), (\ref{accnb}), (%
\ref{mafdc}) and (\ref{mafbc}) into the general formulas for torsion (\ref%
{dt}) and curvature (\ref{dc}) on spaces provided with N--connection
structure.

\subsection{N--connections in metric--affine spaces}

In order to elaborate a unified MAG and generalized Finsler spaces scheme,
it is necessary to explain how the N--connection emerge in a metric--affine
space and/or in more particular cases of Riemann--Cartan and (pseudo)
Riemann geometry.

\subsubsection{Riemann geometry as a Riemann--Cartan geometry with
N--connection}

\label{srgarcg}It is well known \ the interpretation of the Riemann--Cartan
geometry as a generalization of the Riemannian geometry by distorsions (of
the Levi--Civita connection) generated by the torsion tensors \cite{rcg}.
Usually, the Riemann--Cartan geometry is described by certain geometric
relations between the torsion tensor, curvature tensor, metric and the
Levi--Civita connection on effective Riemann spaces. We can establish new
relations between the Riemann and Riemann--Cartan geometry if generic
off--diagonal metrics and anholonomic frames of reference are introduced
into consideration. Roughly speaking, a generic off--diagonal metric induces
alternatively to the well known Riemann spaces a certain class of
Riemann--Cartan geometries, with torsions completely defined by
off--diagonal metric terms and related anholonomic frame structures.

\begin{theorem}
\label{teqrgrcg} Any (pseudo) Riemannian spacetime provided with a generic
off--diagonal metric, defining the torsionless and metric Levi--Civita
connection, can be equivalently modeled as a Riemann--Cartan spacetime
provided with a canonical d--connection adapted to N--connection structure.
\end{theorem}

\textbf{Proof:}

Let us consider how the data for a (pseudo) Riemannian generic off--diagonal
metric $g_{\alpha \beta }$ parametrized in the form (\ref{ansatz}) can
generate a Riemann--Cartan geometry. It is supposed that with respect to any
convenient anholonomic coframes (\ref{ddif}) the metric is transformed into
a diagonalized form of type (\ref{block2}), which gives the possibility to
define $N_{i}^{a}$ and $\mathbf{g}_{\alpha \beta }=\left[ g_{ij},h_{ab}%
\right] $ and to compute the aholonomy coefficients $\mathbf{w}_{\ \alpha
\beta }^{\gamma }$ (\ref{anhc}) and the components of the canonical
d--connection $\widehat{\mathbf{\Gamma }}_{\ \alpha \beta }^{\gamma }=\left(
\widehat{L}_{jk}^{i},\widehat{L}_{bk}^{a},\widehat{C}_{jc}^{i},\widehat{C}%
_{bc}^{a}\right) $ (\ref{candcon}). This connection has nontrivial
d--torsions\ $\widehat{\mathbf{T}}_{.\beta \gamma }^{\alpha },$ see the
Theorem \ref{tdtors} and Corollary \ref{ctlc}. In general, such d--torsions
are not zero being induced by the values $N_{i}^{a}$ and their partial
derivatives, contained in the former off--diagonal components of the metric (%
\ref{ansatz}). So, the former Riemannian geometry, with respect to
anholonomic frames with associated N--connection structure, is equivalently
rewritten in terms of a Riemann--Cartan geometry with nontrivial torsion
structure.

We can provide an inverse construction when a diagonal d--metric (\ref%
{block2}) is given with respect to an anholonomic coframe (\ref{ddif})
defined from nontrivial values of N--connection coefficients, $N_{i}^{a}.$
The related Riemann--Cartan geometry is defined by the canonical
d--connection $\widehat{\mathbf{\Gamma }}_{\ \alpha \beta }^{\gamma }$
possessing nontrivial d--torsions $\widehat{\mathbf{T}}_{.\beta \gamma
}^{\alpha }.$ The data for this geometry with N--connection and torsion can
be directly transformed [even with respect to the same N--adapted (co)
frames] into the data of related (pseudo) Riemannian geometry by using the
relation (\ref{lcsyma}) between the components of $\widehat{\mathbf{\Gamma }}%
_{\ \alpha \beta }^{\gamma }$ and of the Levi--Civita connection $\mathbf{%
\Gamma }_{\bigtriangledown \beta \gamma }^{\tau }.\blacksquare $

\begin{remark}
{\ } \newline
a) Any generic off--diagonal (pseudo) Riemannian metric $g_{\alpha \beta
}[N_{i}^{a}]\rightarrow $ $\mathbf{g}_{\alpha \beta }=\left[ g_{ij},h_{ab}%
\right] $ induces an infinite number of associated Riemann--Cartan
geometries defined by sets of d--connections $\mathbf{D}=\{\mathbf{\Gamma }%
_{\ \alpha \beta }^{\gamma }\}$ which can be constructed according the
Kawaguchi's and, respectively, Miron's Theorems \ref{kmp} and \ref%
{mconnections}.{\ } \newline
b) For any metric d--connection $\mathbf{D}=\{\mathbf{\Gamma }_{\ \alpha
\beta }^{\gamma }\}$ induced by a generic off--diagonal metric (\ref{ansatz}%
), we can define alternatively to the standard (induced by the Levi--Civita
connection) the Ricci d--tensor (\ref{dricci}), $\mathbf{R}_{\alpha \beta },$
and the Einstein d--tensor (\ref{deinst}), $\mathbf{G}_{\alpha \beta }.$
\end{remark}

We emphasize that all Riemann--Cartan geometries induced by metric
d--connections $\mathbf{D}$ are characterized not only by nontrivial induced
torsions $\mathbf{T}_{.\beta \gamma }^{\alpha }$ but also by corresponding
nonsymmetric Ricci d--tensor, $\mathbf{R}_{\alpha \beta },$ and Einstein
d--tensor, $\mathbf{G}_{\alpha \beta },$ for which $\mathbf{D}_{\gamma }%
\mathbf{G}_{\alpha \beta }\neq 0.$ This is not a surprising fact, because we
transferred the geometrical and physical objects on anholonomic spaces, when
the conservation laws should be redefined as to include the anholonomically
imposed constraints.

Finally, we conclude that for any generic off--diagonal (pseudo)\ Riemannian
metric we have two alternatives:\ 1) to choose the approach defined by the
Levi--Civita connection $\bigtriangledown $, with vanishing torsion and
usually defined conservation laws $\mathbf{\bigtriangledown }_{\gamma }%
\mathbf{G}_{\alpha \beta }^{[\bigtriangledown ]}=0,$ or 2) to diagonalize
the metric effectively, by respective anholonomic transforms, and transfer
the geometric and physical objects into effective Riemann--Cartan geometries
defined by corresponding N--connection and d--connection structures. All
types of such geometric constructions are equivalent. Nevertheless, one
could be defined certain priorities for some physical models like
''simplicity'' of field equations and definition of conservation laws and/or
the possibility to construct exact solutions. We note also that a variant
with induced torsions is more appropriate for including in the scheme
various type of generalized Finsler structures and/or models of (super)
string gravity containing nontrivial torsion fields.

\subsubsection{Metric--affine geometry and N--connections}

A general affine (linear) connection $D=\bigtriangledown +Z=\{\Gamma _{\beta
\alpha }^{\gamma }=\Gamma _{\bigtriangledown \beta \alpha }^{\gamma
}+Z_{\beta \alpha }^{\gamma }\}$
\begin{equation}
\Gamma _{\ \alpha }^{\gamma }=\Gamma _{\alpha \beta }^{\gamma }\vartheta
^{\beta },  \label{ac}
\end{equation}%
can always be decomposed into the Riemannian $\Gamma _{\bigtriangledown \
\beta }^{\alpha }$ and post--Riemannian $Z_{\ \beta }^{\alpha }$ parts (see
Refs. \cite{mag} and, for irreducible decompositions to the effective
Einstein theory, see Ref. \cite{oveh}),
\begin{equation}
\Gamma _{\ \beta }^{\alpha }=\Gamma _{\bigtriangledown \ \beta }^{\alpha
}+Z_{\ \beta }^{\alpha }  \label{acc}
\end{equation}%
where the distorsion 1-form $Z_{\ \beta }^{\alpha }$ is expressed in terms
of torsion and nonmetricity,%
\begin{equation}
Z_{\alpha \beta }=e_{\beta }\rfloor T_{\alpha }-e_{\alpha }\rfloor T_{\beta
}+\frac{1}{2}\left( e_{\alpha }\rfloor e_{\beta }\rfloor T_{\gamma }\right)
\vartheta ^{\gamma }+\left( e_{\alpha }\rfloor Q_{\beta \gamma }\right)
\vartheta ^{\gamma }-\left( e_{\beta }\rfloor Q_{\alpha \gamma }\right)
\vartheta ^{\gamma }+\frac{1}{2}Q_{\alpha \beta },  \label{dista}
\end{equation}
$T_{\alpha }$ is defined as (\ref{dta}) and $Q_{\alpha \beta }\doteqdot
-Dg_{\alpha \beta }.$ \footnote{%
We note that our $\Gamma _{\ \alpha }^{\gamma }$ \ and $Z_{\ \beta }^{\alpha
}$ are respectively the $\Gamma _{\ \alpha }^{\gamma }$ and $N_{\alpha \beta
}$ from Ref. \cite{oveh}; in our works we use the symbol $N$ for
N--connections.}\ For $Q_{\beta \gamma }=0,$ we obtain from (\ref{dista})
just the distorsion for the Riemannian--Cartan geometry \cite{rcg}.

By substituting arbitrary (co) frames, metrics and linear connections into
N--adapted ones (i. e. performing changes
\begin{equation*}
e_{\alpha }\rightarrow \mathbf{e}_{\alpha },\vartheta ^{\beta }\rightarrow
\mathbf{\vartheta }^{\beta },g_{\alpha \beta }\rightarrow \mathbf{g}_{\alpha
\beta }=\left( g_{ij},h_{ab}\right) ,\Gamma _{\ \alpha }^{\gamma
}\rightarrow \mathbf{\Gamma }_{\ \alpha }^{\gamma }
\end{equation*}%
with $\mathbf{Q}_{\alpha \beta }=\mathbf{Q}_{\gamma \alpha \beta }\mathbf{%
\vartheta }^{\gamma }$ and $\mathbf{T}^{\alpha }$ as in (\ref{dt})) into
respective formulas (\ref{ac}), (\ref{acc}) and (\ref{dista}), \ we can
define an affine connection $\mathbf{D=\bigtriangledown +Z}=\{\mathbf{\Gamma
}_{\ \beta \alpha }^{\gamma }\}$ with respect to N--adapted (co) frames,
\begin{equation}
\mathbf{\Gamma }_{\ \ \alpha }^{\gamma }=\mathbf{\Gamma }_{\ \alpha \beta
}^{\gamma }\mathbf{\vartheta }^{\beta },  \label{acn}
\end{equation}%
with
\begin{equation}
\mathbf{\Gamma }_{\ \beta }^{\alpha }=\mathbf{\Gamma }_{\bigtriangledown \
\beta }^{\alpha }+\mathbf{Z}_{\ \ \beta }^{\alpha },  \label{accn}
\end{equation}%
$\mathbf{\Gamma }_{\bigtriangledown \ \beta }^{\alpha }$ being expressed as (%
\ref{christa}) (equivalently, defined by (\ref{lcsym})) and $\mathbf{Z}_{\ \
\beta }^{\alpha }$ expressed as
\begin{equation}
\mathbf{Z}_{\alpha \beta }=\mathbf{e}_{\beta }\rfloor \mathbf{T}_{\alpha }-%
\mathbf{e}_{\alpha }\rfloor \mathbf{T}_{\beta }+\frac{1}{2}\left( \mathbf{e}%
_{\alpha }\rfloor \mathbf{e}_{\beta }\rfloor \mathbf{T}_{\gamma }\right)
\mathbf{\vartheta }^{\gamma }+\left( \mathbf{e}_{\alpha }\rfloor \mathbf{Q}%
_{\beta \gamma }\right) \mathbf{\vartheta }^{\gamma }-\left( \mathbf{e}%
_{\beta }\rfloor \mathbf{Q}_{\alpha \gamma }\right) \mathbf{\vartheta }%
^{\gamma }+\frac{1}{2}\mathbf{Q}_{\alpha \beta }.  \label{distan}
\end{equation}%
The h-- and v--components of $\mathbf{\Gamma }_{\ \beta }^{\alpha }$ from (%
\ref{accn}) consists from the components of $\mathbf{\Gamma }%
_{\bigtriangledown \ \beta }^{\alpha }$ (considered for (\ref{christa})) and
of $\mathbf{Z}_{\alpha \beta }$ with $\mathbf{Z}_{\ \ \gamma \beta }^{\alpha
}=\left( Z_{jk}^{i},Z_{bk}^{a},Z_{jc}^{i},Z_{bc}^{a}\right) .$The values
\begin{equation*}
\mathbf{\Gamma }_{\bigtriangledown \gamma \beta }^{\alpha }+\mathbf{Z}_{\ \
\gamma \beta }^{\alpha }=\left( L_{\bigtriangledown
jk}^{i}+Z_{jk}^{i},L_{\bigtriangledown
bk}^{a}+Z_{bk}^{a},C_{\bigtriangledown
jc}^{i}+Z_{jc}^{i},C_{\bigtriangledown bc}^{a}+Z_{bc}^{a}\right)
\end{equation*}%
are defined correspondingly
\begin{eqnarray*}
L_{\bigtriangledown jk}^{i}+Z_{\ jk}^{i} &=&\left[ (\bigtriangledown
_{k}+Z_{k})\delta _{j}\right] \rfloor d^{i},\quad L_{\bigtriangledown
bk}^{a}+Z_{\ bk}^{a}=\left[ (\bigtriangledown _{k}+Z_{k})\partial _{b}\right]
\rfloor \delta ^{a}, \\
\ C_{\bigtriangledown jc}^{i}+Z_{\ jc}^{i} &=&\left[ \left( \bigtriangledown
_{c}+Z_{c}\right) \delta _{j}\right] \rfloor d^{i},\quad C_{\bigtriangledown
bc}^{a}+Z_{\ bc}^{a}=\left[ \left( \bigtriangledown _{c}+Z_{c}\right)
\partial _{b}\right] \rfloor \delta ^{a}.
\end{eqnarray*}%
and related to (\ref{distan}) via h- and v--splitting of N--frames $\mathbf{e%
}_{\gamma }=\left( \delta _{i}=\partial _{i}-N_{i}^{a}\partial _{a},\partial
_{a}\right) $ and N--coframes $\mathbf{\vartheta }^{\beta }=\left(
dx^{i},\delta y^{a}=dy^{a}+N_{i}^{a}dx^{i}\right) $ and d--metric $\mathbf{g}%
=\left( g_{ij,}h_{ab}\right) .$

We note that for $\mathbf{Q}_{\alpha \beta }=0,$ the distorsion 1--form $%
\mathbf{Z}_{\alpha \beta }$ defines a Riemann--Cartan geometry adapted to
the N--connection structure.

Let us briefly outline the procedure of definition of N--connections in a
metric--affine space $V^{n+m}$ \ with arbitrary metric and connection
structures $\left( g^{[od]}=\{g_{\alpha \beta }\},\underline{\Gamma }_{\
\beta \alpha }^{\gamma }\right) $ and show how the geometric objects may be
adapted to the N--connection structure.

\begin{proposition}
\label{pmasnc}Every metric--affine space provided with a generic
off--diagonal metric structure admits nontrivial N--connections.
\end{proposition}

\textbf{Proof:} We give an explicit example how to introduce the
N--connection structure. We write the metric with respect to a local
coordinate basis,
\begin{equation*}
g^{[od]}=g_{\underline{\alpha }\underline{\beta }}du^{\underline{\alpha }%
}\otimes du^{\underline{\beta }},
\end{equation*}%
where the matrix $g_{\underline{\alpha }\underline{\beta }}$ contains a
non--degenerated $\left( m\times m\right) $ submatrix $h_{ab},$ for instance
like in ansatz (\ref{ansatz}). Having fixed the block $h_{ab},$ labeled by
running of indices $a,b,...=n+1,n+2,...,n+m,$ we can define the $\left(
n\times n\right) $ bloc $g_{\underline{i}\underline{j}}$ with indices $%
\underline{i},\underline{j},...=1,2,...n.$ The next step is to find any
nontrivial $N_{i}^{a}$ (the set of coefficients has being defined, we may
omit underlying) and find $N_{\underline{j}}^{\underline{e}}$ from the $%
\left( n\times m\right) $ block relations $g_{\underline{j}\underline{a}}=N_{%
\underline{j}}^{\underline{e}}h_{\underline{a}\underline{e}}.$ This is
always possible if $g_{\underline{\alpha }\underline{\beta }}$ is generic
off--diagonal. The next step is to compute $g_{ij}=g_{\underline{i}%
\underline{j}}-N_{\underline{i}}^{\underline{a}}N_{\underline{j}}^{%
\underline{e}}h_{\underline{a}\underline{e}}$ which gives the possibility to
transform equivalently
\begin{equation*}
g^{[od]}\rightarrow \mathbf{g}=g_{ij}\mathbf{\vartheta }^{i}\otimes \mathbf{%
\vartheta }^{j}+h_{ab}\mathbf{\vartheta }^{a}\otimes \mathbf{\vartheta }^{b}
\end{equation*}%
where
\begin{equation*}
\mathbf{\vartheta }^{i}\doteqdot dx^{i},\ \mathbf{\vartheta }^{a}\doteqdot
\delta y^{a}=dy^{a}+N_{i}^{a}\left( u\right) dx^{i}
\end{equation*}%
are just the N--elongated differentials (\ref{ddif}) if the local
coordinates associated to the block $h_{ab}$ are denoted by $y^{a}$ and the
rest ones by $x^{i}.$ We impose a global splitting of the metric--affine
spacetime by stating that all geometric objects are subjected to anholonomic
frame transforms with vielbein coefficients of type (\ref{vt1}) and (\ref%
{vt2}) defined by $\mathbf{N}=\{N_{i}^{a}\}.$ This way, we define on the
metric--affine space a vector/covector bundle structure if the coordinates $%
y^{a}$ are treated as certain local vector/ covector components.$%
\blacksquare $

We note, that having defined the values $\mathbf{\vartheta }^{\alpha
}=\left( \mathbf{\vartheta }^{i},\mathbf{\vartheta }^{b}\right) $ and their
duals $\mathbf{e}_{\alpha }=\left( \mathbf{e}_{i},\mathbf{e}_{a}\right) ,$
we can compute the linear connection coefficients with respect to N--adapted
(co) frames, $\Gamma _{\ \beta \alpha }^{\gamma }\rightarrow \widetilde{%
\mathbf{\Gamma }}_{\ \beta \alpha }^{\gamma }.$ However, $\widetilde{\mathbf{%
\Gamma }}_{\ \beta \alpha }^{\gamma },$ in general, is not a d--connection,
i. e. it is not \ adapted to the global splitting $T\mathbf{V}^{n+m}=h%
\mathbf{V}^{n+m}\oplus v\mathbf{V}^{n+m}$ defined by N--connection, see
Definition \ref{defdcon}. If the metric and linear connection are not
subjected to any field equations, we are free to consider distorsion tensors
in order to be able to apply the Theorems \ref{kmp} and/or \ref{mconnections}
with the aim to transform $\widetilde{\mathbf{\Gamma }}_{\ \beta \alpha
}^{\gamma }$ into a metric d--connection, or even into a Riemann--Cartan
d--connection. Here, we also note that a metric--affine space, in general,
admits different classes of N--connections with various nontrivial global
splittings $n^{\prime }+m^{\prime }=n+m,$ where $n^{\prime }\neq n.$

We can state from the very beginning that a metric--affine space $\mathbf{V}%
^{n+m}$ is \ provided with d--metric (\ref{block2}) and d--connection
strucuture (\ref{dcon1}) adapted to a class of prescribed vielbein
transforms (\ref{vt1}) and (\ref{vt2}) and N--elongated frames (\ref{dder})
and (\ref{ddif}). All constructions can be redefined with respect to
coordinate frames (\ref{pder}) and (\ref{pdif}) with off--diagonal metric
parametrization (\ref{ansatz}) and then subjected to another frame and
coordinate transforms hiding the existing N--connection structure and
distinguished character of geometric objects. Such 'distinguished'
metric--affine spaces are characterized by corresponding N--connection
geometries and admit geometric constructions with distinguished objects.
They form a particular subclass of metric--affine spaces admitting
transformations of the general linear connection $\Gamma _{\ \beta \alpha
}^{\gamma }$ into certain classes of d--connections $\mathbf{\Gamma }_{\
\beta \alpha }^{\gamma }.$

\begin{definition}
\label{ddmas}A \ distinguished metric--affine space $\mathbf{V}^{n+m}$ is a
usual metric--affine space additionally enabled with a N--connection
structure $\mathbf{N}=\{N_{i}^{a}\}$ inducing splitting into respective
irreducible horizontal and vertical subspaces of dimensions $n$ and $m.$
This space is provided with independent d--metric (\ref{block2}) and affine
d--connection (\ref{dcon1}) structures adapted to the N--connection.
\end{definition}

The metric--affine spacetimes with stated N--connection structure are also
characterized by nontrivial anholonomy relations of type (\ref{anhr}) with
anholonomy coefficients (\ref{anhc}). This is a very specific type of
noncommutative symmetry generated by N--adapted (co) frames defining
different anholonomic noncommutative differential calculi (for details with
respect to the Einstein and gauge gravity see Ref. \cite{vnces}).

We construct and analyze explicit examples of metric--affine spacetimes with
associated N--connection (noncommutative) symmetry in Refs. \cite{exsolmag}.
A surprizing fact is that various types of d--metric ansatz (\ref{block2})
with associated N--elongated frame (\ref{dder}) and coframe (\ref{ddif}) (or
equivalently, respective off--diagonal ansatz (\ref{ansatz})) can be defined
as exact solutions in Einstein gravity of different dimensions and in
metric--affine, or Einstein--Cartan gravity and gauge model realizations.
Such solutions model also generalized Finsler structures.

\section{Generalized Finsler--Affine Spaces}

The aim of this section is to demonstrate that any well known type of
locally anistoropic or locally isotropic spaces can be defined as certain
particular cases of distinguished metric--affine spaces. We use the general
term of ''generalized Finsler--affine spaces'' for all type of geometries
modeled in MAG as generalizations of the Riemann--Cartan--Finsler geometry,
in general, containing nonmetricity fields. A complete classification of
such spaces is given by Tables 1--11 in the Appendix.

\subsection{Spaces with vanishing N--connection curvature}

Three examples of such spaces are given by the well known (pseudo) Riemann,
Riemann--Cartan or Kaluza--Klein manifolds of dimension $\left( n+m\right) $
provided with a generic off--diagonal metric structure $\underline{g}%
_{\alpha \beta }$ of type (\ref{ansatz}), of corresponding signature, which
can be reduced equivalently to the block $\left( n\times n\right) \oplus
\left( m\times m\right) $ form (\ref{block2}) via vielbein transforms (\ref%
{vt1}). Their N--connection structures may be restricted by the condition $%
\Omega _{ij}^{a}=0,$ see (\ref{ncurv}).

\subsubsection{Anholonomic (pseudo) Riemannian spaces}

\label{sarg}The (pseudo) Riemannian manifolds, $\mathbf{V}_{R}^{n+m},$
provided with a generic off--diagonal metric and anholonomic frame structrue
effectively diagonalizing such a metric is an anholonomic (pseudo)
Riemannian space. The space admits associated N--connection structures with
coefficients induced by generic off--diagonal terms in the metric (\ref%
{ansatz}). If the N--connection curvature vanishes, the Levi--Civita
connection is closely defined by the same coefficients as the canonical
d--connection (linear connections computed with respect to the N--adapted
(co) frames), see Proposition \ref{lccdc} and related discussions in section %
\ref{lconnections}. Following the Theorem \ref{teqrgrcg}, any (pseudo)
Riemannian space enabled with generic off--diagonal connection structure can
be equivalently modeled as a effective Riemann--Cartan geometry with induced
N--connection and d--torsions.

There were constructed a number of exact 'off--diagonal' solutions of the
Einstein equations \cite{v1,v2,vd}, for instance, in five dimensional
gravity (with various type restrictions to lower dimensions) with nontrivial
N--connection structure with ansatz for metric of type
\begin{eqnarray}
\mathbf{g} &=&\omega \left( x^{i},y^{4}\right)
[g_{1}(dx^{1})^{2}+g_{2}\left( x^{2},x^{3}\right) (dx^{2})^{2}+g_{3}\left(
x^{2},x^{3}\right) (dx^{3})^{2}  \notag \\
&&+h_{4}\left( x^{2},x^{3},y^{4}\right) \left( \delta y^{4}\right)
^{2}+h_{5}\left( x^{2},x^{3},y^{4}\right) \left( \delta y^{5}\right) ^{2}],
\label{pta}
\end{eqnarray}%
for $g_{1}=const,$ where%
\begin{equation*}
\delta y^{a}=dy^{a}+N_{k}^{a}\left( x^{i},y^{4}\right) dy^{a}
\end{equation*}%
with indices $i,j,k...=1,2,3$ and $a=4,5.$ The coefficients $N_{i}^{a}\left(
x^{i},y^{4}\right) $ were searched as a metric ansatz of type (\ref{ansatz})
transforming equivalently into a certain diagonalized block (\ref{block2})
would parametrize generic off--diagonal exact solutions. Such effective
N--connections are contained into a corresponding anholonomic moving or
static configuration of tetrads/ pentads (vierbeins/funfbeins) defining a
conventional splitting of coordinates into $n$ holonomic and $m$ anholonomic
ones, where $n+m=4,5.$ The ansatz (\ref{pta}) results in exact solutions of
vacuum and nonvacuum Einstein equations which can be integrated in general
form. Perhaps, all known at present time exact solutions in 3-5 dimensional
gravity can be included as particular cases in (\ref{pta}) and generalized
to anholonomic configurations with running constants and gravitational and
matter polarizations (in general, anisotropic on variable $y^{4})$ of the
metric and frame coefficients.

The vector/ tangent bundle configurations and/or torsion structures can be
effectively modeled on such (pseudo) Riemannian spaces by prescribing a
corresponding class of anholonomic frames. Such configurations are very
different from those, for instance, defined by Killing symmetries and the
induced torsion vanishes after frame transforms to coordinate bases. For a
corresponding parametrizations of $N_{i}^{a}(u)$ and $g_{\alpha \beta },$ we
can model Finsler like structures even in (pseudo) Riemannian spacetimes or
in gauge gravity \cite{vd,ncgg,vncggf}.

The anholonomic Riemannian spaces $\mathbf{V}_{R}^{n+m}$ consist a subclass
of distinguished metric--affine spaces $\mathbf{V}^{n+m}$ provided with
N--connection structure, characterized by the condition that nonmetricity
d--filed $\mathbf{Q}_{\alpha \beta \gamma }=0$ and that a certain type of
induced torsions $\mathbf{T}_{\beta \gamma }^{\alpha }$ vanish for the
Levi--Civita connection. We can take a generic off--diagonal metric (\ref%
{ansatz}), transform it into a d--metric (\ref{block2}) and compute the h-
and v-components of the canonical d--connection (\ref{dcon1}) and put them
into the formulas for d--torsions (\ref{dtorsions}) and d--curvatures (\ref%
{dcurv}). The vacuum solutions are defined by d--metrics and N--connections
satisfying the condition $\mathbf{R}_{\alpha \beta }=0,$ see the h--,
v--components (\ref{dricci}).

In order to transform certain geometric constructions defined by the
canonical d--connection into similar ones for the Levi--Civita connection,
we have to constrain the N--connection structure as to have vanishing
N--curvature, $\Omega _{ij}^{a}=0,$ or to see the conditions when the
deformation of Levi--Civita connection to any d--connection result in
non--deformations of the Einstein equation. We obtain a (pseudo) Riemannian
vacuum spacetime with anholonomially induced d--torsion components $\widehat{%
T}_{ja}^{i}=-\widehat{T}_{aj}^{i}=\widehat{C}_{.ja}^{i}$ and $\widehat{T}%
_{.bi}^{a}=-\widehat{T}_{.ib}^{a}=\partial N_{i}^{a}/\partial y^{b}-\widehat{%
L}_{.bj}^{a}.$ This torsion can be related algebraicaly to a spin source
like in the usual Riemann--Cartan gravity if we want to give an algebraic
motivation to the N--connection splitting. We emphasize that the
N--connection and d--metric coefficients can be chosen in order to model on $%
\mathbf{V}_{R}^{n+m}$ a special subclass of Finsler/ Lagrange structures
(see discussion in section \ref{mfls}).

\subsubsection{Kaluza--Klein spacetimes}

Such higher dimension generalizations of the Einstein gravity are
characterized by a metric ansatz
\begin{equation}
\underline{g}_{\alpha \beta }=\left[
\begin{array}{cc}
g_{ij}(x^{\kappa })+A_{i}^{a}(x^{\kappa })A_{j}^{b}(x^{\kappa
})h_{ab}(x^{\kappa },y^{a}) & A_{j}^{e}(x^{\kappa })h_{ae}(x^{\kappa },y^{a})
\\
A_{i}^{e}(x^{\kappa })h_{be}(x^{\kappa },y^{a}) & h_{ab}(x^{\kappa },y^{a})%
\end{array}%
\right]  \label{anskk}
\end{equation}%
(a particular case of the metric (\ref{ansatz})) with certain
compactifications on extra dimension coordinates $y^{a}.$ The values $%
A_{i}^{a}(x^{\kappa })$ are considered to define gauge fields after
compactifications (the electromatgnetic potential in the original extension
to five dimensions by Kaluza and Klein, or some non--Abelian gauge fields
for higher dimension generalizations). Perhaps, the ansatz (\ref{anskk}) was
originally introduced in Refs. \cite{salam} (see \cite{ov} as a review of
non--supersymmetry models and \cite{salamsezgin} for supersymmetric
theories).

The coefficients $A_{i}^{a}(x^{\kappa })$ from (\ref{anskk}) are certain
particular parametrizations of the N--connection coefficients $%
N_{i}^{a}(x^{\kappa },y^{a})$ in (\ref{ansatz}). This suggests a physical
interpretation for the N--connection as a specific nonlinear gauge field
depending both on spacetime and extra dimension coordinates (in general,
noncompactified). In the usual Kaluza--Klein (super) theories, there were
not considered anholonomic transforms to block d--metrics (\ref{block2})
containing dependencies on variables $y^{a}.$

In some more general approaches, with additional anholonomic structures on
lower dimensional spacetime, there were constructed a set of exact vacuum
five dimensional solutions by reducing ansatz (\ref{anskk}) and their
generalizations of form (\ref{ansatz}) to d--metric ansatz of type (\ref{pta}%
), see Refs. \cite{v1,v2,ncgg,vncggf,vd,vnces}. Such vacuum and nonvacuum
solutions describe anisotropically polarized Taub--NUT spaces, wormhole/
flux tube confiugurations, moving four dimensional black holes in bulk five
dimensional spacetimes, anisotropically deformed cosmological spacetimes and
various type of locally anisotropic spinor--soliton--dialton interactions in
generalized Kaluza--Klein and string/ brane gravity.

\subsubsection{Teleparallel spaces}

\label{stps}Teleparallel theories are usually defined by two geometrical
constraints \cite{tgm} (here, we introduce them for d--connections and
nonvanishing N--connection structure),%
\begin{equation}
\mathbf{R}_{~\beta }^{\alpha }=\delta \mathbf{\Gamma }_{~\beta }^{\alpha }+%
\mathbf{\Gamma }_{~\gamma }^{\alpha }\wedge \mathbf{\Gamma }_{~\beta
}^{\gamma }=0  \label{dtel1}
\end{equation}%
and
\begin{equation}
\mathbf{Q}_{\alpha \beta }=-\mathbf{Dg}_{\alpha \beta }=-\delta \mathbf{g}%
_{\alpha \beta }+\mathbf{\Gamma }_{~\beta }^{\gamma }\mathbf{g}_{\alpha
\gamma }+\mathbf{\Gamma }_{~\alpha }^{\gamma }\mathbf{g}_{\beta \gamma }=0.
\label{dtel2}
\end{equation}
The conditions (\ref{dtel1}) and (\ref{dtel2}) establish a distant
paralellism in such spaces because the result of a parallel transport of a
vector does not depend on the path (the angles and lengths being also
preserved under parallel transports). It is always possible to find such
anholonomic transforms $e_{\alpha }=A_{\alpha }^{~\underline{\beta }}e_{%
\underline{\beta }}$ and $e_{\underline{\alpha }}=A_{\underline{\alpha }%
}^{~\beta }e_{\beta },$ where $A_{\underline{\alpha }}^{~\beta }$ is inverse
to $A_{\alpha }^{~\underline{\beta }}$ when
\begin{equation*}
\mathbf{\Gamma }_{~\beta }^{\alpha }\rightarrow \mathbf{\Gamma }_{~%
\underline{\beta }}^{\underline{\alpha }}=A_{\underline{\beta }}^{~\beta }%
\mathbf{\Gamma }_{~\beta }^{\alpha }A_{~\alpha }^{\underline{\alpha }%
}+A_{~\gamma }^{\underline{\alpha }}\delta A_{\underline{\beta }}^{~\gamma
}=0
\end{equation*}%
and the transformed local metrics becomes the standard Minkowski,%
\begin{equation*}
g_{\underline{\alpha }\underline{\beta }}=diag\left( -1,+1,....,+1\right)
\end{equation*}%
(it can be fixed any signature). If the (co) frame is considered as the only
dynamical variable, it is called that the space (and choice of gauge) are of
Weitzenbock type. A coframe of type (\ref{ddif})
\begin{equation*}
\mathbf{\vartheta }_{\ }^{\beta }\doteqdot \left( \delta x^{i}=dx^{i},\delta
y^{a}=dy^{a}+N_{i}^{a}\left( u\right) dx^{i}\right)
\end{equation*}%
is defined by N--connection coefficients. If we impose the condition of
vanishing the N--connection curvature, $\Omega _{ij}^{a}=0,$ see (\ref{ncurv}%
), the N--connection defines a specific anholonomic dynamics because of
nontrivial anholonomic relations (\ref{anhr}) with nonzero components (\ref%
{anhc}).

By embedding teleparallel configurations into metric--affine spaces provided
with N--connection structure we state a distinguished class of (co) frame
fields adapted to this structure and open possibilities to include such
spaces into Finsler--affine ones, see section \ref{stpfa}. For vielbein
fields $\mathbf{e}_{\alpha }^{~\underline{\alpha }}$ and their inverses $%
\mathbf{e}_{~\underline{\alpha }}^{\alpha }$ related to the d--metric (\ref%
{block2}),%
\begin{equation*}
\mathbf{g}_{\alpha \beta }=\mathbf{e}_{\alpha }^{~\underline{\alpha }}%
\mathbf{e}_{\beta }^{~\underline{\beta }}g_{\underline{\alpha }\underline{%
\beta }}
\end{equation*}%
we define the Weitzenbock d--connection%
\begin{equation}
~^{[W]}\mathbf{\Gamma }_{~\beta \gamma }^{\alpha }=\mathbf{e}_{~\underline{%
\alpha }}^{\alpha }\delta _{\gamma }\mathbf{e}_{\beta }^{~\underline{\alpha }%
},  \label{wcon}
\end{equation}%
where $\delta _{\gamma }$ is the N--elongated partial derivative (\ref{dder}%
). It transforms in the usual Weitzenbock connection for trivial
N--connections. The torsion of $~^{[W]}\mathbf{\Gamma }_{~\beta \gamma
}^{\alpha }$ is defined
\begin{equation}
~^{[W]}\mathbf{T}_{~\beta \gamma }^{\alpha }=~^{[W]}\mathbf{\Gamma }_{~\beta
\gamma }^{\alpha }-~^{[W]}\mathbf{\Gamma }_{~\gamma \beta }^{\alpha }.
\label{wtors}
\end{equation}%
It posses h-- and v--irreducible components constructed from the components
of a d--metric and N--adapted frames. We can express
\begin{equation*}
~^{[W]}\mathbf{\Gamma }_{~\beta \gamma }^{\alpha }=\mathbf{\Gamma }%
_{\bigtriangledown ~\beta \gamma }^{\alpha }+\mathbf{Z}_{~\beta \gamma
}^{\alpha }
\end{equation*}%
where $\mathbf{\Gamma }_{\bigtriangledown ~\beta \gamma }^{\alpha }$ is the
Levi--Civita connection (\ref{lcsym}) and the contorsion tensor is
\begin{equation*}
\mathbf{Z}_{\alpha \beta }=\mathbf{e}_{\beta }\rfloor ~^{[W]}\mathbf{T}%
_{\alpha }-\mathbf{e}_{\alpha }\rfloor ~^{[W]}\mathbf{T}_{\beta }+\frac{1}{2}%
\left( \mathbf{e}_{\alpha }\rfloor \mathbf{e}_{\beta }\rfloor ~^{[W]}\mathbf{%
T}_{\gamma }\right) \mathbf{\vartheta }^{\gamma }+\left( \mathbf{e}_{\alpha
}\rfloor \mathbf{Q}_{\beta \gamma }\right) \mathbf{\vartheta }^{\gamma
}-\left( \mathbf{e}_{\beta }\rfloor \mathbf{Q}_{\alpha \gamma }\right)
\mathbf{\vartheta }^{\gamma }+\frac{1}{2}\mathbf{Q}_{\alpha \beta }.
\end{equation*}%
In formulation of teleparallel alternatives to the general relativity it is
considered that $\mathbf{Q}_{\alpha \beta }=0.$

\subsection{Finsler and Finsler--Riemann--Cartan spaces}

\label{ssffc}The first approaches to Finsler spaces \cite{fg,rund} were
developed by generalizing the usual Riemannian metric interval
\begin{equation*}
ds=\sqrt{g_{ij}\left( x\right) dx^{i}dx^{j}}
\end{equation*}%
on a manifold $M$ \ of dimension $n$ into a nonlinear one
\begin{equation}
ds=F\left( x^{i},dx^{j}\right)  \label{finint}
\end{equation}%
defined by the Finsler metric $F$ (fundamental function) on $\widetilde{TM}%
=TM\backslash \{0\}$ (it should be noted an ambiguity in terminology used in
monographs on Finsler geometry and on gravity theories with respect to such
terms as Minkowski space, metric function and so on). It is also considered
a quadratic form on $\R^{2}$ with coefficients
\begin{equation}
\ g_{ij}^{[F]}\rightarrow h_{ab}=\frac{1}{2}\frac{\partial ^{2}F^{2}}{%
\partial y^{i}\partial y^{j}}  \label{finm2}
\end{equation}%
defining a positive definite matrix. The local coordinates are denoted $%
u^{\alpha }=(x^{i},y^{a}\rightarrow y^{i}).$ There are satisfied the
conditions: 1) The Finsler metric on a real manifold $M$ is a function $%
F:TM\rightarrow \R$ which on $\widetilde{TM}=TM\backslash \{0\}$ is of class
$C^{\infty }$ and $F$ is only continuous on the image of the null
cross--sections in the tangent bundle to $M.$ 2) $F\left( x,\chi y\right)
=\chi F\left( x,y\right) $ for every $\R_{+}^{\ast }.$ 3) The restriction of
$F$ to $\widetilde{TM}$ is a positive function. 4) $rank\left[
g_{ij}^{[F]}(x,y)\right] =n.$

There were elaborated a number of models of locally anisotropic spacetime
geometry with broken local Lorentz invariance (see, for instance, those
based on Finsler geometries \cite{as,bog}). In result, in the Ref. \cite%
{will}, it was ambigously concluded that Finsler gravity models are very
restricted by experimental data. Recently, the subject concerning Lorentz
symmetry violations was revived for instance in brane gravity \cite{groj}
(see a detailed analysis and references on such theoretical and experimental
researches in \cite{kost}). In this case, the Finsler like geometries
broking the local four dimensional Lorentz invariance can be considered as a
possible alternative direction for investigating physical models both with
local anisotropy and violation of local spacetime symmetries. But it should
be noted here that violations of postulates of general relativity is not a
generic property of the so--called ''Finsler gravity''. A subclass of
Finsler geometries and their generalizations could be induced by anholonomic
frames even in general relativity theory and Riemannian--Cartan or gauge
gravity \cite{vd,ncgg,vncggf,vsp}. The idea is that instead of geometric
constructions based on straightforward applications of derivatives of (\ref%
{finm2}), following from a nonlinear interval (\ref{finint}), we should
consider d--metrics (\ref{block2}) with the coefficients from Finsler
geometry (\ref{finm2}) or their extended variants. In this case, certain
type Finsler configurations can be defined even as exact 'off--diagonal'
solutions in vacuum Einstein gravity or in string gravity.

\subsubsection{Finsler geometry and its almost Kahlerian model}

We outline a modern approach to Finsler geometry \cite{ma} based on the
geometry of nonlinear connections in tangent bundles.

A real (commutative) Finsler space $\mathbf{F}^{n}=\left( M,F\left(
x,y\right) \right) $ can be modeled on a tangent bundle $TM$ enabled with a
Finsler metric $F\left( x^{i},y^{j}\right) $ and a quadratic form $%
g_{ij}^{[F]}$ (\ref{finm2}) satisfying the mentioned conditions and defining
the Christoffel symbols (not those from the usual Riemannian geometry)%
\begin{equation*}
c_{jk}^{\iota }(x,y)=\frac{1}{2}g^{ih}\left( \partial
_{j}g_{hk}^{[F]}+\partial _{k}g_{jh}^{[F]}-\partial _{h}g_{jk}^{[F]}\right) ,
\end{equation*}%
where $\partial _{j}=\partial /\partial x^{j},$ $\ $and the Cartan nonlinear
connection
\begin{equation}
\ ^{[F]}\mathbf{N}_{j}^{i}(x,y)=\frac{1}{4}\frac{\partial }{\partial y^{j}}%
\left[ c_{lk}^{\iota }(x,y)y^{l}y^{k}\right] ,  \label{ncc}
\end{equation}%
where we do not distinguish the v- and h- indices taking on $TM$ \ the same
values.

In Finsler geometry, there were investigated different classes of remarkable
Finsler linear connections introduced by Cartan, Berwald, Matsumoto and
other geometers (see details in Refs. \cite{fg,as,rund}). Here we note that
we can introduce $g_{ij}^{[F]}=g_{ab}$ and $\ ^{[F]}\mathbf{N}_{j}^{i}(x,y)$
in (\ref{ansatz}) and transfer our considerations to a $\left( n\times
n\right) \oplus \left( n\times n\right) $ blocks of type (\ref{block2}) for
a metric--affine space $V^{n+n}.$

A usual Finsler space $\mathbf{F}^{n}=\left( M,F\left( x,y\right) \right) $
is completely defined by its fundamental tensor $g_{ij}^{[F]}(x,y)$ and the
Cartan nonlinear connection $\ ^{[F]}\mathbf{N}_{j}^{i}(x,y)$ and any chosen
d--connection structure (\ref{dcon1}) (see details on different type of
d--connections in section \ref{lconnections}). Additionally, the
N--connection allows us to define an almost complex structure $I$ on $TM$ as
follows%
\begin{equation*}
I\left( \delta _{i}\right) =-\partial /\partial y^{i}\mbox{ and
}I\left( \partial /\partial y^{i}\right) =\delta _{i}
\end{equation*}%
for which $I^{2}=-1.$

The pair $\left( g^{[F]},I\right) $ consisting from a Riemannian metric on a
tangent bundle $TM,$%
\begin{equation}
\mathbf{g}^{[F]}=g_{ij}^{[F]}(x,y)dx^{i}\otimes
dx^{j}+g_{ij}^{[F]}(x,y)\delta y^{i}\otimes \delta y^{j}  \label{dmetricf}
\end{equation}%
and the almost complex structure $I$ defines an almost Hermitian structure
on $\widetilde{TM}$ associated to a 2--form%
\begin{equation*}
\theta =g_{ij}^{[F]}(x,y)\delta y^{i}\wedge dx^{j}.
\end{equation*}%
This model of Finsler geometry is called almost Hermitian and denoted $%
H^{2n} $ and it is proven \cite{ma} that is almost Kahlerian, i. e. the form
$\theta $ is closed. The almost Kahlerian space $\mathbf{K}^{2n}=\left(
\widetilde{TM},\mathbf{g}^{[F]},I\right) $ is also called the almost
Kahlerian model of the Finsler space $F^{n}.$

On Finsler spaces (and their almost Kahlerian models), one distinguishes the
almost Kahler linear connection of Finsler type, $\mathbf{D}^{[I]}$ on $%
\widetilde{TM}$ with the property that this covariant derivation preserves
by parallelism the vertical distribution and is compatible with the almost
Kahler structure $\left( \mathbf{g}^{[F]},I\right) ,$ i.e.
\begin{equation*}
\mathbf{D}_{X}^{[I]}\mathbf{g}^{[F]}=0\mbox{ and }\mathbf{D}_{X}^{[I]}%
\mathbf{I}=0
\end{equation*}%
for \ every d--vector field on $\widetilde{TM}.$ This d--connection is
defined by the data
\begin{equation}
\ ^{[F]}\widehat{\mathbf{\Gamma }}_{\ \beta \gamma }^{\alpha }=\left( \
^{[F]}\widehat{L}_{jk}^{i},\ ^{[F]}\widehat{L}_{jk}^{i},\ ^{[F]}\widehat{C}%
_{jk}^{i},\ ^{[F]}\widehat{C}_{jk}^{i}\right)  \label{dccfs}
\end{equation}%
with $\ ^{[F]}\widehat{L}_{jk}^{i}$ and $\ ^{[F]}\widehat{C}_{jk}^{i}$
computed by similar formulas in (\ref{candcon}) by using $g_{ij}^{[F]}$ as
in (\ref{finm2}) and $\ ^{[F]}N_{j}^{i}$ from (\ref{ncc}).

We emphasize that a Finsler space $\mathbf{F}^{n}$ with a d--metric (\ref%
{dmetricf}) and Cartan's N--connection structure (\ref{ncc}), or the
corresponding almost Hermitian (Kahler) model $\mathbf{H}^{2n},$ can be
equivalently modeled on a space of dimension $2n,$ $\mathbf{V}^{n+n},$
provided with an off--diagonal metric (\ref{ansatz}) and anholonomic frame
structure with associated Cartan's nonlinear connection. Such anholonomic
frame constructions are similar to modeling of the Einstein--Cartan geometry
on (pseudo) Riemannian spaces where the torsion is considered as an
effective tensor field. From this viewpoint a Finsler geometry is a
Riemannian--Cartan geometry defined on a tangent bundle provided with a
respective off--diagonal metric (and a related anholonomic frame structure
with associated N--connection) and with additional prescriptions with
respect to the type of linear connection chosen to be compatible with the
metric and N--connection structures.

\subsubsection{Finsler--Kaluza--Klein spaces}

In Ref. \cite{vncggf} we defined a 'locally anisotropic' toroidal
compactification of the 10 dimensional heterotic string sction \cite{kir}.
We consider here the corresponding anholonomic frame transforms and
off--diagonal metric ansatz. Let $\left( n^{\prime },m^{\prime }\right) $ be
the (holonomic, anholonomic) dimensions of the compactified spacetime (as a
particular case we can state $n^{\prime }+m^{\prime }=4,$ or any integers $%
n^{\prime }+m^{\prime }<10,$ for instance, for brane configurations). There
are used such parametrizations of indices and of vierbeinds: Greek indices $%
\alpha ,\beta ,...\mu ...$ run values for a 10 dimensional spacetime and
split as $\alpha =\left( \alpha ^{\prime },\widehat{\alpha }\right) ,\beta
=\left( \beta ^{\prime },\widehat{\beta }\right) ,...$ when primed indices $%
\alpha ^{\prime },\beta ^{\prime },...\mu ^{\prime }...$ run values for
compactified spacetime and split into h- and v--components like $\alpha
^{\prime }=\left( i^{\prime },a^{\prime }\right) ,$ $\beta ^{\prime }=\left(
j^{\prime },b^{\prime }\right) ,...;$ the frame coefficients are split as
\begin{equation}
e_{\mu }^{~\underline{\mu }}(u)=\left(
\begin{array}{cc}
e_{\alpha ^{\prime }}^{~\underline{\alpha ^{\prime }}}(u^{\beta ^{\prime }})
& A_{\alpha ^{\prime }}^{\widehat{\alpha }}(u^{\beta ^{\prime }})e_{\widehat{%
\alpha }}^{~\underline{\widehat{\alpha }}}(u^{\beta ^{\prime }}) \\
0 & e_{\widehat{\alpha }}^{~\underline{\widehat{\alpha }}}(u^{\beta ^{\prime
}})%
\end{array}%
\right)  \label{vt1a}
\end{equation}%
where $e_{\alpha ^{\prime }}^{~\underline{\alpha ^{\prime }}}(u^{\beta
^{\prime }}),$ in their turn, are taken in the form (\ref{vt1}),
\begin{equation}
e_{\alpha ^{\prime }}^{~\underline{\alpha ^{\prime }}}(u^{\beta ^{\prime
}})=\left(
\begin{array}{cc}
e_{i^{\prime }}^{~\underline{i^{\prime }}}(x^{j^{\prime }},y^{a^{\prime }})
& N_{i^{\prime }}^{a^{\prime }}(x^{j^{\prime }},y^{a^{\prime }})e_{a^{\prime
}}^{~\underline{a^{\prime }}}(x^{j^{\prime }},y^{a^{\prime }}) \\
0 & e_{a^{\prime }}^{~\underline{a^{\prime }}}(x^{j^{\prime }},y^{a^{\prime
}})%
\end{array}%
\right) .  \label{frame8}
\end{equation}%
For the metric%
\begin{equation}
\mathbf{g}=\underline{g}_{\alpha \beta }du^{\alpha }\otimes du^{\beta }
\label{auxm01}
\end{equation}%
we have the recurrent ansatz%
\begin{equation}
\underline{g}_{\alpha \beta }=\left[
\begin{array}{cc}
g_{\alpha ^{\prime }\beta ^{\prime }}(u^{\beta ^{\prime }})+A_{\alpha
^{\prime }}^{\widehat{\alpha }}(u^{\beta ^{\prime }})A_{\beta ^{\prime }}^{%
\widehat{\beta }}(u^{\beta ^{\prime }})h_{\widehat{\alpha }\widehat{\beta }%
}(u^{\beta ^{\prime }}) & h_{\widehat{\alpha }\widehat{\beta }}(u^{\beta
^{\prime }})A_{\alpha ^{\prime }}^{\widehat{\alpha }}(u^{\beta ^{\prime }})
\\
h_{\widehat{\alpha }\widehat{\beta }}(u^{\beta ^{\prime }})A_{\beta ^{\prime
}}^{\widehat{\beta }}(u^{\beta ^{\prime }}) & h_{\widehat{\alpha }\widehat{%
\beta }}(u^{\beta ^{\prime }})%
\end{array}%
\right] ,  \label{metr8a}
\end{equation}%
where%
\begin{equation}
g_{\alpha ^{\prime }\beta ^{\prime }}=\left[
\begin{array}{cc}
g_{i^{\prime }j^{\prime }}(u^{\beta ^{\prime }})+N_{i^{\prime }}^{a^{\prime
}}(u^{\beta ^{\prime }})N_{j^{\prime }}^{b^{\prime }}(u^{\beta ^{\prime
}})h_{a^{\prime }b^{\prime }}(u^{\beta ^{\prime }}) & h_{a^{\prime
}b^{\prime }}(u^{\beta ^{\prime }})N_{i^{\prime }}^{a^{\prime }}(u^{\beta
^{\prime }}) \\
h_{a^{\prime }b^{\prime }}(u^{\beta ^{\prime }})N_{j^{\prime }}^{b^{\prime
}}(u^{\beta ^{\prime }}) & h_{a^{\prime }b^{\prime }}(u^{\beta ^{\prime }})%
\end{array}%
\right] .  \label{metr8}
\end{equation}

After a toroidal compactification on $u^{\widehat{\alpha }}$ with gauge
fields $A_{\alpha ^{\prime }}^{\widehat{\alpha }}(u^{\beta ^{\prime }}),$
generated by the frame transform (\ref{vt1a}), we obtain a metric (\ref%
{auxm01}) like in the usual Kaluza--Klein theory (\ref{anskk}) but
containing the values $g_{\alpha ^{\prime }\beta ^{\prime }}(u^{\beta
^{\prime }}),$\ defined as in (\ref{metr8}) (in a generic off--diagonal form
similar to (\ref{ansatz}), labeled by primed indices), which can be induced
as in Finsler geometry. This is possible if $g_{i^{\prime }j^{\prime
}}(u^{\beta ^{\prime }}),h_{a^{\prime }b^{\prime }}(u^{\beta ^{\prime
}})\rightarrow g_{i^{\prime }j^{\prime }}^{[F]}(x^{\prime },y^{\prime })$
(see (\ref{finm2})) and $N_{i^{\prime }}^{a^{\prime }}(u^{\beta ^{\prime
}})\rightarrow N_{j^{\prime }}^{[F]i^{\prime }}(x^{\prime },y^{\prime })$
(see (\ref{ncc})) inducing a Finsler space with ''primed'' labels for
objects. Such locally anisotropic spacetimes (in this case we emphasized the
Finsler structures) can be generated anisotropic toroidal compactifications
from different models of higher dimension of gravity (string, brane, or
usual Kaluza--Klein theories). \ They define a mixed variant of Finsler and
Kaluza--Klein spaces.

By using the recurent ansatz (\ref{metr8a}) and (\ref{metr8}), we can
generate both nontrivial nonmetricity and prescribed torsion structures
adapted to a corresponding N--connection $N_{i^{\prime }}^{a^{\prime }}.$
For instance, (after topological compactification on higher dimension) we
can prescribe in the lower dimensional spacetime certain torsion fields $%
T_{\ i^{\prime }j^{\prime }}^{k^{\prime }}$ and $T_{\ b^{\prime }c^{\prime
}}^{a^{\prime }}$ (they could have a particular relation to the so called $B$%
--fields in string theory, or connected to other models). The next steps are
to compute $\tau _{\ i^{\prime }j^{\prime }}^{k^{\prime }}$ and $\tau _{\
b^{\prime }c^{\prime }}^{a^{\prime }}$ by using formulas (\ref{tauformulas})
and define
\begin{equation}
^{\lbrack B\tau ]}\mathbf{\Gamma }_{\alpha ^{\prime }\beta ^{\prime
}}^{\gamma ^{\prime }}=\left( L_{j^{\prime }k^{\prime }}^{i^{\prime }}=%
\widehat{L}_{\ j^{\prime }k^{\prime }}^{i^{\prime }}+\tau _{\ j^{\prime
}k^{\prime }}^{i^{\prime }},\ L_{.b^{\prime }k^{\prime }}^{a^{\prime }}=%
\frac{\partial N_{k^{\prime }}^{a^{\prime }}}{\partial y^{b^{\prime }}}%
,C_{.j^{\prime }a^{\prime }}^{i^{\prime }}=0,C_{b^{\prime }c^{\prime
}}^{a^{\prime }}=\widehat{C}_{\ b^{\prime }c^{\prime }}^{a^{\prime }}+\tau
_{\ b^{\prime }c^{\prime }}^{a^{\prime }}\right)  \label{berwprim}
\end{equation}%
as in (\ref{bct}) \ (all formulas being with primed indices and $\widehat{L}%
_{\ j^{\prime }k^{\prime }}^{i^{\prime }}$ and $\widehat{C}_{\ b^{\prime
}c^{\prime }}^{a^{\prime }}$ defined as in (\ref{candcon})). This way we can
generate from Kaluza--Klein/ string theory a Berwald spacetime with
nontrivial N--adapted nonmetricity
\begin{equation*}
^{\lbrack B\tau ]}\mathbf{Q}_{\alpha ^{\prime }\beta ^{\prime }\gamma
^{\prime }}=\ ^{[B\tau ]}\mathbf{Dg}_{\beta ^{\prime }\gamma ^{\prime
}}=\left( ^{[B\tau ]}Q_{c^{\prime }i^{\prime }j^{\prime }},\ ^{[B\tau
]}Q_{i^{\prime }a^{\prime }b^{\prime }}\right)
\end{equation*}%
and torsions \ $^{[B\tau ]}\mathbf{T}_{\ \beta ^{\prime }\gamma ^{\prime
}}^{\alpha ^{\prime }}$ with h- and v-- irreducible components
\begin{eqnarray}
T_{.j^{\prime }k^{\prime }}^{i^{\prime }} &=&-T_{k^{\prime }j^{\prime
}}^{i^{\prime }}=L_{j^{\prime }k^{\prime }}^{i^{\prime }}-L_{k^{\prime
}j^{\prime }}^{i^{\prime }},\quad T_{j^{\prime }a^{\prime }}^{i^{\prime
}}=-T_{a^{\prime }j^{\prime }}^{i^{\prime }}=C_{.j^{\prime }a^{\prime
}}^{i^{\prime }},\ T_{.i^{\prime }j^{\prime }}^{a^{\prime }}=\frac{\delta
N_{i^{\prime }}^{a^{\prime }}}{\delta x^{j^{\prime }}}-\frac{\delta
N_{j^{\prime }}^{a^{\prime }}}{\delta x^{i^{\prime }}},  \notag \\
\quad T_{.b^{\prime }i^{\prime }}^{a^{\prime }} &=&-T_{.i^{\prime }b^{\prime
}}^{a^{\prime }}=\frac{\partial N_{i^{\prime }}^{a^{\prime }}}{\partial
y^{b^{\prime }}}-L_{.b^{\prime }j^{\prime }}^{a^{\prime }},\ T_{.b^{\prime
}c^{\prime }}^{a^{\prime }}=-T_{.c^{\prime }b^{\prime }}^{a^{\prime
}}=C_{b^{\prime }c^{\prime }}^{a^{\prime }}-C_{c^{\prime }b^{\prime
}}^{a^{\prime }}.\   \label{dtorsions}
\end{eqnarray}%
defined by the h- and v--coefficients of (\ref{berwprim}).

We conclude that if toroidal compactifications are locally anisotropic,
defined by a chain of ansatz containing N--connection, the lower dimensional
spacetime can be not only with torsion structure (like in low energy limit
of string theory) but also with nonmetricity. The anholonomy induced by
N--connection gives the possibility to define a more wide class of linear
connections adapted to the h- and v--splitting.

\subsubsection{Finsler--Riemann--Cartan spaces}

\label{mfls} Such spacetimes are modeled as Riemann--Cartan geometries on a
tangent bundle $TM$ when the metric and anholonomic frame structures
distinguished to be of Finsler type (\ref{dmetricf}). Both Finsler and
Riemann--Cartan spaces possess nontrivial torsion structures (see section %
\ref{torscurv} for details on definition and computation torsions of locally
anisotropic spaces and Refs. \cite{rcg} for a review of the Einstein--Cartan
gravity). The fundamental geometric objects defining
Finsler--Riemann--Cartan spaces consists in the triple $\left( \mathbf{g}%
^{[F]},\mathbf{\vartheta }_{[F]}^{\alpha },\mathbf{\Gamma }_{[F]\alpha \beta
}^{\gamma }\right) $ where $\mathbf{g}^{[F]}$ is a d--metric (\ref{dmetricf}%
), $\mathbf{\vartheta }_{[F]}^{\alpha }=\left( dx^{i},\delta
y^{j}=dy^{j}+N_{[F]k}^{j}\left( x^{l},y^{s}\right) dx^{k}\right) $ with $%
N_{[F]k}^{j}\left( x^{l},y^{s}\right) $ of type (\ref{ncc}) and $\mathbf{%
\Gamma }_{[F]\alpha \beta }^{\gamma }$ is an arbitrary d--connection (\ref%
{dcon1}) on $TM$ (we put the label [F] emphasizing that the N--connection is
a Finsler type one). The torsion $\mathbf{T}_{[F]}^{\alpha }$ and curvature $%
\mathbf{R}_{[F]\beta }^{\alpha }$ d--forms are computed following
respectively the formulas (\ref{dt}) and \ (\ref{dc}) but for $\mathbf{%
\vartheta }_{[F]}^{\alpha }$ and $\mathbf{\Gamma }_{[F]\alpha \beta
}^{\gamma }.$

We can consider an inverse modeling of geometries when (roughly speaking)
the Finsler configurations are 'hidden' in Riemann--Cartan spaces. They can
be distinguished for arbitrary Riemann--Cartan manifolds $V^{n+n}$
coventionally splitted into ''horizontal'' and ''vertical'' subspaces and
provided with a metric ansatz of type (\ref{dmetricf}) and with prescribed
procedure of adapting the geometric objects to the Cartan N--connection $%
N_{[F]k}^{j}.$ Of course, the torsion can not be an arbitrary one but
admitting irreducible decompositions with respect to N--frames $\mathbf{e}%
_{\alpha }^{[F]}$ and N--coframes $\mathbf{\vartheta }_{[F]}^{\alpha }$
(see, respectively, the formulas (\ref{dder}) \ and (\ref{ddif}) when $%
N_{i}^{a}\rightarrow N_{[F]i}^{j}).\,$\ There were constructed and
investigated different classes of exact solutions of the Einstein equations
with anholonomic variables characterized by anholonomically induced torsions
and modelling Finsler like geometries in (pseudo)\ Riemannian and
Riemann--Cartan spaces (see Refs. \cite{v1,v2,vd}). All constructions from
Finsler--Riemann--Cartan geometry reduce to Finsler--Riemann configurations
(in general, we can see metrics of arbitrary signatures) if $\mathbf{\Gamma }%
_{[F]\alpha \beta }^{\gamma }$ is changed into the Levi--Civita metric
connection defined with respect to anholonomic frames $\mathbf{e}_{\alpha }$
and coframes $\mathbf{\vartheta }^{\alpha }$ when the N--connection
curvature $\Omega _{jk}^{i}$ and the anholonomically induced torsion vanish.

\subsubsection{Teleparallel generalized Finsler geometry}

\label{stpfa}In Refs. \cite{vargtor} the teleparallel Finsler connections,
the Cartan--Einstein unification in the teleparallel approach and related
moving frames with Finsler structures were investigated. In our analysis of
teleparallel geometry we heavily use the results on N--connection geometry
in order to illustrate how the teleparallel and metric affine gavity \cite%
{tgm} can defined as to include generalized Finsler structures. For a
general metric--affine spaces admitting N--connection structure $N_{i}^{a},$
the curvature $\mathbf{R}_{.\beta \gamma \tau }^{\alpha }$ of an arbitrary
d--connection $\mathbf{\Gamma }_{\alpha \beta }^{\gamma }=\left(
L_{jk}^{i},L_{bk}^{a},C_{jc}^{i},C_{bc}^{a}\right) $ splits into h-- and
v--irreversible components, $\mathbf{R}_{.\beta \gamma \tau }^{\alpha
}=(R_{\ hjk}^{i},R_{\ bjk}^{a},P_{\ jka}^{i},P_{\ bka}^{c},S_{\
jbc}^{i},S_{\ bcd}^{a}),$ see (\ref{dcurv}). In order to include Finsler
like metrics, we state that the N--connection curvature can be nontrivial $%
\Omega _{jk}^{a}\neq 0,$ which is quite different from the condition imposed
in section \ref{stps}. The condition of vanishing of curvature for
teleparallel spaces, see (\ref{dtel1}), is to be stated separately for every
h- v--irreversible component,%
\begin{equation*}
R_{\ hjk}^{i}=0,R_{\ bjk}^{a}=0,P_{\ jka}^{i}=0,P_{\ bka}^{c}=0,S_{\
jbc}^{i}=0,S_{\ bcd}^{a}=0.
\end{equation*}%
We can define certain types of teleparallel Berwald connections (see
sections \ref{berwdcon} and \ref{berwdcona}) with certain nontrivial
components of nonmetricity d--field (\ref{berwnm}) if we modify the metric
compatibility conditions (\ref{dtel2}) into a less strong one when
\begin{equation*}
Q_{kij}=-D_{k}g_{ij}=0\mbox{ and }Q_{abc}=-D_{a}h_{bc}=0
\end{equation*}%
but with nontrivial components
\begin{equation*}
\mathbf{Q}_{\alpha \beta \gamma }=\left( Q_{cij}=-D_{c}g_{ij},\
Q_{iab}=-D_{i}h_{ab}\right) .
\end{equation*}

The class of teleparallel Finsler spaces is distinguished by Finsler
N--connection and d--connection $^{[F]}\mathbf{N}_{j}^{i}(x,y)$ and $^{[F]}%
\widehat{\mathbf{\Gamma }}_{\ \beta \gamma }^{\alpha }=\left( \ ^{[F]}%
\widehat{L}_{jk}^{i},\ ^{[F]}\widehat{L}_{jk}^{i},\ ^{[F]}\widehat{C}%
_{jk}^{i},\ ^{[F]}\widehat{C}_{jk}^{i}\right) ,$ see, respectively, (\ref%
{ncc}) and (\ref{dccfs}) with vanishing d--curvature components,
\begin{equation*}
~^{[F]}R_{\ hjk}^{i}=0,~^{[F]}P_{\ jka}^{i}=0,~^{[F]}S_{\ jbc}^{i}=0.
\end{equation*}%
We can generate teleparallel Finsler affine structures if it is not imposed
the condition of vanishing of nonmetricity d--field. In this case, there are
considered arbitrary d--connections\ $\mathbf{D}_{\alpha }$ that for the
induced Finsler quadratic form (\ref{dmetricf}) $\mathbf{g}^{[F]}$
\begin{equation*}
~^{[F]}\mathbf{Q}_{\alpha \beta \gamma }=-\mathbf{D}_{\alpha }\mathbf{g}%
^{[F]}\neq 0
\end{equation*}%
but $\mathbf{R}_{.\beta \gamma \tau }^{\alpha }\left( \mathbf{D}\right) =0.$

The teleparallel--Finsler configurations are contained as particular cases
of Finsler--affine spaces, see section \ref{stpfa}. For vielbein fields $%
\mathbf{e}_{\alpha }^{~\underline{\alpha }}$ and their inverses $\mathbf{e}%
_{~\underline{\alpha }}^{\alpha }$ related to the d--metric (\ref{dmetricf}),%
\begin{equation*}
\mathbf{g}_{\alpha \beta }^{[F]}=\mathbf{\tilde{e}}_{\alpha }^{~\underline{%
\alpha }}\mathbf{\tilde{e}}_{\beta }^{~\underline{\beta }}g_{\underline{%
\alpha }\underline{\beta }}
\end{equation*}%
we define the Weitzenbock--Finsler d--connection%
\begin{equation}
~^{[WF]}\mathbf{\Gamma }_{~\beta \gamma }^{\alpha }=\mathbf{\tilde{e}}_{~%
\underline{\alpha }}^{\alpha }\delta _{\gamma }\mathbf{\tilde{e}}_{\beta }^{~%
\underline{\alpha }}  \label{wfdcon}
\end{equation}%
where $\delta _{\gamma }$ are the elongated by $^{[F]}\mathbf{N}%
_{j}^{i}(x,y) $ partial derivatives (\ref{dder}). The torsion of $~^{[WF]}%
\mathbf{\Gamma }_{~\beta \gamma }^{\alpha }$ is defined
\begin{equation}
~^{[WF]}\mathbf{T}_{~\beta \gamma }^{\alpha }=~^{[WF]}\mathbf{\Gamma }%
_{~\beta \gamma }^{\alpha }-~^{[WF]}\mathbf{\Gamma }_{~\gamma \beta
}^{\alpha }  \label{wftors}
\end{equation}%
containing h-- and v--irreducible components being constructed from the
components of a d--metric and N--adapted frames. We can express
\begin{equation*}
~^{[WF]}\mathbf{\Gamma }_{~\beta \gamma }^{\alpha }=\mathbf{\Gamma }%
_{\bigtriangledown ~\beta \gamma }^{\alpha }+\mathbf{\hat{Z}}_{~\beta \gamma
}^{\alpha }+\mathbf{Z}_{~\beta \gamma }^{\alpha }
\end{equation*}%
where $\mathbf{\Gamma }_{\bigtriangledown ~\beta \gamma }^{\alpha }$ is the
Levi--Civita connection (\ref{lcsym}), $\mathbf{\hat{Z}}_{~\beta \gamma
}^{\alpha }=^{[F]}\widehat{\mathbf{\Gamma }}_{\ \beta \gamma }^{\alpha }-%
\mathbf{\Gamma }_{\bigtriangledown ~\beta \gamma }^{\alpha },$ and the
contorsion tensor is
\begin{equation*}
\mathbf{Z}_{\alpha \beta }=\mathbf{e}_{\beta }\rfloor ~^{[W]}\mathbf{T}%
_{\alpha }-\mathbf{e}_{\alpha }\rfloor ~^{[W]}\mathbf{T}_{\beta }+\frac{1}{2}%
\left( \mathbf{e}_{\alpha }\rfloor \mathbf{e}_{\beta }\rfloor ~^{[W]}\mathbf{%
T}_{\gamma }\right) \mathbf{\vartheta }^{\gamma }+\left( \mathbf{e}_{\alpha
}\rfloor \mathbf{Q}_{\beta \gamma }\right) \mathbf{\vartheta }^{\gamma
}-\left( \mathbf{e}_{\beta }\rfloor \mathbf{Q}_{\alpha \gamma }\right)
\mathbf{\vartheta }^{\gamma }+\frac{1}{2}\mathbf{Q}_{\alpha \beta }.
\end{equation*}%
In the non-Berwald standard approaches to the Finsler--teleparallel gravity
it is considered that $\mathbf{Q}_{\alpha \beta }=0.$

\subsubsection{Cartan geometry}

\label{scs}The theory of Cartan spaces (see, for instance, \cite{rund,kaw1})
\ can be reformulated as a dual to Finsler geometry \cite{mironc} (see
details and references in \cite{mhss}). The Cartan space is constructed on a
cotangent bundle $T^{\ast }M$ similarly to the Finsler space on the tangent
bundle $TM.$

Consider a real smooth manifold $M,$ the cotangent bundle $\left( T^{\ast
}M,\pi ^{\ast },M\right) $ and the manifold $\widetilde{T^{\ast }M}=T^{\ast
}M\backslash \{0\}.$

\begin{definition}
A Cartan space is a pair $C^{n}=\left( M,K(x,p)\right) $ \ such that $%
K:T^{\ast }M\rightarrow \R$ is a scalar function satisfying the following
conditions:

\begin{enumerate}
\item $K$ is a differentiable function on the manifold $\widetilde{T^{\ast }M%
}$ $=T^{\ast }M\backslash \{0\}$ and continuous on the null section of the
projection $\pi ^{\ast }:T^{\ast }M\rightarrow M;$

\item $K$ is a positive function, homogeneous on the fibers of the $T^{\ast
}M,$ i. e. $K(x,\lambda p)=\lambda F(x,p),\lambda \in \R;$

\item The Hessian of $K^{2}$ with elements
\begin{equation}
\check{g}_{[K]}^{ij}(x,p)=\frac{1}{2}\frac{\partial ^{2}K^{2}}{\partial
p_{i}\partial p_{j}}  \label{carm}
\end{equation}%
is positively defined on $\widetilde{T^{\ast }M}.$
\end{enumerate}
\end{definition}

The function $K(x,y)$ and $\check{g}^{ij}(x,p)$ are called \ respectively
the fundamental function and the fundamental (or metric) tensor of the
Cartan space $C^{n}.$ We use symbols like $"\check{g}"$ as to emphasize that
the geometrical objects are defined on a dual space.

One considers ''anisotropic'' (depending on directions, momenta, $p_{i})$
Christoffel symbols. For simplicity, we write the inverse to (\ref{carm}) as
$g_{ij}^{(K)}=\check{g}_{ij}$ and introduce the coefficients
\begin{equation*}
\check{\gamma}_{~jk}^{i}(x,p)=\frac{1}{2}\check{g}^{ir}\left( \frac{\partial
\check{g}_{rk}}{\partial x^{j}}+\frac{\partial \check{g}_{jr}}{\partial x^{k}%
}-\frac{\partial \check{g}_{jk}}{\partial x^{r}}\right) ,
\end{equation*}%
defining the canonical N--connection $\mathbf{\check{N}=\{}\check{N}_{ij}%
\mathbf{\},}$
\begin{equation}
\check{N}_{ij}^{[K]}=\check{\gamma}_{~ij}^{k}p_{k}-\frac{1}{2}\gamma
_{~nl}^{k}p_{k}p^{l}{\breve{\partial}}^{n}\check{g}_{ij}  \label{nccartan}
\end{equation}%
where $~{\breve{\partial}}^{n}=\partial /\partial p_{n}.$ The N--connection $%
\mathbf{\check{N}}=\{\check{N}_{ij}\}$ can be used for definition of an
almost complex structure like in (\ref{dmetricf}) and introducing on $%
T^{\ast }M$ a d--metric
\begin{equation}
\mathbf{\check{G}}_{[k]}=\check{g}_{ij}(x,p)dx^{i}\otimes dx^{j}+\check{g}%
^{ij}(x,p)\delta p_{i}\otimes \delta p_{j},  \label{dmcar}
\end{equation}%
with $\check{g}^{ij}(x,p)$ taken as (\ref{carm}).

Using the canonical N--connection (\ref{nccartan}) and Finsler metric tensor
(\ref{carm}) (or, equivalently, the d--metric (\ref{dmcar})), we can define
the canonical d--connection $\mathbf{\check{D}=\{\check{\Gamma}}\left(
\check{N}_{[k]}\right) \mathbf{\}}$
\begin{equation*}
\mathbf{\check{\Gamma}}\left( \check{N}_{[k]}\right) =\check{\Gamma}%
_{[k]\beta \gamma }^{\alpha }=\left( \check{H}_{[k]~jk}^{i},\check{H}%
_{[k]~jk}^{i},\check{C}_{[k]~i}^{\quad jk},\check{C}_{[k]~i}^{\quad
jk}\right)
\end{equation*}%
with the coefficients computed
\begin{equation*}
\check{H}_{[k]~jk}^{i}=\frac{1}{2}\check{g}^{ir}\left( \check{\delta}_{j}%
\check{g}_{rk}+\check{\delta}_{k}\check{g}_{jr}-\check{\delta}_{r}\check{g}%
_{jk}\right) ,\ \check{C}_{[k]~i}^{\quad jk}=\check{g}_{is}{\breve{\partial}}%
^{s}\check{g}^{jk}.
\end{equation*}%
The d--connection $\mathbf{\check{\Gamma}}\left( \check{N}_{(k)}\right) $
satisfies the metricity conditions both for the horizontal and vertical
components, i. e. $\mathbf{\check{D}}_{\alpha }\mathbf{\check{g}}_{\beta
\gamma }=0.$

The d--torsions (\ref{dtorsions}) \ and d--curvatures (\ref{dcurv}) are
computed like in Finsler geometry but starting from the coefficients in (\ref%
{nccartan}) and (\ref{dmcar}), when the indices $a,b,c...$ run the same
values as indices $i,j,k,...$ and the geometrical objects are modeled as on
the dual tangent bundle. It should be emphasized that in this case all
values $\check{g}_{ij,}\ \check{\Gamma}_{[k]\beta \gamma }^{\alpha }$ and $%
\check{R}_{[k]\beta \gamma \delta }^{.\alpha }$ are defined by a fundamental
function $K\left( x,p\right) .$

In general, we can consider that a Cartan space is provided with a metric $%
\check{g}^{ij}=\partial ^{2}K^{2}/2\partial p_{i}\partial p_{j},$ but the
N--connection and d--connection could be defined in a different manner, even
not be determined by $K.$ If a Cartan space is modeled in a metric--affine
space $V^{n+n},$ with local coordinates $\left( x^{i},y^{k}\right) ,$ we
have to define a procedure of dualization of vertical coordinates, $%
y^{k}\rightarrow p_{k}.$

\subsection{Generalized Lagrange and Hamilton geometries}

The notion of Finsler spaces was extended by J. Kern \cite{ker} and R. Miron %
\cite{mironlg}. It is was elaborated in vector bundle spaces in Refs. \cite%
{ma} and generalized to superspaces \cite{vsup}. We illustrate how such
geometries can be modeled on a space $\mathbf{V}^{n+n}$ provided with
N--connection structure.

\subsubsection{Lagrange geometry and generalizations}

\label{ssslgg}The idea of generalization of the Finsler geometry was to
consider instead of the homogeneous fundamental function $F(x,y)$ in a
Finsler space a more general one, a Lagrangian $L\left( x,y\right) $,
defined as a differentiable mapping $L:(x,y)\in TV^{n+n}\rightarrow
L(x,y)\in \R,$ of class $C^{\infty }$ on manifold $\widetilde{TV}^{n+n}$ and
continuous on the null section $0:V^{n}\rightarrow \widetilde{TV}^{n+n}$ of
the projection $\pi :\widetilde{TV}^{n+n}\rightarrow V^{n}.$ A Lagrangian is
regular if it is differentiable and the Hessian
\begin{equation}
g_{ij}^{[L]}(x,y)=\frac{1}{2}\frac{\partial ^{2}L^{2}}{\partial
y^{i}\partial y^{j}}  \label{lagm}
\end{equation}%
is of rank $n$ on $V^{n}.$

\begin{definition}
\label{defls}A Lagrange space is a pair $\mathbf{L}^{n}=\left(
V^{n},L(x,y)\right) $ where $V^{n}$ is a smooth real $n$--dimensional
manifold provided with regular Lagrangian \ $L(x,y)$ structure $%
L:TV^{n}\rightarrow \R$ $\ $for which $g_{ij}(x,y)$ from (\ref{lagm}) has a
constant signature over the manifold $\widetilde{TV}^{n+n}.$
\end{definition}

The fundamental Lagrange function $L(x,y)$ defines a canonical
N--con\-nec\-ti\-on
\begin{equation}
\ ^{[cL]}N_{~j}^{i}=\frac{1}{2}\frac{\partial }{\partial y^{j}}\left[
g^{ik}\left( \frac{\partial ^{2}L^{2}}{\partial y^{k}\partial y^{h}}y^{h}-%
\frac{\partial L}{\partial x^{k}}\right) \right]  \label{cncls}
\end{equation}%
as well a d-metric
\begin{equation}
\mathbf{g}_{[L]}=g_{ij}(x,y)dx^{i}\otimes dx^{j}+g_{ij}(x,y)\delta
y^{i}\otimes \delta y^{j},  \label{dmlag}
\end{equation}%
with $g_{ij}(x,y)$ taken as (\ref{lagm}). As well we can introduce an almost
K\"{a}hlerian structure and an almost Hermitian model of $\mathbf{L}^{n},$
denoted as $\mathbf{H}^{2n}$ as in the case of Finsler spaces but with a
proper fundamental Lagrange function and metric tensor $g_{ij}.$ The
canonical metric d--connection $\widehat{\mathbf{D}}_{[L]}$ is defined by
the coefficients
\begin{equation}
\ ^{[L]}\widehat{\mathbf{\Gamma }}_{\beta \gamma }^{\alpha }=\left( \ ^{[L]}%
\widehat{L}_{~jk}^{i},\ ^{[L]}\widehat{L}_{~jk}^{i},\ ^{[L]}\widehat{C}%
_{~jk}^{i},\ ^{[L]}\widehat{C}_{~jk}^{i}\right)  \label{lagdcon}
\end{equation}%
computed for $N_{[cL]~j}^{i}$ and by respective formulas (\ref{candcon})
with $h_{ab}\rightarrow g_{ij}^{[L]}$ and $\widehat{C}_{bc}^{a}\rightarrow $
$\widehat{C}_{ij}^{i}.$ The d--torsions (\ref{dtorsions}) and d--curvatures (%
\ref{dcurv}) are determined, in this case, by $\ ^{[L]}\widehat{L}_{~jk}^{i}$
and $\ ^{[L]}\widehat{C}_{~jk}^{i}.$ We also note that instead of $\
^{[cL]}N_{~j}^{i}$ and $\ ^{[L]}\widehat{\mathbf{\Gamma }}_{\ \beta \gamma
}^{\alpha }$ we can consider on a $L^{n}$--space different N--connections $%
N_{~j}^{i},$ d--connections $\mathbf{\Gamma }_{\ \beta \gamma }^{\alpha }$
which are not defined only by $L(x,y)$ and $g_{ij}^{[L]}$ but can be metric,
or non--metric with respect to the Lagrange metric.

The next step of generalization \cite{mironlg} is to consider an arbitrary
metric $g_{ij}\left( x,y\right) $ on $\mathbf{TV}^{n+n}$ (we use boldface
symbols in order to emphasize that the space is enabled with N--connection
structure) instead of (\ref{lagm}) which is the second derivative of
''anisotropic'' coordinates $y^{i}$ of a Lagrangian.

\begin{definition}
\label{defgls}A generalized Lagrange space is a pair $\mathbf{GL}^{n}=\left(
V^{n},g_{ij}(x,y)\right) $ where $g_{ij}(x,y)$ is a covariant, symmetric and
N--adapted d--tensor field of rank $n$ and of constant signature on $%
\widetilde{TV}^{n+n}.$
\end{definition}

One can consider different classes of N-- and d--connections on $TV^{n+n},$
which are compatible (metric) or non compatible with (\ref{dmlag}) for
arbitrary $g_{ij}(x,y)$ and arbitrary d--metric
\begin{equation}
\mathbf{g}_{[gL]}=g_{ij}(x,y)dx^{i}\otimes dx^{j}+g_{ij}(x,y)\delta
y^{i}\otimes \delta y^{j},  \label{dmgls}
\end{equation}%
We can apply all formulas for d--connections, N--curvatures, d--torsions and
d--curvatures as in sections \ref{dlcms} and \ref{torscurv} but
reconsidering them on $\mathbf{TV}^{n+n},$ by changing \
\begin{equation*}
h_{ab}\rightarrow g_{ij}(x,y),\widehat{C}_{bc}^{a}\rightarrow \widehat{C}%
_{ij}^{i}\mbox{ and } N_{i}^{a}\rightarrow N_{~i}^{k}.
\end{equation*}
Prescribed torsions $T_{\ jk}^{i}$ and $S_{\ jk}^{i}$ can be introduced on $%
\mathbf{GL}^{n}$ by using the d--connection%
\begin{equation}
\ \widehat{\mathbf{\Gamma }}_{\beta \gamma }^{\alpha }=\left( \widehat{L}%
_{[gL]~jk}^{i}+\tau _{\ jk}^{i},\widehat{L}_{[gL]~jk}^{i}+\tau _{\ jk}^{i},%
\widehat{C}_{[gL]~jk}^{i}+\sigma _{\ jk}^{i},\widehat{C}_{[gL]~jk}^{i}+%
\sigma _{\ jk}^{i}\right)  \label{ptdcls}
\end{equation}%
with
\begin{equation}
\tau _{\ jk}^{i}=\frac{1}{2}g^{il}\left(
g_{kh}T_{.lj}^{h}+g_{jh}T_{.lk}^{h}-g_{lh}T_{\ jk}^{h}\right) \mbox{ and }%
\sigma _{\ jk}^{i}=\frac{1}{2}g^{il}\left(
g_{kh}S_{.lj}^{h}+g_{jh}S_{.lk}^{h}-g_{lh}S_{\ jk}^{h}\right)  \notag
\end{equation}%
like we have performed for the Berwald connections (\ref{bct}) with (\ref%
{tauformulas}) and (\ref{berwprim}) but in our case
\begin{equation}
\ ^{[aL]}\widehat{\mathbf{\Gamma }}_{\beta \gamma }^{\alpha }=\left(
\widehat{L}_{[gL]~jk}^{i},\widehat{L}_{[gL]~jk}^{i},\widehat{C}%
_{[gL]~jk}^{i},\widehat{C}_{[gL]~jk}^{i}\right)  \label{dccls}
\end{equation}%
is metric compatible being modeled like on a tangent bundle and with the
coefficients computed as in (\ref{candcon}) with $h_{ab}\rightarrow
g_{ij}^{[L]}$ and $\widehat{C}_{bc}^{a}\rightarrow $ $\widehat{C}_{ij}^{i},$
by using the d--metric $\mathbf{G}_{[gL]}$ (\ref{dmgls}). The connection (%
\ref{ptdcls}) is a Riemann--Cartan one modeled on effective tangent bundle
provided with N--connection structure.

\subsubsection{Hamilton geometry and generalizations}

\label{shgg}The geometry of Hamilton spaces was defined and investigated by
R. Miron in the papers \cite{mironh} (see details and additional references
in \cite{mhss}). It was developed on the cotangent bundle as a dual geometry
to the geometry of Lagrange spaces. Here we consider their modeling on
couples of spaces $\left( V^{n},\ ^{\ast }V^{n}\right) ,$ or cotangent
bundle $T^{\ast }M,$ where $\ ^{\ast }V^{n}$ is considered as a 'dual'
manifold defined by local coordinates satisfying a duality condition with
respect to coordinates on $V^{n}.$\ We start with the definition of
generalized Hamilton spaces and then consider the particular cases.

\begin{definition}
A generalized Hamilton space is a pair $\mathbf{GH}^{n}=\left( V^{n},\check{g%
}^{ij}(x,p)\right) $ where $V^{n}$ is a real $n$--dimensional manifold and $%
\check{g}^{ij}(x,p)$ is a contravariant, symmetric, nondegenerate of rank $n$
and of constant signature on $\widetilde{T^{\ast }V}^{n+n}.$
\end{definition}

The value $\check{g}^{ij}(x,p)$ is called the fundamental (or metric) tensor
of the space $\mathbf{GH}^{n}.$ One can define such values for every
paracompact manifold $V^{n}.$ In general, a N--connection on $\mathbf{GH}%
^{n} $ is not determined by $\check{g}^{ij}.$ Therefore we can consider an
arbitrary N--connection $\mathbf{\check{N}}=\{\check{N}_{ij}\left(
x,p\right) \}$ and define on $T^{\ast }V^{n+n}$ a d--metric similarly to (%
\ref{block2}) and/or (\ref{dmlag})
\begin{equation}
{\breve{\mathbf{G}_{[gH]}}}={\breve{g}}_{\alpha \beta }\left( {\breve{u}}%
\right) {\breve{\delta}}^{\alpha }\otimes {\breve{\delta}}^{\beta }={%
\breve{g}}_{ij}\left( {\breve{u}}\right) d^{i}\otimes d^{j}+{\check{g}}%
^{ij}\left( {\breve{u}}\right) {\breve{\delta}}_{i}\otimes {\breve{\delta}}%
_{j},  \label{dmghs}
\end{equation}%
The N--coefficients $\check{N}_{ij}\left( x,p\right) $ and the d--metric
structure (\ref{dmghs}) define an almost K\"{a}hler model of generalized
Hamilton spaces provided with canonical d--connections, d--torsions and
d-curvatures (see respectively the formulas d--torsions (\ref{dtorsions})
and d--curvatures (\ref{dcurv}) with the fiber coefficients redefined for
the cotangent bundle $T^{\ast }V$ $^{n+n}$).

A generalized Hamilton space $\mathbf{GH}^{n}$ is called reducible to a
Hamilton one if there exists a Hamilton function $H\left( x,p\right) $ on $%
T^{\ast }V$ $^{n+n}$ such that
\begin{equation}
\check{g}_{[H]}^{ij}(x,p)=\frac{1}{2}\frac{\partial ^{2}H}{\partial
p_{i}\partial p_{j}}.  \label{hsm}
\end{equation}

\begin{definition}
A Hamilton space is a pair $\mathbf{H}^{n}=\left( V^{n},H(x,p)\right) $ \
such that $H:T^{\ast }V^{n}\rightarrow \R$ is a scalar function which
satisfy the following conditions:

\begin{enumerate}
\item $H$ is a differentiable function on the manifold $\widetilde{T^{\ast }V%
}^{n+n}$ $=T^{\ast }V^{n+n}\backslash \{0\}$ and continuous on the null
section of the projection $\pi ^{\ast }:T^{\ast }V^{n+n}\rightarrow V^{n};$

\item The Hessian of $H$ with elements (\ref{hsm}) is positively defined on $%
\widetilde{T^{\ast }V}^{n+n}$ and $\check{g}^{ij}(x,p)$ is nondegenerate
matrix of rank $n$ and of constant signature.
\end{enumerate}
\end{definition}

For Hamilton spaces, the canonical N--connection (defined by $H$ and its
Hessian) is introduced as
\begin{equation}
\ ^{[H]}\check{N}_{ij}=\frac{1}{4}\{\check{g}_{ij},H\}-\frac{1}{2}\left(
\check{g}_{ik}\frac{\partial ^{2}H}{\partial p_{k}\partial x^{j}}+\check{g}%
_{jk}\frac{\partial ^{2}H}{\partial p_{k}\partial x^{i}}\right) ,
\label{ncchs}
\end{equation}%
where the Poisson brackets, for arbitrary functions $f$ and $g$ on $T^{\ast
}V^{n+n},$ act as
\begin{equation*}
\{f,g\}=\frac{\partial f}{\partial p_{i}}\frac{\partial g}{\partial x^{i}}-%
\frac{\partial g}{\partial p_{i}}\frac{\partial p}{\partial x^{i}}.
\end{equation*}%
The canonical metric d--connection $\ ^{[H]}\widehat{\mathbf{D}}$ is defined
by the coefficients
\begin{equation*}
\ ^{[H]}\widehat{\mathbf{\Gamma }}_{\ \beta \gamma }^{\alpha }=\left( \
^{[c]}\check{H}_{~jk}^{i},\ ^{[c]}\check{H}_{~jk}^{i},\ ^{[c]}\check{C}%
_{~jk}^{i},\ ^{[c]}\check{C}_{~jk}^{i}\right)
\end{equation*}%
computed for $\ ^{[H]}\check{N}_{ij}$ and by respective formulas (\ref%
{candcon}) with $g_{ij}\rightarrow {\breve{g}}_{ij}\left( {\breve{u}}\right)
,$ $h_{ab}\rightarrow {\check{g}}^{ij}$ and $\widehat{L}_{~jk}^{i}%
\rightarrow \ ^{[c]}\widehat{H}_{~jk}^{i},$ $\ \widehat{C}%
_{bc}^{a}\rightarrow $ $\ ^{[c]}\check{C}_{~i}^{\quad jk}$ when
\begin{equation*}
\ ^{[c]}\check{H}_{~jk}^{i}=\frac{1}{2}\check{g}^{is}\left( \check{\delta}%
_{j}\check{g}_{sk}+\check{\delta}_{k}\check{g}_{js}-\check{\delta}_{s}\check{%
g}_{jk}\right) \mbox{ and }\ ^{[c]}\check{C}_{~i}^{\quad jk}=-\frac{1}{2}%
\check{g}_{is}\check{\partial}^{j}\check{g}^{sk}.
\end{equation*}%
In result, we can compute the d--torsions and d--curvatures like on Lagrange
\ or on Cartan spaces. On Hamilton spaces all such objects are defined by
the Hamilton function $H(x,p)$ and indices have to be reconsidered for
co--fibers of the cotangent bundle.

We note that there were elaborated various type of higher order
generalizations (on the higher order tangent and contangent bundles) of the
Finsler--Cartan and Lagrange--Hamilton geometry \cite{mat1} and on higher
order supersymmetric (co) vector bundles in Ref. \cite{vsup}. We can
generalize the d--connection $\ ^{[H]}\widehat{\mathbf{\Gamma }}_{\beta
\gamma }^{\alpha }$ to any d--connection in $\mathbf{H}^{n}$ with prescribed
torsions, like we have done in previous section for Lagrange spaces, see (%
\ref{ptdcls}). This type of Riemann--Cartan geometry is modeled like on a
dual tangent bundle by a Hamilton metric structure (\ref{hsm}),
N--connection $\ ^{[H]}\check{N}_{ij},$ and d--connection coefficients $\
^{[c]}\check{H}_{~jk}^{i}$ and $\ ^{[c]}\check{C}_{~i}^{\quad jk}.$

\subsection{ Nonmetricity and generalized Finsler--affine spaces}

The generalized Lagrange and Finsler geometry may be defined on tangent
bundles by using d--connections and d--metrics satisfying metric
compatibility conditions \cite{ma}. Nonmetricity components can be induced
if Berwald type d--connections are introduced into consideration on
different type of manifolds provided with N--connection structure, see
formulas (\ref{berw}), (\ref{bct}), (\ref{mafbc}) and (\ref{berwprim}).

We define such spaces as generalized Finsler spaces with nonmetricity.

\begin{definition}
\label{defglas}A generalized Lagrange--affine space $\mathbf{GLa}^{n}=\left(
V^{n},g_{ij}(x,y),\ ^{[a]}\mathbf{\Gamma }_{\ \beta }^{\alpha }\right) $ is
defined on manifold $\mathbf{TV}^{n+n},$ provided with an arbitrary
nontrivial N--connection structure $\mathbf{N}=\{N_{j}^{i}\},$\ as a general
Lagrange space $\mathbf{GL}^{n}=\left( V^{n},g_{ij}(x,y)\right) $ (see
Definition \ref{defgls}) enabled with a d--connection structure $\ ^{[a]}%
\mathbf{\Gamma }_{\ \ \alpha }^{\gamma }=\ ^{[a]}\mathbf{\Gamma }_{\ \alpha
\beta }^{\gamma }\mathbf{\vartheta }^{\beta }$ distorted by arbitrary
torsion, $\mathbf{T}_{\beta },$ and nonmetricity, $\mathbf{Q}_{\beta \gamma
},$ d--fields, \
\begin{equation}
\ \ ^{[a]}\mathbf{\Gamma }_{\ \beta }^{\alpha }=\ \ ^{[aL]}\widehat{\mathbf{%
\Gamma }}_{\ \beta }^{\alpha }\ +\ \ ^{[a]}\ \mathbf{Z}_{\ \ \beta }^{\alpha
},  \label{dcglma}
\end{equation}%
where$\ ^{[L]}\widehat{\mathbf{\Gamma }}_{\beta }^{\alpha }$ is the
canonical \ generalized Lagrange d--connection (\ref{dccls}) and
\begin{equation*}
\ \ ^{[a]}\ \mathbf{Z}_{\ \alpha \beta }=\mathbf{e}_{\beta }\rfloor \
\mathbf{T}_{\alpha }-\mathbf{e}_{\alpha }\rfloor \ \mathbf{T}_{\beta }+\frac{%
1}{2}\left( \mathbf{e}_{\alpha }\rfloor \mathbf{e}_{\beta }\rfloor \ \mathbf{%
T}_{\gamma }\right) \mathbf{\vartheta }^{\gamma }+\left( \mathbf{e}_{\alpha
}\rfloor \ \mathbf{Q}_{\beta \gamma }\right) \mathbf{\vartheta }^{\gamma
}-\left( \mathbf{e}_{\beta }\rfloor \ \mathbf{Q}_{\alpha \gamma }\right)
\mathbf{\vartheta }^{\gamma }+\frac{1}{2}\ \mathbf{Q}_{\alpha \beta }.
\end{equation*}
\end{definition}

The d--metric structure on $\mathbf{GLa}^{n}$ is stated by an arbitrary
N--adapted form (\ref{block2}) but on $\mathbf{TV}^{n+n},$
\begin{equation}
\mathbf{g}_{[a]}=g_{ij}(x,y)dx^{i}\otimes dx^{j}+g_{ij}(x,y)\delta
y^{i}\otimes \delta y^{j}.  \label{dmglas}
\end{equation}

The torsions and curvatures on $\mathbf{GLa}^{n}$ are computed by using
formulas (\ref{dt}) and (\ref{dc}) with $\mathbf{\Gamma }_{\ \beta }^{\gamma
}\rightarrow \ ^{[a]}\mathbf{\Gamma }_{\ \beta }^{\alpha },$

\begin{equation}
\ \ ^{[a]}\mathbf{T}^{\alpha }\doteqdot \ ^{[a]}\mathbf{D\vartheta }^{\alpha
}=\delta \mathbf{\vartheta }^{\alpha }+\ ^{[a]}\mathbf{\Gamma }_{\ \beta
}^{\gamma }\wedge \mathbf{\vartheta }^{\beta }  \label{tglma}
\end{equation}%
and
\begin{equation}
\ \ ^{[a]}\mathbf{R}_{\ \beta }^{\alpha }\doteqdot \ ^{[a]}\mathbf{D(\ }%
^{[a]}\mathbf{\Gamma }_{\ \beta }^{\alpha })=\delta (\ ^{[a]}\mathbf{\Gamma }%
_{\ \beta }^{\alpha })-\ ^{[a]}\mathbf{\Gamma }_{\ \beta }^{\gamma }\wedge \
^{[a]}\mathbf{\Gamma }_{\ \ \gamma }^{\alpha }.  \label{cglma}
\end{equation}%
Modeling in $V^{n+n},$ with local coordinates $u^{\alpha }=\left(
x^{i},y^{k}\right) ,$ a tangent bundle structure, we redefine the operators (%
\ref{ddif}) and (\ref{dder}) respectively as
\begin{equation}
\mathbf{e}_{\alpha }\doteqdot \delta _{\alpha }=\left( \delta _{i},\tilde{%
\partial}_{k}\right) \equiv \frac{\delta }{\delta u^{\alpha }}=\left( \frac{%
\delta }{\delta x^{i}}=\partial _{i}-N_{i}^{a}\left( u\right) \partial _{a},%
\frac{\partial }{\partial y^{k}}\right)  \label{ddert}
\end{equation}%
and the N--elongated differentials (in brief, N--differentials)
\begin{equation}
\mathbf{\vartheta }_{\ }^{\beta }\doteqdot \delta \ ^{\beta }=\left( d^{i},%
\tilde{\delta}^{k}\right) \equiv \delta u^{\alpha }=\left( \delta
x^{i}=dx^{i},\delta y^{k}=dy^{k}+N_{i}^{k}\left( u\right) dx^{i}\right)
\label{ddift}
\end{equation}%
where Greek indices run the same values, $i,j,...=1,2,...n$ (we shall use
the symbol ''$\sim $'' if one would be necessary to distinguish operators
and coordinates defined on h-- and v-- subspaces).

Let us define the h-- and v--irreducible components of the d--connection $\
^{[a]}\mathbf{\Gamma }_{\ \beta }^{\alpha }$ like in (\ref{hcov}) and (\ref%
{vcov}),%
\begin{equation*}
\ ^{[a]}\widehat{\mathbf{\Gamma }}_{\ \beta \gamma }^{\alpha }=\left( \
^{[L]}\widehat{L}_{~jk}^{i}+\ z_{~jk}^{i},\ ^{[L]}\widehat{L}_{~jk}^{i}\ +\
z_{~jk}^{i},\ ^{[L]}\widehat{C}_{~jk}^{i}+c_{~jk}^{i},\ ^{[L]}\widehat{C}%
_{~jk}^{i}+c_{~jk}^{i}\right)
\end{equation*}%
with the distorsion d--tensor $\ $%
\begin{equation*}
\ ^{[a]}\ \mathbf{Z}_{\ \ \beta }^{\alpha }=\left( z_{~jk}^{i},\
z_{~jk}^{i},c_{~jk}^{i},c_{~jk}^{i}\right)
\end{equation*}%
defined as on a tangent bundle
\begin{eqnarray*}
\ ^{[a]}L_{jk}^{i} &=&\left( ^{[a]}\mathbf{D}_{\delta _{k}}\delta
_{j}\right) \rfloor \delta ^{i}=\left( ^{[L]}\widehat{\mathbf{D}}_{\delta
_{k}}\delta _{j}+\ ^{[a]}\ \mathbf{Z}_{\delta _{k}}\delta _{j}\right)
\rfloor \delta ^{i}=\ ^{[L]}\widehat{L}_{~jk}^{i}+\ z_{~jk}^{i},\quad \\
\ ^{[a]}\tilde{L}_{jk}^{i} &=&\left( ^{[a]}\mathbf{D}_{\delta _{k}}\tilde{%
\partial}_{j}\right) \rfloor \tilde{\partial}^{i}=\left( ^{[L]}\widehat{%
\mathbf{D}}_{k}\tilde{\partial}_{j}+\ ^{[a]}\ \mathbf{Z}_{k}\tilde{\partial}%
_{j}\right) \rfloor \tilde{\partial}^{i}=\ ^{[L]}\widehat{L}_{~jk}^{i}+\
z_{~jk}^{i}, \\
\ \ ^{[a]}C_{jk}^{i} &=&\left( ^{[a]}\mathbf{D}_{\tilde{\partial}_{k}}\delta
_{j}\right) \rfloor \delta ^{i}=\left( ^{[L]}\widehat{\mathbf{D}}_{\tilde{%
\partial}_{k}}\delta _{j}+\ ^{[a]}\ \mathbf{Z}_{\tilde{\partial}_{k}}\delta
_{j}\right) \rfloor \delta ^{i}=\ ^{[L]}\widehat{C}_{~jk}^{i}+c_{~jk}^{i}, \\
\ \ ^{[a]}\tilde{C}_{jk}^{i} &=&\left( ^{[a]}\mathbf{D}_{\tilde{\partial}%
_{k}}\tilde{\partial}_{j}\right) \rfloor \tilde{\partial}^{i}=\left( ^{[L]}%
\widehat{\mathbf{D}}_{\tilde{\partial}_{k}}\tilde{\partial}_{j}+\ ^{[a]}\
\mathbf{Z}_{\tilde{\partial}_{k}}\tilde{\partial}_{j}\right) \rfloor \tilde{%
\partial}^{i}=\ ^{[L]}\widehat{C}_{~jk}^{i}+c_{~jk}^{i},
\end{eqnarray*}%
where for 'lifts' from the h--subspace to the v--subspace we consider that $%
\ ^{[a]}L_{jk}^{i}=\ ^{[a]}\tilde{L}_{jk}^{i}$ and $\ ^{[a]}C_{jk}^{i}=\
^{[a]}\tilde{C}_{jk}^{i}.$ As a consequence, on spaces with modeled tangent
space structure, the d--connections are distinguished as\ $\ \mathbf{\Gamma }%
_{\beta \gamma }^{\alpha }=\left( L_{jk}^{i},C_{jk}^{i}\right) .$

\begin{theorem}
\label{ttgfls}The torsion $\ \ ^{[a]}\mathbf{T}^{\alpha }$ $\ $(\ref{tglma})
of a d--connection $^{[a]}\mathbf{\Gamma }_{\ \beta }^{\alpha }=\left( \
^{[a]}L_{jk}^{i},\ ^{[a]}C_{jc}^{i}\right) $ (\ref{dcglma}) has as
irreducible h- v--components, $\ \ ^{[a]}\mathbf{T}^{\alpha }=\left(
T_{jk}^{i},\tilde{T}_{jk}^{i}\right) ,$ the d--torsions
\begin{eqnarray}
T_{.jk}^{i} &=&-T_{kj}^{i}=\ ^{[L]}\widehat{L}_{~jk}^{i}+\ z_{~jk}^{i}-\
^{[L]}\widehat{L}_{~kj}^{i}-\ z_{~kj}^{i},\quad  \label{dtorsgla} \\
\quad \ \tilde{T}_{jk}^{i} &=&-\ \tilde{T}_{kj}^{i}=\ ^{[L]}\widehat{C}%
_{~jk}^{i}+c_{~jk}^{i}-\ ^{[L]}\widehat{C}_{~kj}^{i}-c_{~kj}^{i}.\   \notag
\end{eqnarray}
\end{theorem}

The proof of this Theorem consists from a standard calculus for
metric--affine spaces of $\ ^{[a]}\mathbf{T}^{\alpha }$ \cite{mag} but with
N--adapted frames. We note that in $\ z_{~jk}^{i}$ and $c_{~kj}^{i}$ it is
possible to include any prescribed values of the d--torsions.

\begin{theorem}
\label{tcgfls}The curvature $\ \ ^{[a]}\mathbf{R}_{\ \beta }^{\alpha }$ $\ $(%
\ref{cglma}) of a d--connection $^{[a]}\mathbf{\Gamma }_{\ \beta }^{\alpha
}=\left( \ ^{[a]}L_{jk}^{i},\ ^{[a]}C_{jc}^{i}\right) $ (\ref{dcglma}) has
as irreducible h- v--components, $^{[a]}\mathbf{R}_{.\beta \gamma \tau
}^{\alpha }=\{\ ^{[a]}R_{\ hjk}^{i},\ ^{[a]}P_{\ jka}^{i},\ ^{[a]}S_{\
jbc}^{i}\}$ the d--curvatures
\begin{eqnarray*}
\ ^{[a]}R_{\ hjk}^{i} &=&\frac{\delta \ ^{[a]}L_{.hj}^{i}}{\delta x^{h}}-%
\frac{\delta \ ^{[a]}L_{.hk}^{i}}{\delta x^{j}}+\ ^{[a]}L_{.hj}^{m}\
^{[a]}L_{mk}^{i}-\ ^{[a]}L_{.hk}^{m}\ ^{[a]}L_{mj}^{i}-\
^{[a]}C_{.ho}^{i}\Omega _{.jk}^{o}, \\
\ ^{[a]}P_{\ jks}^{i} &=&\frac{\partial \ ^{[a]}L_{.jk}^{i}}{\partial y^{s}}%
-\left( \frac{\partial \ ^{[a]}C_{.js}^{i}}{\partial x^{k}}+\
^{[a]}L_{.lk}^{i}\ ^{[a]}C_{.js}^{l}-\ ^{[a]}L_{.jk}^{l}\
^{[a]}C_{.ls}^{i}-\ ^{[a]}L_{.sk}^{p}\ ^{[a]}C_{.jp}^{i}\right) \\
&&+\ ^{[a]}C_{.jp}^{i}\ ^{[a]}P_{.ks}^{p}, \\
\ ^{[a]}S_{\ jlm}^{i} &=&\frac{\partial \ ^{[a]}C_{.jl}^{i}}{\partial y^{m}}-%
\frac{\partial \ ^{[a]}C_{.jm}^{i}}{\partial y^{l}}+\ ^{[a]}C_{.jl}^{h}\
^{[a]}C_{.hm}^{i}-\ ^{[a]}C_{.jm}^{h}\ ^{[a]}C_{hl}^{i},
\end{eqnarray*}%
where\ $^{[a]}L_{.hk}^{m}=\ ^{[L]}\widehat{L}_{~jk}^{i}+\ z_{~jk}^{i},\
^{[a]}C_{.jk}^{i}=\ ^{[L]}\widehat{C}_{~jk}^{i}+c_{~jk}^{i},\ \Omega
_{.jk}^{o}=\delta _{j}N_{i}^{o}-\delta _{i}N_{j}^{o}$ and $\
^{[a]}P_{.ks}^{p}=\partial N_{i}^{p}/\partial y^{s}-\ ^{[a]}L_{.ks}^{p}.$
\end{theorem}

The proof consists from a straightforward calculus.

\begin{remark}
\label{rlafs}As a particular case of $\mathbf{GLa}^{n}$, we can define a
Lagrange--affine space $\mathbf{La}^{n}=\left( V^{n},g_{ij}^{[L]}(x,y),\
^{[b]}\mathbf{\Gamma }_{\ \beta }^{\alpha }\right) ,$ provided with a
Lagrange quadratic form $g_{ij}^{[L]}(x,y)$ (\ref{lagm}) inducing the
canonical N--connection structure $^{[cL]}\mathbf{N}=\{\ ^{[cL]}N_{j}^{i}\}$
(\ref{cncls}) \ as in a Lagrange space $\mathbf{L}^{n}=\left(
V^{n},g_{ij}(x,y)\right) $ (see Definition \ref{defls})) but with a
d--connection structure $\ ^{[b]}\mathbf{\Gamma }_{\ \ \alpha }^{\gamma }=\
^{[b]}\mathbf{\Gamma }_{\ \alpha \beta }^{\gamma }\mathbf{\vartheta }^{\beta
}$ distorted by arbitrary torsion, $\mathbf{T}_{\beta },$ and nonmetricity, $%
\mathbf{Q}_{\beta \gamma },$ d--fields, \
\begin{equation*}
\ \ ^{[b]}\mathbf{\Gamma }_{\ \beta }^{\alpha }=\ ^{[L]}\widehat{\mathbf{%
\Gamma }}_{\beta }^{\alpha }+\ \ ^{[b]}\ \mathbf{Z}_{\ \ \beta }^{\alpha },
\end{equation*}%
where$\ ^{[L]}\widehat{\mathbf{\Gamma }}_{\beta }^{\alpha }$ is the
canonical Lagrange d--connection (\ref{lagdcon}),
\begin{equation*}
\ \ ^{[b]}\ \mathbf{Z}_{\ \ \beta }^{\alpha }=\mathbf{e}_{\beta }\rfloor \
\mathbf{T}_{\alpha }-\mathbf{e}_{\alpha }\rfloor \ \mathbf{T}_{\beta }+\frac{%
1}{2}\left( \mathbf{e}_{\alpha }\rfloor \mathbf{e}_{\beta }\rfloor \ \mathbf{%
T}_{\gamma }\right) \mathbf{\vartheta }^{\gamma }+\left( \mathbf{e}_{\alpha
}\rfloor \ \mathbf{Q}_{\beta \gamma }\right) \mathbf{\vartheta }^{\gamma
}-\left( \mathbf{e}_{\beta }\rfloor \ \mathbf{Q}_{\alpha \gamma }\right)
\mathbf{\vartheta }^{\gamma }+\frac{1}{2}\ \mathbf{Q}_{\alpha \beta },
\end{equation*}%
and the (co) frames $\mathbf{e}_{\beta }$ and $\mathbf{\vartheta }^{\gamma }$
are respectively constructed as in (\ref{dder}) and (\ref{ddif}) by using $%
^{[cL]}N_{j}^{i}.$
\end{remark}

\begin{remark}
\label{rfafs}The Finsler--affine spaces $\mathbf{Fa}^{n}=\left(
V^{n},F\left( x,y\right) ,\ ^{[f]}\mathbf{\Gamma }_{\ \beta }^{\alpha
}\right) $ can be introduced by further restrictions of $\mathbf{La}^{n}$ to
a quadratic form $g_{ij}^{[F]}$ (\ref{finm2}) constructed from a Finsler
metric $F\left( x^{i},y^{j}\right) $ inducing the canonical N--connection
structure $\ ^{[F]}\mathbf{N}=\{\ ^{[F]}N_{j}^{i}\}$ (\ref{ncc})\ as in a
Finsler space $\mathbf{F}^{n}=\left( V^{n},F\left( x,y\right) \right) $ but
with a d--connection structure $\ ^{[f]}\mathbf{\Gamma }_{\ \ \alpha
}^{\gamma }=\ ^{[f]}\mathbf{\Gamma }_{\ \alpha \beta }^{\gamma }\mathbf{%
\vartheta }^{\beta }$ distorted by arbitrary torsion, $\mathbf{T}_{\beta },$
and nonmetricity, $\mathbf{Q}_{\beta \gamma },$ d--fields, \
\begin{equation*}
\ \ ^{[f]}\mathbf{\Gamma }_{\ \beta }^{\alpha }=\ ^{[F]}\widehat{\mathbf{%
\Gamma }}_{\ \beta }^{\alpha }\ +\ \ ^{[f]}\ \mathbf{Z}_{\ \ \beta }^{\alpha
},
\end{equation*}%
where $\ ^{[F]}\widehat{\mathbf{\Gamma }}_{\ \beta }^{\alpha }$ is the
canonical Finsler d--connection (\ref{dccfs}),
\begin{equation*}
\ \ ^{[f]}\ \mathbf{Z}_{\ \ \beta }^{\alpha }=\mathbf{e}_{\beta }\rfloor \
\mathbf{T}_{\alpha }-\mathbf{e}_{\alpha }\rfloor \ \mathbf{T}_{\beta }+\frac{%
1}{2}\left( \mathbf{e}_{\alpha }\rfloor \mathbf{e}_{\beta }\rfloor \ \mathbf{%
T}_{\gamma }\right) \mathbf{\vartheta }^{\gamma }+\left( \mathbf{e}_{\alpha
}\rfloor \ \mathbf{Q}_{\beta \gamma }\right) \mathbf{\vartheta }^{\gamma
}-\left( \mathbf{e}_{\beta }\rfloor \ \mathbf{Q}_{\alpha \gamma }\right)
\mathbf{\vartheta }^{\gamma }+\frac{1}{2}\ \mathbf{Q}_{\alpha \beta },
\end{equation*}%
and the (co) frames $\mathbf{e}_{\beta }$ and $\mathbf{\vartheta }^{\gamma }$
are respectively constructed as in (\ref{dder}) and (\ref{ddif}) by using $%
^{[F]}N_{j}^{i}.$
\end{remark}

\begin{remark}
\label{rghas}By similar geometric constructions (see Remarks \ref{rlafs} and %
\ref{rfafs}) on spaces modeling cotangent bundles, we can define generalized
Hamilton--affine spaces $\mathbf{GHa}^{n}=\left( V^{n},\check{g}^{ij}(x,p),\
^{[a]}\mathbf{\check{\Gamma}}_{\ \beta }^{\alpha }\right) $ and theirs
restrictions to Hamilton--affine $\mathbf{Ha}^{n}=(V^{n},\check{g}%
_{[H]}^{ij}(x,p),$ $\ ^{[b]}\mathbf{\check{\Gamma}}_{\ \beta }^{\alpha })$
and Cartan--affine spaces $\mathbf{Ca}^{n}=\left( V^{n},\check{g}%
_{[K]}^{ij}(x,p),\ ^{[c]}\mathbf{\check{\Gamma}}_{\ \beta }^{\alpha }\right)
$ (see sections \ref{shgg} and \ref{scs}) as to contain distorsions induced
by nonmetricity $\mathbf{\check{Q}}_{\alpha \gamma }.$ The geometric objects
have to be adapted to the corresponding N--connection and d--metric/
quadratic form structures (arbitrary $\check{N}_{ij}\left( x,p\right) $ and
d--metric (\ref{dmghs}), $\ ^{[H]}\check{N}_{ij}\left( x,p\right) $ (\ref%
{ncchs}) and quadratic form $\check{g}_{[H]}^{ij}$ (\ref{hsm}) and $\check{N}%
_{ij}^{[K]}$ (\ref{nccartan}) and $\check{g}_{[K]}^{ij}$ (\ref{carm}).
\end{remark}

Finally, in this section, we note that Theorems \ref{ttgfls} and \ref{tcgfls}
can be reformulated in the forms stating procedures of computing d--torsions
and d--curvatures on every type of spaces with nonmetricity and local
anisotropy by adapting the abstract symbol and/or coordinate calculations
with respect to corresponding N--connection, d--metric and canonical
d--connection structures.

\section{Conclusions}

The method of moving anholonomic frames with associated nonlinear connection
(N--connection) structure elaborated in this work on metric--affine spaces
provides a general framework to deal with any possible model of locally
isotropic and/or anisotropic interactions and geometries defined effectively
in the presence of generic off--diagonal metric and linear connection
configurations, in general, subjected to certain anholonomic constraints. As
it has been pointed out, the metric--affine gravity (MAG) contains various
types of generalized Finsler--Lagrange--Hamilton--Cartan geometries which
can be distinguished by a corresponding N--connection structure and metric
and linear connection adapted to the N--connection structure.

As far as the anholonomic frames, nonmetricity and torsion are considered as
fundamental quantities, all mentioned geometries can be included into a
unique scheme which can be developed on arbitrary manifolds, vector and
tangent bundles and their dual bundles (co-bundles) or restricted to
Riemann--Cartan and (pseudo) Riemannian spaces. We observe that a generic
off--diagonal metric (which can not be diagonalized by any coordinate
transform) defining a (pseudo) Riemannian space induces alternatively
various type of Riemann--Cartan and Finsler like configuratons modeled by
respective frame structures. The constructions are generalized if the linear
connection structures are not constrained to metricity conditions. One can
regard this as extensions to metric--affine spaces provided with
N--connection structure modeling also bundle structures and generalized
noncommutative symmetries of metrics and anholonomic frames.

In this paper we have studied the general properties of metric--affine
spaces provided with N--connection structure. We formulated and proved the
main theorems concerning general metric and nonlinear and linear connections
in MAG. There were stated the criteria when the spaces with local isotropy
and/or local anisotropy can be modeled in metric--affine spaces and on
vector/ tangent bundles. We elaborated the concept of generalized
Finsler--affine geometry as a sinthesis of metric--affine (with nontrivial
torsion and nonmetricity) and Finsler like configurations (with nontrivial
N--connection structure and locally anisotropic metrics and connections).

In a general sense, we note that the generalized Finsler--affine geometries
are contained as anhlonomic and noncommutative configurations in extra
dimension gravity models (string and brane models and certain limits to the
Einstein and gauge gravity defined by off--diagonal metrics and anholonomic
constraints). We would like to stress that the N--connection formalism
developed for the metric--affine spaces relates the bulk geometry in string
and/or MAG to gauge theories in vector/tangent bundles and to various type
of non--Riemannian gravity models.

The approach presented here could be advantageous in a triple sense. First,
it provides a uniform treatment of all metric and connection geometries, in
general, with vector/tangent bundle structures which arise in various type
of string and brane gravity models. Second, it defines a complete
classification of the generalized Finsler--affine geometries stated in
Tables 1-11 from the Appendix. Third, it states a new geometric method of
constructing exact solutions with generic off--diagonal metric ansatz,
torsions and nonmetricity, depending on 2--5 variables, in string and
metric--affine gravity, with limits to the Einstein gravity, see Refs. \cite%
{exsolmag}.

\section{Classification of Generalized Finsler--Affine Spaces}

We outline and give a brief characterization of the main classes of
generalized Finsler--affine spaces (see Tables \ref{tablegs}--\ref{tablets}%
). A unified approach to such very different locally isotropic and
anisotropic geometries, defined in the framework of the metric--affine
geometry, can be elaborated only by introducing the concept on N--connection
(see Definition \ref{dnlc}).

The N--connection curvature is computed following the formula $\Omega
_{ij}^{a}=\delta _{\lbrack i}N_{j]}^{a},$ see (\ref{ncurv}), for any
N--connection $N_{i}^{a}.$ A d--connection $\mathbf{D}=[\mathbf{\Gamma }%
_{\beta \gamma }^{\alpha }]=[L_{\ jk}^{i},L_{\ bk}^{a},C_{\ jc}^{i},C_{\
bc}^{a}]$ (see Definition \ref{defdcon}) defines nontrivial d--torsions $%
\mathbf{T}_{\ \beta \gamma }^{\alpha }=[L_{[\ jk]}^{i},C_{\ ja}^{i},\Omega
_{ij}^{a},T_{\ bj}^{a},C_{\ [bc]}^{a}]$ and d--curvatures $\mathbf{R}_{\
\beta \gamma \tau }^{\alpha }=[R_{\ jkl}^{i},R_{\ bkl}^{a},P_{\
jka}^{i},P_{\ bka}^{c},S_{\ jbc}^{i},S_{\ dbc}^{a}]$ adapted to the
N--connection structure (see, respectively, the formulas (\ref{dtorsb})\ and
(\ref{dcurv})). It is considered that a generic off--diagonal metric $%
g_{\alpha \beta }$ (see Remark \ref{rgod}) is associated to a N--connection
structure and reprezented as a d--metric $\mathbf{g}_{\alpha \beta
}=[g_{ij},h_{ab}]$ (see formula (\ref{block2})). The components of a
N--connection and a d--metric define the canonical d--connection $\mathbf{D}%
=[\widehat{\mathbf{\Gamma }}_{\beta \gamma }^{\alpha }]=[\widehat{L}_{\
jk}^{i},\widehat{L}_{\ bk}^{a},\widehat{C}_{\ jc}^{i},\widehat{C}_{\
bc}^{a}] $ (see (\ref{candcon})) with the corresponding values of
d--torsions $\widehat{\mathbf{T}}_{\ \beta \gamma }^{\alpha }$ and
d--curvatures $\widehat{\mathbf{R}}_{\ \beta \gamma \tau }^{\alpha }.$ The
nonmetricity d--fields are computed by using formula $\mathbf{Q}_{\alpha
\beta \gamma }=-\mathbf{D}_{\alpha }\mathbf{g}_{\beta \gamma
}=[Q_{ijk},Q_{iab},Q_{ajk},Q_{abc}],$ see (\ref{nmf}).

\subsection{Generalized Lagrange--affine spaces}

The Table \ref{tablegs} outlines seven classes of geometries modeled in the
framework of metric--affine geometry as spaces with nontrivial N--connection
structure. There are emphasized the configurations:

\begin{enumerate}
\item Metric--affine spaces (in brief, MA) are those stated by Definition %
\ref{defmas} as certain manifolds $V^{n+m}$ of necessary smoothly class\
provided with arbitrary metric, $g_{\alpha \beta },$ and linear connection, $%
\Gamma _{\beta \gamma }^{\alpha },$ structures. For generic off--diagonal
metrics, a MA\ space always admits nontrivial N--connection structures (see
Proposition \ref{pmasnc}). Nevertheless, in general, only the metric field $%
g_{\alpha \beta }$ can be transformed into a d--metric one $\mathbf{g}%
_{\alpha \beta }=[g_{ij},h_{ab}],$ but \ $\Gamma _{\beta \gamma }^{\alpha }$
can be not adapted to the N--connection structure. As a consequence, the
general strength fields $\left( T_{\ \beta \gamma }^{\alpha },R_{\ \beta
\gamma \tau }^{\alpha },Q_{\alpha \beta \gamma }\right) $ can be also not
N--adapted. By using the Kawaguchi's metrization process and Miron's
procedure stated by Theorems \ref{kmp} and \ref{mconnections} we can
consider alternative geometries with d--connections $\mathbf{\Gamma }_{\beta
\gamma }^{\alpha }$ (see Definition \ref{defdcon}) derived from the
components of N--connection and d--metric. Such geometries are adapted to
the N--connection structure. They are characterized by d--torsion $\mathbf{T}%
_{\ \beta \gamma }^{\alpha },$ d--curvature $\mathbf{R}_{\ \beta \gamma \tau
}^{\alpha },$ and nonmetricity d--field $\mathbf{Q}_{\alpha \beta \gamma }.$

\item Distinguished metric--affine spaces (DMA) are defined (see Definition %
\ref{ddmas}) as manifolds $\mathbf{V}^{n+m}$ $\ $provided with N--connection
structure $N_{i}^{a},$ d--metric field (\ref{block2}) and arbitrary
d--connection $\mathbf{\Gamma }_{\beta \gamma }^{\alpha }.$ In this case,
all strenghs $\left( \mathbf{T}_{\ \beta \gamma }^{\alpha },\mathbf{R}_{\
\beta \gamma \tau }^{\alpha },\mathbf{Q}_{\alpha \beta \gamma }\right) $ are
N--adapted.

\item Berwald--affine spaces (BA, see section \ref{berwdcon}) are
metric--affine spaces provided with generic off--diagonal metrics with
associated N--connection structure and with a Ber\-wald d--connection $^{[B]}%
\mathbf{D}=[\ ^{[B]}\mathbf{\Gamma }_{\beta \gamma }^{\alpha }]=[\widehat{L}%
_{\ jk}^{i},\partial _{b}N_{k}^{a},0,\widehat{C}_{\ bc}^{a}]$, see (\ref%
{berw}), for with the d--torsions $\ ^{[B]}\mathbf{T}_{\ \beta \gamma
}^{\alpha }=[\ ^{[B]}L_{\ [jk]}^{i},0,\Omega _{ij}^{a},T_{\ bj}^{a},C_{\
[bc]}^{a}]$ and d--curvatures
\begin{equation*}
^{\lbrack B]}\mathbf{R}_{\ \beta \gamma \tau }^{\alpha }=^{[B]}[R_{\
jkl}^{i},R_{\ bkl}^{a},P_{\ jka}^{i},P_{\ bka}^{c},S_{\ jbc}^{i},S_{\
dbc}^{a}]
\end{equation*}%
\ are computed by introducing the components of $^{[B]}\mathbf{\Gamma }%
_{\beta \gamma }^{\alpha },$ respectively, in formulas (\ref{dtorsb})\ and (%
\ref{dcurv}). By definition, this space satisfies the metricity conditions
on the h- and v--subspaces, $Q_{ijk}=0$ and $Q_{abc}=0,$ but, in general,
there are nontrivial nonmetricity d--fields because $Q_{iab}$ and $Q_{ajk}$
are not vanishing (see formulas (\ref{berwnm})).

\item Berwald--affine spaces with prescribed torsion (BAT, see sections \ref%
{berwdcon} and \ref{berwdcona}) are described by a more general class of
d--connection $^{[BT]}\mathbf{\Gamma }_{\beta \gamma }^{\alpha }=[L_{\
jk}^{i},\partial _{b}N_{k}^{a},0,C_{\ bc}^{a}],$ with more general h-- and
v--components, $\ \widehat{L}_{\ jk}^{i}\rightarrow L_{\ jk}^{i}$ and $%
\widehat{C}_{\ bc}^{a}\rightarrow C_{\ bc}^{a},$ inducing prescribed values $%
\tau _{\ jk}^{i}$ and $\tau _{\ bc}^{a}$ in d--torsion\
\begin{equation*}
^{\lbrack BT]}\mathbf{T}_{\ \beta \gamma }^{\alpha }=[L_{\ [jk]}^{i},+\tau
_{\ jk}^{i},0,\Omega _{ij}^{a},T_{\ bj}^{a},C_{\ [bc]}^{a}+\tau _{\ bc}^{a}],
\end{equation*}%
see (\ref{tauformulas})$.$ The components of curvature $^{[BT]}\mathbf{R}_{\
\beta \gamma \tau }^{\alpha }$ have to be computed by introducing $^{[BT]}%
\mathbf{\Gamma }_{\beta \gamma }^{\alpha }$ into (\ref{dcurv}). There are
nontrivial components of nonmetricity d--fields, $^{[B\tau ]}\mathbf{Q}%
_{\alpha \beta \gamma }=\left( ^{[B\tau ]}Q_{cij},\ ^{[B\tau
]}Q_{iab}\right) .$

\item Generalized Lagrange--affine spaces (GLA, see Definition \ref{defglas}%
), $\mathbf{GLa}^{n}=(V^{n},g_{ij}(x,y),$ $\ ^{[a]}\mathbf{\Gamma }_{\ \beta
}^{\alpha })$ $,$ are modeled as distinguished metric--affine spaces of
odd--dimension, $\mathbf{V}^{n+n},$ provided with generic off--diagonal
metrics with associated N--connection inducing a tangent bundle structure.
The d--metric $\mathbf{g}_{[a]}$ (\ref{dmglas}) and the d--connection $\ \
^{[a]}\mathbf{\Gamma }_{\ \alpha \beta }^{\gamma }$ $=\left( \
^{[a]}L_{jk}^{i},\ ^{[a]}C_{jc}^{i}\right) $ (\ref{dcglma}) are similar to
those for the usual Lagrange spaces (see Definition \ref{defgls}) but with
distorsions $\ ^{[a]}\ \mathbf{Z}_{\ \ \beta }^{\alpha }$ inducing general
nontrivial nonmetricity d--fields $^{[a]}\mathbf{Q}_{\alpha \beta \gamma }.$
The components of d--torsions $^{[a]}\mathbf{T}^{\alpha }=\left( T_{jk}^{i},%
\tilde{T}_{jk}^{i}\right) $and d--curvatures $^{[a]}\mathbf{R}_{.\beta
\gamma \tau }^{\alpha }=\{\ ^{[a]}R_{\ hjk}^{i},\ ^{[a]}P_{\ jka}^{i},\
^{[a]}S_{\ jbc}^{i}\}$ are computed following Theorems \ref{ttgfls} and \ref%
{tcgfls}.

\item Lagrange--affine spaces (LA, see Remark \ \ref{rlafs}), $\mathbf{La}%
^{n}=(V^{n},g_{ij}^{[L]}(x,y),$ $\ ^{[b]}\mathbf{\Gamma }_{\ \beta }^{\alpha
}),$ are provided with a Lagrange quadratic form $g_{ij}^{[L]}(x,y)=\frac{1}{%
2}\frac{\partial ^{2}L^{2}}{\partial y^{i}\partial y^{j}}$ (\ref{lagm})
inducing the canonical N--connection structure $^{[cL]}\mathbf{N}=\{\
^{[cL]}N_{j}^{i}\}$ (\ref{cncls})\ for a Lagrange space $\mathbf{L}%
^{n}=\left( V^{n},g_{ij}(x,y)\right) $ (see Definition \ref{defls})) but
with a d--connection structure $\ ^{[b]}\mathbf{\Gamma }_{\ \ \alpha
}^{\gamma }=\ ^{[b]}\mathbf{\Gamma }_{\ \alpha \beta }^{\gamma }\mathbf{%
\vartheta }^{\beta }$ distorted by arbitrary torsion, $\mathbf{T}_{\beta },$
and nonmetricity d--fields,$\ \mathbf{Q}_{\beta \gamma \alpha },$ when $%
^{[b]}\mathbf{\Gamma }_{\ \beta }^{\alpha }=\ ^{[L]}\widehat{\mathbf{\Gamma }%
}_{\beta }^{\alpha }+\ \ ^{[b]}\ \mathbf{Z}_{\ \ \beta }^{\alpha }.$ This is
a particular case of GLA spaces with prescribed types of N--connection $%
^{[cL]}N_{j}^{i}$ and d--metric to be like in Lagrange geometry.

\item Finsler--affine spaces (FA, see Remark \ref{rfafs}), $\mathbf{Fa}%
^{n}=\left( V^{n},F\left( x,y\right) ,\ ^{[f]}\mathbf{\Gamma }_{\ \beta
}^{\alpha }\right) ,$ in their turn are introduced by further restrictions
of $\mathbf{La}^{n}$ to a quadratic form $g_{ij}^{[F]}=\frac{1}{2}\frac{%
\partial ^{2}F^{2}}{\partial y^{i}\partial y^{j}}$ (\ref{finm2}) constructed
from a Finsler metric $F\left( x^{i},y^{j}\right) .$ It is induced the
canonical N--connection structure $\ ^{[F]}\mathbf{N}=\{\ ^{[F]}N_{j}^{i}\}$
(\ref{ncc})\ as in the Finsler space $\mathbf{F}^{n}=\left( V^{n},F\left(
x,y\right) \right) $ but with a d--connection structure $\ ^{[f]}\mathbf{%
\Gamma }_{\ \alpha \beta }^{\gamma }$ distorted by arbitrary torsion, $%
\mathbf{T}_{\beta \gamma }^{\alpha },$ and nonmetricity, $\mathbf{Q}_{\beta
\gamma \tau },$ d--fields, $\ ^{[f]}\mathbf{\Gamma }_{\ \beta }^{\alpha }=\
^{[F]}\widehat{\mathbf{\Gamma }}_{\ \beta }^{\alpha }\ +\ \ ^{[f]}\ \mathbf{Z%
}_{\ \ \beta }^{\alpha },$where $\ ^{[F]}\widehat{\mathbf{\Gamma }}_{\ \beta
\gamma }^{\alpha }$ is the canonical Finsler d--connection (\ref{dccfs}).
\end{enumerate}

\subsection{Generalized Hamilton--affine spaces}

The Table \ref{tableghs} outlines geometries modeled in the framework of
metric--affine geometry as spaces with nontrivial N--connection structure
splitting the space into any conventional a horizontal subspace and vertical
subspace being isomorphic to a dual vector space provided with respective
dual coordinates. We can use respectively the classification from Table \ref%
{tablegs} when the v--subspace is transformed into dual one as we noted in
Remark \ref{rghas} For simplicity, we label such spaces with symbols like $%
\check{N}_{ai}$ instead $N_{i}^{a}$ where ''inverse hat'' points that the
geometric object is defined for a space containing a dual subspaces. The
local h--coordinates are labeled in the usual form, $x^{i},$ with $%
i=1,2,...,n$ but the v--coordinates are certain dual vectors $\check{y}%
^{a}=p_{a}$ with $a=n+1,n+2,...,n+m.$ The local coordinates are denoted $%
\check{u}^{\alpha }=\left( x^{i},\check{y}^{a}\right) =\left(
x^{i},p_{a}\right) .$ The curvature of a N--connection $\check{N}_{ai}$ is
computed as $\check{\Omega}_{iaj}=\delta _{\lbrack i}\check{N}_{j]a}.$ The
h-- v--irreducible components of a general d--connection are parametrized $%
\mathbf{\check{D}}=[\mathbf{\check{\Gamma}}_{\beta \gamma }^{\alpha
}]=[L_{~jk}^{i},L_{a\ \ k}^{\ b},\check{C}_{\ j}^{i\ c},\check{C}_{a}^{\
bc}],$ the d--torsions are $\mathbf{\check{T}}_{\ \beta \gamma }^{\alpha
}=[L_{~[jk]}^{i},L_{a\ \ k}^{\ b},\check{C}_{\ j}^{i\ c},\check{C}_{a}^{\
[bc]}]$ and the d--curvatures $^{[B]}\mathbf{\check{R}}_{\ \beta \gamma \tau
}^{\alpha }=$ $[R_{\ jkl}^{i},\check{R}_{a\ kl}^{~b},\check{P}_{\ jk}^{i~~a},%
\check{P}_{c~k}^{~b~a},\check{S}_{\ j}^{i~bc},\check{S}_{a\ }^{~dbc}].$ The
nonmetricity d--fields are stated $\mathbf{\check{Q}}_{\alpha \beta \gamma
}=-\mathbf{\check{D}}_{\alpha }\mathbf{\check{g}}_{\beta \gamma }=[Q_{ijk},%
\check{Q}_{i}^{~ab},\check{Q}_{~jk}^{a},\check{Q}^{abc}].$ There are also
considered additional labels for the Berwald, Cartan and another type
d--connections.

\begin{enumerate}
\item Metric--dual--affine spaces (in brief, MDA) are usual metric--affine
spaces with a prescribed structure of ''dual'' local cooridnates.

\item Distinguished metric--dual-affine spaces (DMDA) are provided with
d--metric and d--connection structures adapted to a N--connection $\check{N}%
_{ai}$ defining a global splitting into a usual h--subspace and a
v--dual--subspace being dual to a usual v--subspace.

\item Berwald--dual--affine spaces (BDA) are Berwald--affine spaces with a
dual v--subspace. Their Berwald d--connection is stated in the form $\ $%
\begin{equation*}
^{\lbrack B]}\mathbf{\check{D}}=[^{[B]}\mathbf{\check{\Gamma}}_{\beta \gamma
}^{\alpha }]=[\widehat{L}_{\ jk}^{i},\partial _{b}\check{N}_{ai},0,\check{C}%
_{a}^{~[bc]}]
\end{equation*}%
with induced d--torsions $\ ^{[B]}\mathbf{\check{T}}_{\ \beta \gamma
}^{\alpha }=[L_{[jk]}^{i},0,\check{\Omega}_{iaj},\check{T}_{a~j}^{~b},%
\check{C}_{a}^{~[bc]}]$ and d--curvatures
\begin{equation*}
^{\lbrack B]}\mathbf{\check{R}}_{\ \beta \gamma \tau }^{\alpha }=[R_{\
jkl}^{i},\check{R}_{a\ kl}^{~b},,\check{P}_{\ jk}^{i~~a},\check{P}%
_{c~k}^{~b~a},\check{S}_{\ j}^{i~bc},\check{S}_{a\ }^{~dbc}]
\end{equation*}%
computed by introducing the components of $^{[B]}\mathbf{\check{\Gamma}}%
_{\beta \gamma }^{\alpha },$ respectively, in formulas (\ref{dtorsb})\ and (%
\ref{dcurv}) re--defined for dual v--subspaces. By definition, this
d--connection satisfies the metricity conditions in the h- and v--subspaces,
$Q_{ijk}=0$ and $\check{Q}^{abc}=0$ but with nontrivial components of $^{[B]}%
\mathbf{\check{Q}}_{\alpha \beta \gamma }=-\ ^{[B]}\mathbf{\check{D}}%
_{\alpha }\mathbf{\check{g}}_{\beta \gamma }=$ $[Q_{ijk}=0,\check{Q}%
_{i}^{~ab},\check{Q}_{~jk}^{a},\check{Q}^{abc}=0].$

\item Berwald--dual--affine spaces with prescribed torsion (BDAT) are
described by a more general class of d--connections $\ ^{[BT]}\mathbf{\check{%
\Gamma}}_{\beta \gamma }^{\alpha }=[L_{\ jk}^{i},\partial _{b}\check{N}%
_{ai},0,\check{C}_{a}^{~bc}]$ $,$ inducing prescribed values $\tau _{\
jk}^{i}$ and $\check{\tau}_{a\ }^{~bc}$ for d--torsions\
\begin{equation*}
^{\lbrack BT]}\mathbf{\check{T}}_{\ \beta \gamma }^{\alpha }=[L_{\
[jk]}^{i}+\tau _{\ jk}^{i},0,\check{\Omega}_{iaj}=\delta _{\lbrack i}%
\check{N}_{j]a},T_{a~j}^{~b},\check{C}_{a}^{~[bc]}+\check{\tau}_{a\ }^{~bc}].
\end{equation*}%
The components of d--curvatures
\begin{equation*}
^{\lbrack BT]}\mathbf{\check{R}}_{\ \beta \gamma \tau }^{\alpha }=\ [R_{\
jkl}^{i},\check{R}_{a~\ kl}^{~b},\check{P}_{\ jk}^{i~a},\check{P}_{c\
k}^{~b~a},\check{S}_{\ j}^{i~bc},\check{S}_{a\ }^{~dbc}]
\end{equation*}%
have to be computed by introducing $^{[BT]}\mathbf{\check{\Gamma}}_{\beta
\gamma }^{\alpha }$ into dual form of formulas (\ref{dcurv}). There are
nontrivial components of nonmetricity d--field, $^{[B\tau ]}\mathbf{Q}%
_{\alpha \beta \gamma }=-\ ^{[BT]}\mathbf{\check{D}}_{\alpha }\mathbf{\check{%
g}}_{\beta \gamma }=(Q_{ijk}=0,\check{Q}_{i}^{~ab},\check{Q}_{~jk}^{a},$ $%
\check{Q}^{abc}=0).$

\item Generalized Hamilton--affine spaces (GHA), $\mathbf{GHa}^{n}=\left(
V^{n},\check{g}^{ij}(x,p),\ ^{[a]}\mathbf{\check{\Gamma}}_{\ \beta }^{\alpha
}\right) ,$ are modeled as distinguished metric--affine spaces of
odd--dimension, $\mathbf{V}^{n+n},$ provided with generic off--diagonal
metrics with associated N--connection inducing a cotangent bundle structure.
The d--metric $\mathbf{\check{g}}_{[a]}=[g_{ij},\check{h}^{ab}]$ and the
d--connection $\ \ ^{[a]}\mathbf{\check{\Gamma}}_{\ \alpha \beta }^{\gamma }$
$=(\ ^{[a]}L_{jk}^{i},$ $\ ^{[a]}\check{C}_{i}^{~jc})$ are similar to those
for usual Hamilton spaces (see section \ref{shgg}) but with distorsions $\
^{[a]}\ \mathbf{\check{Z}}_{\ \ \beta }^{\alpha }$ inducing general
nontrivial nonmetricity d--fields $^{[a]}\mathbf{\check{Q}}_{\alpha \beta
\gamma }.$ The components of d--torsion and d--curvature, respectively, $%
^{[a]}\mathbf{\check{T}}_{\ \beta \gamma }^{\alpha }=[L_{\ [jk]}^{i},\check{%
\Omega}_{iaj},\check{C}_{a}^{\ [bc]}]$ and $^{[a]}\mathbf{\check{R}}_{.\beta
\gamma \tau }^{\alpha }=[R_{\ jkl}^{i},\check{P}_{\ jk}^{i~a},\check{S}_{a\
}^{\ dbc}],$ are computed following Theorems \ref{ttgfls} and \ref{tcgfls}
reformulated for cotangent bundle structures.

\item Hamilton--affine spaces (HA, see Remark \ \ref{rghas}), $\mathbf{Ha}%
^{n}=(V^{n},$ $\check{g}_{[H]}^{ij}(x,p),$ $\ ^{[b]}\mathbf{\check{\Gamma}}%
_{\ \beta }^{\alpha }),$ are provided with Hamilton N--connection $^{[H]}%
\check{N}_{ij}\left( x,p\right) $ (\ref{ncchs}) and quadratic form $\check{g}%
_{[H]}^{ij}$ (\ref{hsm}) for a Hamilton space $\mathbf{H}^{n}=\left(
V^{n},H(x,p)\right) $ (see section \ref{shgg})) but with a d--connection
structure $^{[H]}\mathbf{\check{\Gamma}}_{\ \alpha \beta }^{\gamma }=\
^{[H]}[L_{\ jk}^{i},\check{C}_{a}^{\ bc}]$ distorted by arbitrary torsion, $%
\mathbf{\check{T}}_{\ \ \beta \gamma }^{\alpha },$ and nonmetricity
d--fields,$\ \mathbf{\check{Q}}_{\beta \gamma \alpha },$ when $\mathbf{%
\check{\Gamma}}_{\ \beta }^{\alpha }=\ ^{[H]}\widehat{\mathbf{\check{\Gamma}}%
}_{\beta }^{\alpha }+\ \ ^{[H]}\ \mathbf{\check{Z}}_{\ \ \beta }^{\alpha }.$
This is a particular case of GHA spaces with prescribed types of
N--connection $\ ^{[H]}\check{N}_{ij}$ and d--metric$\ \mathbf{\check{g}}%
_{\alpha \beta }^{[H]}=[g_{[H]}^{ij}=\frac{1}{2}\frac{\partial ^{2}H}{%
\partial p_{i}\partial p_{i}}]$ to be like in the Hamilton geometry.

\item Cartan--affine spaces (CA, see Remark \ref{rghas}), $\mathbf{Ca}%
^{n}=\left( V^{n},\check{g}_{[K]}^{ij}(x,p),\ ^{[c]}\mathbf{\check{\Gamma}}%
_{\ \beta }^{\alpha }\right) ,$ are dual to the Finsler spaces $\mathbf{Fa}%
^{n}=\left( V^{n},F\left( x,y\right) ,\ ^{[f]}\mathbf{\Gamma }_{\ \beta
}^{\alpha }\right) .$ The CA spaces are introduced by further restrictions
of $\mathbf{Ha}^{n}$ to a quadratic form $\check{g}_{[C]}^{ij}$ (\ref{carm})
and canonical N--connection $\check{N}_{ij}^{[C]}$ (\ref{nccartan}). They
are like usual Cartan spaces,\ see section \ref{scs}) but contain
distorsions induced by nonmetricity $\mathbf{\check{Q}}_{\alpha \beta \gamma
}.$ The d--metric is parametrized $\mathbf{\check{g}}_{\alpha \beta
}^{[C]}=[g_{[C]}^{ij}=\frac{1}{2}\frac{\partial ^{2}K^{2}}{\partial
p_{i}\partial p_{i}}]$ and the curvature $\ ^{[C]}\check{\Omega}_{iaj}$ of
N--connection $\ ^{[C]}\check{N}_{ia}$ is computed $\ ^{[C]}\check{\Omega}%
_{iaj}=\delta _{\lbrack i}\ ^{[C]}\check{N}_{j]a}.$ The Cartan's
d--connection \ $^{[C]}\mathbf{\check{\Gamma}}_{\ \alpha \beta }^{\gamma }=\
^{[C]}[L_{\ jk}^{i},L_{\ jk}^{i},\check{C}_{a}^{\ bc},\check{C}_{a}^{\ bc}]$
possess nontrivial d--torsions $^{[C]}\mathbf{\check{T}}_{\alpha \ }^{\
\beta \gamma }=[L_{\ [jk]}^{i},\check{\Omega}_{iaj},\check{C}_{a}^{\ [bc]}]$
and d--curvatures $^{[C]}\mathbf{\check{R}}_{.\beta \gamma \tau }^{\alpha
}=[R_{\ jkl}^{i},$ $\check{P}_{\ jk}^{i~a},$ $\check{S}_{a\ }^{\ dbc}]$
computed following Theorems \ref{ttgfls} and \ref{tcgfls} reformulated on
cotangent bundles with explicit type of N--connection $\check{N}_{ij}^{[C]}$
d--metric $\mathbf{\check{g}}_{\alpha \beta }^{[C]}$ and d--connection $%
^{[C]}\mathbf{\check{\Gamma}}_{\ \alpha \beta }^{\gamma }.$ The nonmetricity
d--fields are not trivial for such spaces, $\ ^{[C]}\mathbf{\check{Q}}%
_{\alpha \beta \gamma }=-\ ^{[C]}\mathbf{\check{D}}_{\alpha }\mathbf{\check{g%
}}_{\beta \gamma }$ $=$ $[Q_{ijk},$ $\check{Q}_{i}^{~ab},$ $\check{Q}%
_{~jk}^{a},\check{Q}^{abc}].$
\end{enumerate}

\subsection{Teleparallel Lagrange--affine spaces}

We considered the main properties of teleparallel Finsler--affine spaces in
section \ref{stpfa} (see also section \ref{stps} on locally isotropic
teleparallel spaces). Every type of teleparallel spaces is distinguished by
the condition that the curvature tensor vanishes but the torsion plays a
cornstone role. Modeling generalized Finsler structures on metric--affine
spaces, we do not impose the condition on vanishing nonmetricity (which is
stated for usual teleparallel spaces). For $\mathbf{R}_{\ \beta \gamma \tau
}^{\alpha }=0,$ the classification of spaces from Table \ref{tablegs}
trasforms in that from Table \ref{tabletls}.

\begin{enumerate}
\item Teleparallel metric--affine spaces (in brief, TMA) are usual
metric--affine ones but with vanishing curvature, modeled on manifolds $%
V^{n+m}$ of necessary smoothly class\ provided, for instance, with the
Weitzenbock connection $^{[W]}\Gamma _{\beta \gamma }^{\alpha }$ (\ref{wcon}%
). For generic off--diagonal metrics, a TMA\ space always admits nontrivial
N--connection structures (see Proposition \ref{pmasnc}). We can model
teleparallel geometries with local anisotropy by distorting the Levi--Civita
or the canonical d--connection $\mathbf{\Gamma }_{\beta \gamma }^{\alpha }$
(see Definition \ref{defdcon}) both constructed from the components of
N--connection and d--metric. In general, such geometries are characterized
by d--torsion $\mathbf{T}_{\ \beta \gamma }^{\alpha }$ and nonmetricity
d--field $\mathbf{Q}_{\alpha \beta \gamma }$ both constrained to the
condition to result in zero d--curvatures.

\item Distinguished teleparallel metric--affine spaces (DTMA) are manifolds $%
\mathbf{V}^{n+m}$ provided with N--connection structure $N_{i}^{a},$
d--metric field (\ref{block2}) and d--connection $\mathbf{\Gamma }_{\beta
\gamma }^{\alpha }$ with vanishing d--curvatures defined by
Weitzenbock--affine d--connection $^{[Wa]}\mathbf{\Gamma }_{\beta \gamma
}^{\alpha }=\mathbf{\Gamma }_{\bigtriangledown ~\beta \gamma }^{\alpha }+%
\mathbf{\hat{Z}}_{~\beta \gamma }^{\alpha }+\mathbf{Z}_{~\beta \gamma
}^{\alpha }$ with distorsions by nonmetricity d--fields preserving the
condition of zero values for d--curvatures.

\item Teleparallel Berwald--affine spaces (TBA) are defined by distorsions
of the Weitzenbock connection to any Berwald like strucutre, $\ ^{[WB]}%
\mathbf{\Gamma }_{\beta \gamma }^{\alpha }=\mathbf{\Gamma }%
_{\bigtriangledown ~\beta \gamma }^{\alpha }+\mathbf{\hat{Z}}_{~\beta \gamma
}^{\alpha }+\mathbf{Z}_{~\beta \gamma }^{\alpha }$ satisfying the condition
that the curvature is zero. All constructions with generic off--diagonal
metrics can be adapted to the N--connection and considered for d--objects.
By definition, such spaces satisfy the metricity conditions in the h- and
v--subspaces, $Q_{ijk}=0$ and $Q_{abc}=0,$ but, in general, there are
nontrivial nonmetricity d--fields because $Q_{iab}$ and $Q_{ajk}$ are not
vanishing (see formulas (\ref{berwnm})).

\item Teleparallel Berwald--affine spaces with prescribed torsion (TBAT) are
defined by a more general class of distorsions resulting in the Weitzenbock
type d--connections, $\ ^{[WB\tau ]}\mathbf{\Gamma }_{\beta \gamma }^{\alpha
}=\mathbf{\Gamma }_{\bigtriangledown ~\beta \gamma }^{\alpha }+\mathbf{\hat{Z%
}}_{~\beta \gamma }^{\alpha }+\mathbf{Z}_{~\beta \gamma }^{\alpha },$ with
more general h-- and v--components, $\ \widehat{L}_{\ jk}^{i}\rightarrow
L_{\ jk}^{i}$ and $\widehat{C}_{\ bc}^{a}\rightarrow C_{\ bc}^{a},$ having
prescribed values $\tau _{\ jk}^{i}$ and $\tau _{\ bc}^{a}$ in d--torsion\ $%
^{[WB]}\mathbf{T}_{\ \beta \gamma }^{\alpha }=[L_{\ [jk]}^{i},+\tau _{\
jk}^{i},0,\Omega _{ij}^{a},T_{\ bj}^{a},C_{\ [bc]}^{a}+\tau _{\ bc}^{a}]$
and characterized by the condition $^{[WB\tau ]}\mathbf{R}_{\ \beta \gamma
\tau }^{\alpha }=0$ with notrivial components of nonmetricity $^{[WB\tau ]}%
\mathbf{Q}_{\alpha \beta \gamma }=\left( Q_{cij},\ Q_{iab}\right) .$

\item Teleparallel generalized Lagrange--affine spaces (TGLA) are
distinguished metric--affine spaces of odd--dimension, $\mathbf{V}^{n+n},$
provided with generalized Lagrange d--metric and associated N--connection
inducing a tangent bundle structure with zero d--cur\-va\-tu\-re. The\
Weitzenblock--Lagrange d--connection$\ ^{[Wa]}\mathbf{\Gamma }_{\ \alpha
\beta }^{\gamma }$ $=\left( \ ^{[Wa]}L_{jk}^{i},\ ^{[Wa]}C_{jc}^{i}\right)
,~\ $\ where $^{[WaL]}\mathbf{\Gamma }_{\beta \gamma }^{\alpha }=\mathbf{%
\Gamma }_{\bigtriangledown ~\beta \gamma }^{\alpha }+\mathbf{\hat{Z}}%
_{~\beta \gamma }^{\alpha }+\mathbf{Z}_{~\beta \gamma }^{\alpha }$ is
defined by a d--metric $\mathbf{g}_{[a]}$ (\ref{dmglas}) $\mathbf{Z}_{\ \
\beta }^{\alpha }$ inducing general nontrivial nonmetricity d--fields $^{[a]}%
\mathbf{Q}_{\alpha \beta \gamma }$ and $^{[Wa]}\mathbf{R}_{\ \beta \gamma
\tau }^{\alpha }=0.$

\item Teleparallel Lagrange--affine spaces (TLA\ \ref{rlafs}) consist a
subclass of spaces $\mathbf{La}^{n}=\left( V^{n},g_{ij}^{[L]}(x,y),\ ^{[b]}%
\mathbf{\Gamma }_{\ \beta }^{\alpha }\right) $ provided with a Lagrange
quadratic form $g_{ij}^{[L]}(x,y)=\frac{1}{2}\frac{\partial ^{2}L^{2}}{%
\partial y^{i}\partial y^{j}}$ (\ref{lagm}) inducing the canonical
N--connection structure $^{[cL]}\mathbf{N}=\{\ ^{[cL]}N_{j}^{i}\}$ (\ref%
{cncls})\ for a Lagrange space $\mathbf{L}^{n}=\left(
V^{n},g_{ij}(x,y)\right) $ but with vanishing d--curvature. The
d--connection structure $\ ^{[WL]}\mathbf{\Gamma }_{\ \alpha \beta }^{\gamma
}$ (of Weitzenblock--Lagrange type) is the generated as a distortion by the
Weitzenbock d--torsion, $^{[W]}\mathbf{T}_{\beta },$ and nonmetricity
d--fields,$\ \mathbf{Q}_{\beta \gamma \alpha },$ when $^{[WL]}\mathbf{\Gamma
}_{\ \alpha \beta }^{\gamma }=\mathbf{\Gamma }_{\bigtriangledown ~\beta
\gamma }^{\alpha }+\mathbf{\hat{Z}}_{~\beta \gamma }^{\alpha }+\mathbf{Z}%
_{~\beta \gamma }^{\alpha }.$ This is a generalization of teleparallel
Finsler affine spaces (see section (\ref{stpfa})) when $g_{ij}^{[L]}(x,y)$
is considered instead of $g_{ij}^{[F]}(x,y).$

\item Teleparallel Finsler--affine spaces (TFA) are particular cases of \
spaces of type $\mathbf{Fa}^{n}=(V^{n},F\left( x,y\right) ,\ ^{[f]}\mathbf{%
\Gamma }_{\ \beta }^{\alpha }),$ defined by a quadratic form $g_{ij}^{[F]}=%
\frac{1}{2}\frac{\partial ^{2}F^{2}}{\partial y^{i}\partial y^{j}}$ (\ref%
{finm2}) constructed from a Finsler metric $F\left( x^{i},y^{j}\right) .$
They are provided with a canonical N--connection structure $\ ^{[F]}\mathbf{N%
}=\{\ ^{[F]}N_{j}^{i}\}$ (\ref{ncc})\ as in the Finsler space $\mathbf{F}%
^{n}=\left( V^{n},F\left( x,y\right) \right) $ but with a
Finsler--Weitzenbock d--connection structure $\ ^{[WF]}\mathbf{\Gamma }_{\
\alpha \beta }^{\gamma },$ respective d--torsion, $^{[WF]}\mathbf{T}_{\beta
},$ and nonmetricity, $\mathbf{Q}_{\beta \gamma \tau },$ d--fields, $\
^{[WF]}\mathbf{\Gamma }_{\ \alpha \beta }^{\gamma }=\mathbf{\Gamma }%
_{\bigtriangledown ~\beta \gamma }^{\alpha }+\mathbf{\hat{Z}}_{~\beta \gamma
}^{\alpha }+\mathbf{Z}_{~\beta \gamma }^{\alpha },$where $\ \mathbf{\hat{Z}}%
_{~\beta \gamma }^{\alpha }$ contains distorsions from the canonical Finsler
d--connection (\ref{dccfs}). Such distorsions are constrained to satisfy the
condition of vanishing curvature d--tensors (see section (\ref{stpfa})).
\end{enumerate}

\subsection{Teleparallel Hamilton--affine spaces}

This class of metric--affine spaces is similar to that outlined in previous
subsection, see Table \ref{tabletls} but derived on spaces with dual vector
bundle structure and induced generalized Hamilton--Cartan geometry (section %
\ref{shgg} and Remark \ref{rghas}). We outline the main denotations for such
spaces and note that they are characterized by the condition $\mathbf{%
\check{R}}_{\ \beta \gamma \tau }^{\alpha }=0.$

\begin{enumerate}
\item Teleparallel metric dual affine spaces (in brief, TMDA) define
teleparalles structures on metric--affine spaces provided with generic
off--diagonal metrics and associated N--connections modeling splittings with
effective dual vector bundle structures.

\item Distinguished teleparallel metric dual affine spaces (DTMDA) are
spaces provided with independent d--metric, d--connection structures adapted
to a N--connection in an effective dual vector bundle and resulting in zero
d--curvatures.

\item Teleparallel Berwald dual affine spaces (TBDA) .

\item Teleparallel dual Berwald--affine spaces with prescribed torsion
(TDBAT).

\item Teleparallel dual generalized Hamilton--affine spaces (TDGHA).

\item Teleparallel dual Hamilton--affine spaces (TDHA, see\ section \ref%
{rlafs}).

\item Teleparallel dual Cartan--affine spaces (TDCA).
\end{enumerate}

\subsection{Generalized Finsler--Lagrange spaces}

This class of geometries is modeled on vector/tangent bundles \cite{ma} (see
subsections \ref{ssffc} and \ref{ssslgg}) or on metric--affine spaces
provided with N--connection structure. There are also alternative variants
when metric--affine structures are defined for vector/tangent bundles with
independent generic off--diagonal metrics and linear connection structures.
The standard approaches to generalized Finsler geometries emphasize the
connections satisfying the metricity conditions. Nevertherless, the Berwald
type connections admit certain nonmetricity d--fields. The classification
stated in Table \ref{tablegfls} is similar to that from Table \ref{tablegs}
with that difference that the spaces are defined from the very beginning to
be any vector or tangent bundles. The local coordinates $x^{i}$ are
considered for base subspaces and $y^{a}$ are for fiber type subspaces. We
list the short denotations and main properties of such spaces:

\begin{enumerate}
\item Metric affine vector bundles (in brief, MAVB) are provieded with
arbitrary metric $g_{\alpha \beta }~$ and linear connection $\Gamma _{\beta
\gamma }^{\alpha }$ structure. For generic off--diagonal metrics, we can
introduce associated nontrivial N--connection structures. In general, only
the metric field $g_{\alpha \beta }$ can be transformed into a d--metric $%
\mathbf{g}_{\alpha \beta }=[g_{ij},h_{ab}],$ but \ $\Gamma _{\beta \gamma
}^{\alpha }$ may be not adapted to the N--connection structure. As a
consequence, the general strength fields $\left( T_{\ \beta \gamma }^{\alpha
},R_{\ \beta \gamma \tau }^{\alpha },Q_{\alpha \beta \gamma }=0\right) ,$
defined in the total space of the vector bundle are also not N--adapted. We
can consider a metric--affine (MA) structure on the total space if $%
Q_{\alpha \beta \gamma }\neq 0.$

\item Distinguished metric--affine vector bundles (DMAVB) are provided with
N--connection structure $N_{i}^{a},$ d--metric field and arbitrary
d--connection $\mathbf{\Gamma }_{\beta \gamma }^{\alpha }.$ In this case,
all strengths $(\mathbf{T}_{\ \beta \gamma }^{\alpha },\mathbf{R}_{\ \beta
\gamma \tau }^{\alpha },\mathbf{Q}_{\alpha \beta \gamma }=0) $ are
N--adapted. A distinguished metric--affine (DMA) structure on the total
space is considered if $\mathbf{Q}_{\alpha \beta \gamma }\neq 0.$

\item Berwald metric--affine tangent bundles (BMATB) are provided with
Berwald d--con\-nec\-ti\-on structure $^{[B]}\mathbf{\Gamma }.$ By
definition, this space satisfies the metricity conditions in the h- and
v--subspaces, $Q_{ijk}=0$ and $Q_{abc}=0,$ but, in general, there are
nontrivial nonmetricity d--fields because $Q_{iab}$ and $Q_{ajk}$ do not
vanish (see formulas (\ref{berwnm})).

\item Berwald metric--affine bundles with prescribed torsion (BMATBT) are
described by a more general class of d--connection $^{[BT]}\mathbf{\Gamma }%
_{\beta \gamma }^{\alpha }=[L_{\ jk}^{i},\partial _{b}N_{k}^{a},0,C_{\
bc}^{a}]$ inducing prescribed values $\tau _{\ jk}^{i}$ and $\tau _{\
bc}^{a} $ in d--torsion\
\begin{equation*}
^{\lbrack BT]}\mathbf{T}_{\ \beta \gamma }^{\alpha }=[L_{\ [jk]}^{i},+\tau
_{\ jk}^{i},0,\Omega _{ij}^{a},T_{\ bj}^{a},C_{\ [bc]}^{a}+\tau _{\ bc}^{a}],
\end{equation*}%
see (\ref{tauformulas})$.$ There are nontrivial nonmetricity d--fields, $%
^{[B\tau ]}\mathbf{Q}_{\alpha \beta \gamma }=(Q_{cij},Q_{iab}).$

\item Generalized Lagrange metric--affine bundles (GLMAB) are modeled as $%
\mathbf{GLa}^{n}=(V^{n},$ $g_{ij}(x,y),\ ^{[a]}\mathbf{\Gamma }_{\ \beta
}^{\alpha })$ spaces on tangent bundles provided with generic off--diagonal
metrics with associated N--connection. If the d--connection is a canonical
one, $\widehat{\mathbf{\Gamma }}_{\beta \gamma }^{\alpha },$ the
nonmetricity vanish. But we can consider arbitrary d--connections $\mathbf{%
\Gamma }_{\beta \gamma }^{\alpha }$ with nontrivial nonmetricity d--fields.

\item Lagrange metric--affine bundles (LMAB) are defined on tangent bundles
as spaces $\mathbf{La}^{n}=\left( V^{n},g_{ij}^{[L]}(x,y),\ ^{[b]}\mathbf{%
\Gamma }_{\ \beta }^{\alpha }\right) $ provided with a Lagrange quadratic
form $g_{ij}^{[L]}(x,y)=\frac{1}{2}\frac{\partial ^{2}L^{2}}{\partial
y^{i}\partial y^{j}}$ inducing the canonical N--connection structure $^{[cL]}%
\mathbf{N}=\{\ ^{[cL]}N_{j}^{i}\}$ for a Lagrange space $\mathbf{L}%
^{n}=\left( V^{n},g_{ij}(x,y)\right) $ (see Definition \ref{defls})) but
with a d--connection structure $\ ^{[b]}\mathbf{\Gamma }_{\ \ \alpha
}^{\gamma }=\ ^{[b]}\mathbf{\Gamma }_{\ \alpha \beta }^{\gamma }\mathbf{%
\vartheta }^{\beta }$ distorted by arbitrary torsion, $\mathbf{T}_{\beta },$
and nonmetricity d--fields,$\ \mathbf{Q}_{\beta \gamma \alpha },$ when $%
^{[b]}\mathbf{\Gamma }_{\ \beta }^{\alpha }=\ ^{[L]}\widehat{\mathbf{\Gamma }%
}_{\beta }^{\alpha }+\ \ ^{[b]}\ \mathbf{Z}_{\ \ \beta }^{\alpha }.$ This is
a particular case of GLA spaces with prescribed types of N--connection $%
^{[cL]}N_{j}^{i}$ and d--metric to be like in Lagrange geometry.

\item Finsler metric--affine bundles (FMAB), are modeled on tangent bundles
as spaces $\mathbf{Fa}^{n}=\left( V^{n},F\left( x,y\right) ,\ ^{[f]}\mathbf{%
\Gamma }_{\ \beta }^{\alpha }\right) $ with quadratic form $g_{ij}^{[F]}=%
\frac{1}{2}\frac{\partial ^{2}F^{2}}{\partial y^{i}\partial y^{j}}$ (\ref%
{finm2}) constructed from a Finsler metric $F\left( x^{i},y^{j}\right) .$ It
is induced the canonical N--connection structure $\ ^{[F]}\mathbf{N}=\{\
^{[F]}N_{j}^{i}\}$ as in the Finsler space $\mathbf{F}^{n}=\left(
V^{n},F\left( x,y\right) \right) $ but with a d--connection structure $\
^{[f]}\mathbf{\Gamma }_{\ \alpha \beta }^{\gamma }$ distorted by arbitrary
torsion, $\mathbf{T}_{\beta \gamma }^{\alpha },$ and nonmetricity, $\mathbf{Q%
}_{\beta \gamma \tau },$ d--fields, $\ ^{[f]}\mathbf{\Gamma }_{\ \beta
}^{\alpha }=\ ^{[F]}\widehat{\mathbf{\Gamma }}_{\ \beta }^{\alpha }\ +\ \
^{[f]}\ \mathbf{Z}_{\ \ \beta }^{\alpha },$where $\ ^{[F]}\widehat{\mathbf{%
\Gamma }}_{\ \beta \gamma }^{\alpha }$ is the canonical Finsler
d--connection (\ref{dccfs}).
\end{enumerate}

\subsection{Generalized Hamilton--Cartan spaces}

Such spaces are modeled on vector/tangent dual bundles (see sections
subsections \ref{shgg} and \ref{scs}) as metric--affine spaces provided with
N--connection structure. The classification stated in Table \ref{tableghcs}
is similar to that from Table \ref{tableghs} with that difference that the
geometry is modeled from the very beginning as vector or tangent dual
bundles. The local coordinates $x^{i}$ are considered for base subspaces and
$y^{a}=p_{a}$ are for cofiber type subspaces. So, the spaces from Table \ref%
{tableghcs} are dual to those from Table \ref{tabletfls}, when the
respective Lagrange--Finsler structures are changed into Hamilton--Cartan
structures. We list the short denotations and main properties of such spaces:

\begin{enumerate}
\item The metric--affine dual vector bundles (in brief, MADVB) are defined
by metric--affine independent metric and linear connection structures stated
on dual vector bundles. For generic off--diagonal metrics, there are
nontrivial N--connection structures. The linear connection may be not
adapted to the N--connection structure.

\item Distinguished metric-affine dual vector bundles (DMADVB) are provided
with d--metric and d--connection structures adapted to a N--connection $%
\check{N}_{ai}.$

\item Berwald metric--affine dual bundles (BMADB) are provided with a
Berwald d--con\-nec\-ti\-on
\begin{equation*}
\ ^{[B]}\mathbf{\check{D}}=[\ ^{[B]}\mathbf{\check{\Gamma}}_{\beta \gamma
}^{\alpha }]=[\widehat{L}_{\ jk}^{i},\partial _{b}\check{N}_{ai},0,\check{C}%
_{a}^{~[bc]}].
\end{equation*}%
By definition, on such spaces, there are satisfied the metricity conditions
in the h- and v--subspaces, $Q_{ijk}=0$ and $\check{Q}^{abc}=0$ but with
nontrivial components of $^{[B]}\mathbf{\check{Q}}_{\alpha \beta \gamma }=-\
^{[B]}\mathbf{\check{D}}_{\alpha }\mathbf{\check{g}}_{\beta \gamma
}=[Q_{ijk}=0,\check{Q}_{i}^{~ab},\check{Q}_{~jk}^{a},\check{Q}^{abc}=0].$

\item Berwald metricl--affine dual bundles with prescribed torsion (BMADBT)
are described by a more general class of d--connections $\ ^{[BT]}\mathbf{%
\check{\Gamma}}_{\beta \gamma }^{\alpha }=[L_{\ jk}^{i},\partial _{b}%
\check{N}_{ai},0,\check{C}_{a}^{~bc}]$ inducing prescribed values $\tau _{\
jk}^{i}$ and $\check{\tau}_{a\ }^{~bc}$ for d--torsions\
\begin{equation*}
^{\lbrack BT]}\mathbf{\check{T}}_{\ \beta \gamma }^{\alpha }=[L_{\
[jk]}^{i}+\tau _{\ jk}^{i},0,\check{\Omega}_{iaj}=\delta _{\lbrack i}%
\check{N}_{j]a},T_{a~j}^{~b},\check{C}_{a}^{~[bc]}+\check{\tau}_{a\ }^{~bc}].
\end{equation*}%
There are nontrivial components of nonmetricity d--field, $^{[B\tau ]}%
\mathbf{Q}_{\alpha \beta \gamma }=\ ^{[BT]}\mathbf{\check{D}}_{\alpha }%
\mathbf{\check{g}}_{\beta \gamma }=\left( Q_{ijk}=0,\check{Q}_{i}^{~ab},%
\check{Q}_{~jk}^{a},\check{Q}^{abc}=0\right) .$

\item Generalized metric--affine Hamilton bundles (GMAHB) are modeled on
dual vector bundles as spaces $\mathbf{GHa}^{n}=\left( V^{n},\check{g}%
^{ij}(x,p),\ ^{[a]}\mathbf{\check{\Gamma}}_{\ \beta }^{\alpha }\right) ,$
provided with generic off--diagonal metrics with associated N--connection
inducing a cotangent bundle structure. The d--metric $\mathbf{\check{g}}%
_{[a]}=[g_{ij},\check{h}^{ab}]$ and the d--connection$\ ^{[a]}\mathbf{\check{%
\Gamma}}_{\ \alpha \beta }^{\gamma }$ $=\left( \ ^{[a]}L_{jk}^{i},\ ^{[a]}%
\check{C}_{i}^{~jc}\right) $ are similar to those for usual Hamilton spaces,
with distorsions $\ ^{[a]}\ \mathbf{\check{Z}}_{\ \ \beta }^{\alpha }$
inducing general nontrivial nonmetricity d--fields $^{[a]}\mathbf{\check{Q}}%
_{\alpha \beta \gamma }.$ For canonical configurations, $^{[GH]}\mathbf{%
\check{\Gamma}}_{\ \alpha \beta }^{\gamma },$ we obtain $^{[GH]}\mathbf{%
\check{Q}}_{\alpha \beta \gamma }=0.$

\item Metric--affine Hamilton bundles (MAHB) are defied on dual bundles as
spaces $\mathbf{Ha}^{n}=\left( V^{n},\check{g}_{[H]}^{ij}(x,p),\ ^{[b]}%
\mathbf{\check{\Gamma}}_{\ \beta }^{\alpha }\right) ,$ provided with
Hamilton N--connection $^{[H]}\check{N}_{ij}\left( x,p\right) $ and
qua\-drat\-ic form $\check{g}_{[H]}^{ij}$ for a Hamilton space $\mathbf{H}%
^{n}=\left( V^{n},H(x,p)\right) $ (see section \ref{shgg}) with a
d--connection structure $^{[H]}\mathbf{\check{\Gamma}}_{\ \alpha \beta
}^{\gamma }=\ ^{[H]}[L_{\ jk}^{i},\check{C}_{a}^{\ bc}]$ distorted by
arbitrary torsion, $\mathbf{\check{T}}_{\ \ \beta \gamma }^{\alpha },$ and
nonmetricity d--fields,$\ \mathbf{\check{Q}}_{\beta \gamma \alpha },$ when $%
\mathbf{\check{\Gamma}}_{\ \beta }^{\alpha }=\ ^{[H]}\widehat{\mathbf{\check{%
\Gamma}}}_{\beta }^{\alpha }+\ \ ^{[H]}\ \mathbf{\check{Z}}_{\ \ \beta
}^{\alpha }.$ This is a particular case of GMAHB spaces with prescribed
types of N--connection $\ ^{[H]}\check{N}_{ij}$ and d--metric$\ \mathbf{%
\check{g}}_{\alpha \beta }^{[H]}=[g_{[H]}^{ij}=\frac{1}{2}\frac{\partial
^{2}H}{\partial p_{i}\partial p_{i}}]$ to be like in the Hamilton geometry
but with nontrivial nonmetricity.

\item Metric--affine Cartan bundles (MACB) are modeled on dual tangent
bundles as spaces $\mathbf{Ca}^{n}=\left( V^{n},\check{g}_{[K]}^{ij}(x,p),\
^{[c]}\mathbf{\check{\Gamma}}_{\ \beta }^{\alpha }\right) $ being dual to
the Finsler spaces\textbf{.} They are like usual Cartan spaces,\ see section %
\ref{scs}) but may contain distorsions induced by nonmetricity $\mathbf{%
\check{Q}}_{\alpha \beta \gamma }.$ The d--metric is parametrized $\mathbf{%
\check{g}}_{\alpha \beta }^{[C]}=[g_{[C]}^{ij}=\frac{1}{2}\frac{\partial
^{2}K^{2}}{\partial p_{i}\partial p_{i}}]$ and the curvature $\ ^{[C]}\check{%
\Omega}_{iaj}$ of N--connection $\ ^{[C]}\check{N}_{ia}$ is computed $\
^{[C]}\check{\Omega}_{iaj}=\delta _{\lbrack i}\ ^{[C]}\check{N}_{j]a}.$ The
Cartan's d--connection \ $^{[C]}\mathbf{\check{\Gamma}}_{\ \alpha \beta
}^{\gamma }=\ ^{[C]}[L_{\ jk}^{i},L_{\ jk}^{i},\check{C}_{a}^{\ bc},\check{C}%
_{a}^{\ bc}]$ possess nontrivial d--torsions $^{[C]}\mathbf{\check{T}}%
_{\alpha \ }^{\ \beta \gamma }=[L_{\ [jk]}^{i},\check{\Omega}_{iaj},\check{C}%
_{a}^{\ [bc]}]$ and d--curvatures $^{[C]}\mathbf{\check{R}}_{.\beta \gamma
\tau }^{\alpha }=~^{[C]}[R_{\ jkl}^{i},\check{P}_{\ jk}^{i~a},\check{S}_{a\
}^{\ dbc}]$ computed following Theorems \ref{ttgfls} and \ref{tcgfls}
reformulated on cotangent bundles with explicit type of N--connection $%
\check{N}_{ij}^{[C]}$ d--metric $\mathbf{\check{g}}_{\alpha \beta }^{[C]}$
and d--connection $^{[C]}\mathbf{\check{\Gamma}}_{\ \alpha \beta }^{\gamma
}. $ Distorsions result in d--connection $\mathbf{\check{\Gamma}}_{\ \beta
\gamma }^{\alpha }=\ ^{[C]}\mathbf{\check{\Gamma}}_{\beta \gamma }^{\alpha
}+\ ^{[C]}\ \mathbf{\check{Z}}_{\ \ \beta \gamma }^{\alpha }.$ The
nonmetricity d--fields are not trivial for such spaces.
\end{enumerate}

\subsection{Teleparallel Finsler--Lagrange spaces}

The teleparallel configurations can be modeled on vector and tangent bundles
(the teleparallel Finsler--affine spaces are defined in section \ref{stpfa},
see also section \ref{stps} on locally isotropic teleparallel spaces) were
constructed as subclasses of metric--affine spaces on manifolds of necessary
smoothly class. The classification from Table \ref{tabletfls} is a similar
to that from Table \ref{tabletls} but for direct vector/ tangent bundle
configurations with vanishing nonmetricity. Nevertheless, certain nonzero
nonmetricity d--fields can be present if the Berwald d--connection is
considered or if we consider a metric--affine geometry in bundle spaces.

\begin{enumerate}
\item Teleparallel vector bundles (in brief, TVB) are provided with
independent metric and linear connection structures like in metric--affine
spaces satisfying the condition of vanishing curvature. The N--connection is
associated to generic off--diagonal metrics. The TVB spaces can be provided
with a Weitzenbock connection $^{[W]}\Gamma _{\beta \gamma }^{\alpha }$(\ref%
{wcon}) which can be transformed in a d--connection one with respect to
N--adapted frames. We can model teleparallel geometries with local
anisotropy by distorting the Levi--Civita or the canonical d--connection $%
\mathbf{\Gamma }_{\beta \gamma }^{\alpha }$ (see Definition \ref{defdcon})
both constructed from the components of N--connection and d--metric. In
general, such vector (in particular cases, tangent) bundle geometries are
characterized by d--torsions $\mathbf{T}_{\ \beta \gamma }^{\alpha }$ and
nonmetricity d--fields $\mathbf{Q}_{\alpha \beta \gamma }$ both constrained
to the condition to result in zero d--curvatures.

\item Distinguished teleparallel vector bundles (DTVB, or vect. b.) are
provided with N--connection structure $N_{i}^{a},$ d--metric field (\ref%
{block2}) and arbitrary d--connection $\mathbf{\Gamma }_{\beta \gamma
}^{\alpha }$ with vanishing d--curvatures. The geometric constructions are
stated by the Weitzenbock--affine d--connection $^{[Wa]}\mathbf{\Gamma }%
_{\beta \gamma }^{\alpha }=\mathbf{\Gamma }_{\bigtriangledown ~\beta \gamma
}^{\alpha }+\mathbf{\hat{Z}}_{~\beta \gamma }^{\alpha }+\mathbf{Z}_{~\beta
\gamma }^{\alpha }$ with distorsions by nonmetricity d--fields preserving
the condition of zero values for d--curvatures. The standard constructions
from Finsler geometry and generalizations are with vanishing nonmetricity.

\item Teleparallel Berwald vector bundles (TBVB) are defined by Weitzenbock
connections of Berwald type strucutre, $\ ^{[WB]}\mathbf{\Gamma }_{\beta
\gamma }^{\alpha }=\mathbf{\Gamma }_{\bigtriangledown ~\beta \gamma
}^{\alpha }+\mathbf{\hat{Z}}_{~\beta \gamma }^{\alpha }+\mathbf{Z}_{~\beta
\gamma }^{\alpha }$ satisfying the condition that the curvature is zero. By
definition, such spaces satisfy the metricity conditions in the h- and
v--subspaces, $Q_{ijk}=0$ and $Q_{abc}=0,$ but, in general, there are
nontrivial nonmetricity d--fields because $Q_{iab}$ and $Q_{ajk}$ do not
vanish (see formulas (\ref{berwnm})).

\item Teleparallel Berwald vector bundles with prescribed torsion (TBVBT)
are defined by a more general class of distorsions alsow resulting in the
Weitzenbock d--connection, $\ ^{[WB\tau ]}\mathbf{\Gamma }_{\beta \gamma
}^{\alpha }=\mathbf{\Gamma }_{\bigtriangledown ~\beta \gamma }^{\alpha }+%
\mathbf{\hat{Z}}_{~\beta \gamma }^{\alpha }+\mathbf{Z}_{~\beta \gamma
}^{\alpha }$ with prescribed values $\tau _{\ jk}^{i}$ and $\tau _{\ bc}^{a}$
in d--torsion,\
\begin{equation*}
^{\lbrack WB]}\mathbf{T}_{\ \beta \gamma }^{\alpha }=[L_{\ [jk]}^{i},+\tau
_{\ jk}^{i},0,\Omega _{ij}^{a},T_{\ bj}^{a},C_{\ [bc]}^{a}+\tau _{\ bc}^{a}],
\end{equation*}%
characterized by the condition $^{[WB\tau ]}\mathbf{R}_{\ \beta \gamma \tau
}^{\alpha }=0$ and nontrivial components of nonmetricity d--field, $%
^{[WB\tau ]}\mathbf{Q}_{\alpha \beta \gamma }=\left( Q_{cij},Q_{iab}\right)
. $

\item Teleparallel generalized Lagrange spaces (TGL) are modeled on tangent
bundles (tang. b.) provided with generalized Lagrange d--metric and
associated N--connection inducing a tangent bundle structure being enabled
with zero d--curvature. The\ Weitzenblock--Lagrange d--connections$\ ^{[Wa]}%
\mathbf{\Gamma }_{\ \alpha \beta }^{\gamma }$ $=\left( \ ^{[Wa]}L_{jk}^{i},\
^{[Wa]}C_{jc}^{i}\right) ,~\ $\ $^{[WaL]}\mathbf{\Gamma }_{\beta \gamma
}^{\alpha }$ $=\mathbf{\Gamma }_{\bigtriangledown ~\beta \gamma }^{\alpha }+%
\mathbf{\hat{Z}}_{~\beta \gamma }^{\alpha }+\mathbf{Z}_{~\beta \gamma
}^{\alpha }$ are defined by a d--metric $\mathbf{g}_{[a]}$ (\ref{dmglas}) $%
\mathbf{Z}_{\ \ \beta }^{\alpha }$ inducing $^{[Wa]}\mathbf{R}_{\ \beta
\gamma \tau }^{\alpha }=0.$ For simplicity, we consider the configurations
when nonmetricity d--fields $^{[Wa]}\mathbf{Q}_{\alpha \beta \gamma }=0$.

\item Teleparallel Lagrange spaces (TL) are modeled on tangent bundles
provided with a Lagrange quadratic form $g_{ij}^{[L]}(x,y)=\frac{1}{2}\frac{%
\partial ^{2}L^{2}}{\partial y^{i}\partial y^{j}}$ (\ref{lagm}) inducing the
canonical N--connection structure $^{[cL]}\mathbf{N}=\{\ ^{[cL]}N_{j}^{i}\}$
(\ref{cncls})\ for a Lagrange space $\mathbf{L}^{n}=\left(
V^{n},g_{ij}(x,y)\right) $ but with vanishing d--curvature. The
d--connection structure $\ ^{[WL]}\mathbf{\Gamma }_{\ \alpha \beta }^{\gamma
}$ \ (of Weitzenblock--Lagrange type) is the generated as a distortion by
the Weitzenbock d--torsion, $^{[W]}\mathbf{T}_{\beta }$ when $^{[WL]}\mathbf{%
\Gamma }_{\ \alpha \beta }^{\gamma }=\mathbf{\Gamma }_{\bigtriangledown
~\beta \gamma }^{\alpha }+\mathbf{\hat{Z}}_{~\beta \gamma }^{\alpha }+%
\mathbf{Z}_{~\beta \gamma }^{\alpha }.$ For simplicity, we can consider
configurations with zero nonmetricity d--fields, $\ \mathbf{Q}_{\beta \gamma
\alpha }.$

\item Teleparallel Finsler spaces (TF) are modeled on tangent bundles
provided with a quadratic form $g_{ij}^{[F]}=\frac{1}{2}\frac{\partial
^{2}F^{2}}{\partial y^{i}\partial y^{j}}$ (\ref{finm2}) constructed from a
Finsler metric $F\left( x^{i},y^{j}\right) .$ They are also enabled with a
canonical N--connection structure $\ ^{[F]}\mathbf{N}=\{\ ^{[F]}N_{j}^{i}\}$
(\ref{ncc})\ as in the Finsler space $\mathbf{F}^{n}=\left( V^{n},F\left(
x,y\right) \right) $ but with a Finsler--Weitzenbock d--connection structure$%
\ ^{[WF]}\mathbf{\Gamma }_{\ \alpha \beta }^{\gamma },$ respective
d--torsion, $^{[WF]}\mathbf{T}_{\beta }.$ We can write $\ ^{[WF]}\mathbf{%
\Gamma }_{\ \alpha \beta }^{\gamma }=\mathbf{\Gamma }_{\bigtriangledown
~\beta \gamma }^{\alpha }+\mathbf{\hat{Z}}_{~\beta \gamma }^{\alpha }+%
\mathbf{Z}_{~\beta \gamma }^{\alpha },$where $\ \mathbf{\hat{Z}}_{~\beta
\gamma }^{\alpha }$ contains distorsions from the canonical Finsler
d--connection (\ref{dccfs}). Such distorsions are constrained to satisfy the
condition of vanishing curvature d--tensors (see section (\ref{stpfa})) and,
for simplicity, of vanishing nonmetricity, $\mathbf{Q}_{\beta \gamma \tau
}=0.$
\end{enumerate}

\subsection{Teleparallel Hamilton--Cartan spaces}

This subclass of Hamilton--Cartan spaces is modeled on dual vector/ tangent
bundles being similar to that outlined in Table \ref{tableths} (on
generalized Hamilton--Cartan geometry, see section \ref{shgg} and Remark \ref%
{rghas}) and dual to the subclass outlined in Table \ref{tabletfls}. We
outline the main denotations and properties of such spaces and note that
they are characterized by the condition $\mathbf{\check{R}}_{\ \beta \gamma
\tau }^{\alpha }=0$ and $\mathbf{\check{Q}}_{\ \beta \gamma }^{\alpha }=0$
with that exception that there are nontrivial nonmetricity d--fields for
Berwald configuratons.

\begin{enumerate}
\item Teleparallel dual vector bundles (TDVB, or d. vect. b.) are provided
with generic off--diagonal metrics and associated N--connections. In
general, $\check{Q}_{\ \beta \gamma }^{\alpha }\neq 0.$

\item Distinguished teleparallel dual vector bundles spaces (DTDVB) are
provided with independent d--metric, d--connection structures adapted to a
N--connection in an effective dual vector bundle and resulting in zero
d--curvatures. In general, $\mathbf{\check{Q}}_{\ \beta \gamma }^{\alpha
}\neq 0.$

\item Teleparallel Berwald dual vector bundles (TBDVB) are provided with
Berwald--Weitz\-enbock d--connection structure resulting in vanishing
d--curvature.

\item Teleparallel Berwald dual vector bundles with prescribed d--torsion
(TBDVB) are with d--connections $\ ^{[BT]}\mathbf{\check{\Gamma}}_{\beta
\gamma }^{\alpha }=[L_{\ jk}^{i},\partial _{b}\check{N}_{ai},0,\check{C}%
_{a}^{~bc}]$ inducing prescribed values $\tau _{\ jk}^{i}$ and $\check{\tau}%
_{a\ }^{~bc}$ for d--torsions\ $^{[BT]}\mathbf{\check{T}}_{\ \beta \gamma
}^{\alpha }=[L_{\ [jk]}^{i}+\tau _{\ jk}^{i},0,\check{\Omega}_{iaj}=\delta
_{\lbrack i}\check{N}_{j]a},T_{a~j}^{~b},\check{C}_{a}^{~[bc]}+\check{\tau}%
_{a\ }^{~bc}].$ They are described by certain distorsions to a Weitzenbock
d--connection.

\item Teleparallel generalized Hamilton spaces (TGH) consist a subclass of
generalized Hamilton spaces with vanishing d--curvature structure, defined
on dual tangent bundles (d. tan. b.). They are described by distorsions to a
Weitzenbock d--connection $^{[Wa]}\mathbf{\check{\Gamma}}_{\ \alpha \beta
}^{\gamma }.$ In the simplest case, we consider $^{[Wa]}\mathbf{\check{Q}}%
_{\alpha \beta \gamma }=0.$

\item Teleparallel Hamilton spaces (TH, see section\ \ref{rlafs}), as a
particular subclass of TGH, are provided with d--connection and
N--connection structures corresponding to Hamilton configurations.

\item Teleparallel Cartan spaces (TC) are particular Cartan configurations
with absolut teleparallelism.
\end{enumerate}

\subsection{Distinguished Riemann--Cartan spaces}

A wide class of generalized Finsler geometries can be modeled on
Riemann--Cartan spaces by using generic off--diagonal metrics and associated
N--connection structures. The locally anisotropic metric--affine
configurations from Table \ref{tablegs} transform into a Riemann--Cartan
ones if we impose the condition of metricity. For the Berwald type
connections one could be certain nontrivial nonmetricity d--fields on
intersection of h- and v--subspaces. The local coordinates $x^{i}$ are
considered as certain holonomic ones and $y^{a}$ are anholonomic. We list
the short denotations and main properties of such spaces:

\begin{enumerate}
\item Riemann--Cartan spaces (in brief, RC, see related details in section %
\ref{srgarcg}) are certain manifolds $V^{n+m}$ of necessary smoothly class\
provided with metric structure $g_{\alpha \beta }~$ and linear connection
structure $\Gamma _{\beta \gamma }^{\alpha }$ (constructed as a distorsion
by torsion of the Levi--Civita connection) both satisfying the conditions of
metric compatibility, $Q_{\alpha \beta \gamma }=0.$ For generic
off--diagonal metrics, a RC\ space always admits nontrivial N--connection
structures (see Proposition \ref{pmasnc} reformulated for the case of
vanishing nonmetricity). In general, only the metric field $g_{\alpha \beta
} $ can be transformed into a d--metric one, $\mathbf{g}_{\alpha \beta
}=[g_{ij},h_{ab}],$ but \ $\Gamma _{\beta \gamma }^{\alpha }$ may be not
adapted to the N--connection structure.

\item Distinguished Riemann--Cartan spaces (DRC) are manifolds $\mathbf{V}%
^{n+m}$ provided with N--connection structure $N_{i}^{a},$ d--metric field (%
\ref{block2}) and d--connection $\mathbf{\Gamma }_{\beta \gamma }^{\alpha }$
(a distorsion of the Levi--Civita connection, or of the canonical
d--connection) satisfying the condition $\mathbf{Q}_{\alpha \beta \gamma
}=0. $ In this case, the strengths $\left( \mathbf{T}_{\ \beta \gamma
}^{\alpha },\mathbf{R}_{\ \beta \gamma \tau }^{\alpha }\right) $ are
N--adapted.

\item Berwald Riemann--Cartan (BRC) are modeled if a N--connection structure
is defined in a Riemann--Cartan space and distorting the connection to a
Berwald d--connection $^{[B]}\mathbf{D}=[\ ^{[B]}\mathbf{\Gamma }_{\beta
\gamma }^{\alpha }]=[\widehat{L}_{\ jk}^{i},\partial _{b}N_{k}^{a},0,%
\widehat{C}_{\ bc}^{a}]$, see (\ref{berw}). By definition, this space
satisfies the metricity conditions in the h- and v--subspaces, $Q_{ijk}=0$
and $Q_{abc}=0,$ but, in general, there are nontrivial nonmetricity
d--fields because $Q_{iab}$ and $Q_{ajk}$ are not vanishing (see formulas (%
\ref{berwnm})). Nonmetricities vanish with respect to holonomic frames.

\item Berwald Riemann--Cartan spaces with prescribed torsion (BRCT) are
defined by a more general class of d--connection $^{[BT]}\mathbf{\Gamma }%
_{\beta \gamma }^{\alpha }=[L_{\ jk}^{i},\partial _{b}N_{k}^{a},0,C_{\
bc}^{a}]$ inducing prescribed values $\tau _{\ jk}^{i}$ and $\tau _{\
bc}^{a} $ in d--torsion\ $^{[BT]}\mathbf{T}_{\ \beta \gamma }^{\alpha
}=[L_{\ [jk]}^{i},+\tau _{\ jk}^{i},0,\Omega _{ij}^{a},T_{\ bj}^{a},C_{\
[bc]}^{a}+\tau _{\ bc}^{a}],$ see (\ref{tauformulas}). The nontrivial
components of nonmetricity d--fields are $^{[B\tau ]}\mathbf{Q}_{\alpha
\beta \gamma }=\left( Q_{cij},Q_{iab}\right) .$ Such components vanish with
respect to holonomic frames.

\item Generalized Lagrange Riemann--Cartan spaces (GLRC) are modeled as
distinguished Riemann--Cartan spaces of odd--dimension, $\mathbf{V}^{n+n},$
provided with generic off--diagonal metrics with associated N--connection
inducing a tangent bundle structure. The d--metric $\mathbf{g}_{[a]}$ (\ref%
{dmglas}) and the d--connection $\ \ ^{[a]}\mathbf{\Gamma }_{\ \alpha \beta
}^{\gamma }$ $=\left( \ ^{[a]}L_{jk}^{i},\ ^{[a]}C_{jc}^{i}\right) $ (\ref%
{dcglma}) are those for the usual Lagrange spaces (see Definition \ref%
{defgls}).

\item Lagrange Riemann--Cartan spaces (LRC, see Remark \ \ref{rlafs}) are
provided with a Lagrange quadratic form $g_{ij}^{[L]}(x,y)=\frac{1}{2}\frac{%
\partial ^{2}L^{2}}{\partial y^{i}\partial y^{j}}$ (\ref{lagm}) inducing the
canonical N--connection structure $^{[cL]}\mathbf{N}=\{\ ^{[cL]}N_{j}^{i}\}$
(\ref{cncls})\ for a Lagrange space $\mathbf{L}^{n}=\left(
V^{n},g_{ij}(x,y)\right) $ (see Definition \ref{defls})) and, for instance,
with a canonical d--connection structure $\ ^{[b]}\mathbf{\Gamma }_{\ \
\alpha }^{\gamma }=\ ^{[b]}\mathbf{\Gamma }_{\ \alpha \beta }^{\gamma }%
\mathbf{\vartheta }^{\beta }$ satisfying metricity conditions for the
d--metric defined by $g_{ij}^{[L]}(x,y).$

\item Finsler Riemann--Cartan spaces (FRC, see Remark \ref{rfafs}) are
defined by a quadratic form $g_{ij}^{[F]}=\frac{1}{2}\frac{\partial ^{2}F^{2}%
}{\partial y^{i}\partial y^{j}}$ (\ref{finm2}) constructed from a Finsler
metric $F\left( x^{i},y^{j}\right) .$ It is induced the canonical
N--connection structure $\ ^{[F]}\mathbf{N}=\{\ ^{[F]}N_{j}^{i}\}$ (\ref{ncc}%
)\ as in the Finsler space $\mathbf{F}^{n}=\left( V^{n},F\left( x,y\right)
\right) $ with $\ ^{[F]}\widehat{\mathbf{\Gamma }}_{\ \beta \gamma }^{\alpha
}$ being the canonical Finsler d--connection (\ref{dccfs}).
\end{enumerate}

\subsection{Distinguished (pseudo) Riemannian spaces}

Sections \ref{srgarcg}\ and \ref{sarg} are devoted to modeling of locally
anisotropic geometric configurations in (pseudo) Riemannian spaces enabled
with generic off--diagonal metrics and associated N--connection structure.
Different classes of generalized Finsler metrics can be embedded in (pseudo)
Riemannian spaces as certain anholonomic frame configurations. Every such
space is characterized by a corresponding off--diagonal metric ansatz and
Levi--Civita connection stated with respect to coordinate frames or,
alternatively (see Theorem \ref{teqrgrcg}), by certain N--connection and
induced d--metric and d--connection structures related to the Levi--Cevita
connection with coefficients defined with respect to N--adapted anholonomic
(co) frames. We characterize every such type of (pseudo) Riemannian spaces
both by Levi--Civita and induced canonical/or Berwald d--connections which
contain also induced (by former off--diagonal metric terms) nontrivial
d--torsion and/or nonmetricity d--fields.

\begin{enumerate}
\item (Pseudo) Riemann spaces (in brief, pR) are certain manifolds $V^{n+m}$
of necessary smoothly class\ provided with generic off--diagonal metric
structure $g_{\alpha \beta }$ of arbitrary signature inducing the unique
torsionless and metric Levi--Civita connection $\Gamma _{\bigtriangledown
\beta \gamma }^{\alpha }.$ We can effectively diagonalize such metrics by
anholonomic frame transforms with associated N--connection structure. We can
also consider alternatively the canonical d--connection $\widehat{\mathbf{%
\Gamma }}_{\beta \gamma }^{\alpha }=[L_{\ jk}^{i},L_{\ bk}^{a},C_{\
jc}^{i},C_{\ bc}^{a}]$ (\ref{candcon})defined by the coefficients of
d--metric $\mathbf{g}_{\alpha \beta }=[g_{ij},h_{ab}]$ and N--connection $%
N_{i}^{a}.$ We have nontrivial d--torsions $\widehat{\mathbf{T}}_{\beta
\gamma }^{\alpha },$ but $T_{\bigtriangledown \beta \gamma }^{\alpha
}=0,Q_{\alpha \beta \gamma }^{\bigtriangledown }=0$ and $\mathbf{\hat{Q}}%
_{\alpha \beta \gamma }=0.$ The simplest anholonomic configurations are
characterized by associated N--connections with vanishing N--connection
curvature, $\Omega _{ij}^{a}=\delta _{\lbrack i}N_{j]}^{a}=0$. The
d--torsions $\widehat{\mathbf{T}}_{\ \beta \gamma }^{\alpha }=[\widehat{L}%
_{[\ jk]}^{i},\widehat{C}_{\ ja}^{i},\Omega _{ij}^{a},\widehat{T}_{\ bj}^{a},%
\widehat{C}_{\ [bc]}^{a}]$ and d--curvatures $\widehat{\mathbf{R}}_{\ \beta
\gamma \tau }^{\alpha }=[\widehat{R}_{\ jkl}^{i},\widehat{R}_{\ bkl}^{a},%
\widehat{P}_{\ jka}^{i},\widehat{P}_{\ bka}^{c},\widehat{S}_{\ jbc}^{i},%
\widehat{S}_{\ dbc}^{a}]$ \ are computed by introducing the components of $%
\widehat{\mathbf{\Gamma }}_{\beta \gamma }^{\alpha },$ respectively, in
formulas (\ref{dtorsb})\ and (\ref{dcurv}).

\item Distinguished (pseudo) Riemannian spaces (DpR) are defined as
manifolds $\mathbf{V}^{n+m}$ $\ $provided with N--connection structure $%
N_{i}^{a},$ d--metric field and d--connection $\mathbf{\Gamma }_{\beta
\gamma }^{\alpha }$ (a distorsion of the Levi--Civita connection, or of the
canonical d--connection) satisfying the condition $\mathbf{Q}_{\alpha \beta
\gamma }=0.$

\item Berwald (pseudo) Riemann spaces (pRB) are modeled if a N--connection
structure is defined by a generic off--diagonal metric. The Levi--Civita
connection is distorted to a Berwald d--connection $^{[B]}\mathbf{D}=[\
^{[B]}\mathbf{\Gamma }_{\beta \gamma }^{\alpha }]=[\widehat{L}_{\
jk}^{i},\partial _{b}N_{k}^{a},0,\widehat{C}_{\ bc}^{a}]$, see (\ref{berw}).
By definition, this space satisfies the metricity conditions in the h- and
v--subspaces, $Q_{ijk}=0$ and $Q_{abc}=0,$ but, in general, there are
nontrivial nonmetricity d--fields because $Q_{iab}$ and $Q_{ajk}$ are not
vanishing (see formulas (\ref{berwnm})). Such nonmetricities vanish with
respect to holonomic frames. The torsion is zero for the Levi--Civita
connection but $^{[B]}\mathbf{T}_{\ \beta \gamma }^{\alpha }=[L_{\
[jk]}^{i},0,\Omega _{ij}^{a},T_{\ bj}^{a},C_{\ [bc]}^{a}]$ is not trivial.

\item Berwald (pseudo) Riemann spaces with prescribed d--torsion (pRBT) are
defined by a more general class of d--connection $^{[BT]}\mathbf{\Gamma }%
_{\beta \gamma }^{\alpha }=[L_{\ jk}^{i},\partial _{b}N_{k}^{a},0,C_{\
bc}^{a}]$ inducing prescribed values $\tau _{\ jk}^{i}$ and $\tau _{\
bc}^{a} $ in d--torsion\ $^{[BT]}\mathbf{T}_{\ \beta \gamma }^{\alpha
}=[L_{\ [jk]}^{i},+\tau _{\ jk}^{i},0,\Omega _{ij}^{a},T_{\ bj}^{a},C_{\
[bc]}^{a}+\tau _{\ bc}^{a}],$ see (\ref{tauformulas}). The nontrivial
components of nonmetricity d--fields are $^{[B\tau ]}\mathbf{Q}_{\alpha
\beta \gamma }=\left( ^{[B\tau ]}Q_{cij},\ ^{[B\tau ]}Q_{iab}\right) .$ Such
components vanish with respect to holonomic frames.

\item Generalized Lagrange (pseudo) Riemannian spaces (pRGL) are modeled as
distinguished Riemann spaces of odd--dimension, $\mathbf{V}^{n+n},$ provided
with generic off--diagonal metrics with associated N--connection inducing a
tangent bundle structure. The d--metric $\mathbf{g}_{[a]}$ (\ref{dmglas})
and the d--connection $\ \ ^{[a]}\mathbf{\Gamma }_{\ \alpha \beta }^{\gamma
} $ $=\left( \ ^{[a]}L_{jk}^{i},\ ^{[a]}C_{jc}^{i}\right) $ (\ref{dcglma})
are those for the usual Lagrange spaces (see Definition \ref{defgls}) but on
a (pseudo) Riemann manifold with prescribed N--connection structure.

\item Lagrange (pseudo) Riemann spaces (pRL) are provided with a Lagrange
quadratic form $g_{ij}^{[L]}(x,y)=\frac{1}{2}\frac{\partial ^{2}L^{2}}{%
\partial y^{i}\partial y^{j}}$ (\ref{lagm}) inducing the canonical
N--connection structure $^{[cL]}\mathbf{N}=\{\ ^{[cL]}N_{j}^{i}\}$ (\ref%
{cncls})\ for a Lagrange space $\mathbf{L}^{n}=\left(
V^{n},g_{ij}(x,y)\right) $ and, for instance, provided with a canonical
d--connection structure $\ ^{[b]}\mathbf{\Gamma }_{\ \ \alpha }^{\gamma }=\
^{[b]}\mathbf{\Gamma }_{\ \alpha \beta }^{\gamma }\mathbf{\vartheta }^{\beta
}$ satisfying metricity conditions for the d--metric defined by $%
g_{ij}^{[L]}(x,y).$ There is an alternative construction with Levi--Civita
connection.

\item Finsler (pseudo) Riemann (FpR) are defined by a quadratic form $%
g_{ij}^{[F]}=\frac{1}{2}\frac{\partial ^{2}F^{2}}{\partial y^{i}\partial
y^{j}}$ (\ref{finm2}) constructed from a Finsler metric $F\left(
x^{i},y^{j}\right) .$ It is induced the canonical N--connection structure $\
^{[F]}\mathbf{N}=\{\ ^{[F]}N_{j}^{i}\}$ (\ref{ncc})\ as in the Finsler space
$\mathbf{F}^{n}=\left( V^{n},F\left( x,y\right) \right) $ with $\ ^{[F]}%
\widehat{\mathbf{\Gamma }}_{\ \beta \gamma }^{\alpha }$ being the canonical
Finsler d--connection (\ref{dccfs}).
\end{enumerate}

\subsection{Teleparallel spaces}

Teleparallel spaces were considered in sections \ref{stps} and \ref{stpfa}.
Here we classify what type of locally isotropic and anisotropic structures
can be modeled in by anholonomic transforms of (pseudo) Riemannian spaces to
teleparallel ones. The anholonomic frame structures are with associated
N--connection with the components defined by the off--diagonal metric
coefficients.

\begin{enumerate}
\item Teleparallel spaces (in brief, T) are usual ones with vanishing
curvature, modeled on manifolds $V^{n+m}$ of necessary smoothly class\
provided, for instance, with the Weitzenbock connection $^{[W]}\Gamma
_{\beta \gamma }^{\alpha }$(\ref{wcon}) which can be transformed in a
d--connection one with respect to N--adapted frames. In general, such
geometries are characterized by torsion $^{[W]}T_{\ \beta \gamma }^{\alpha }$
constrained to the condition to result in zero d--curvatures. The simplest
theories are with vanishing nonmetricity.

\item Distinguished teleparallel spaces (DT) are manifolds $\mathbf{V}^{n+m}$
provided with N--connecti\-on structure $N_{i}^{a},$ d--metric field (\ref%
{block2}) and arbitrary d--connection $\mathbf{\Gamma }_{\beta \gamma
}^{\alpha }$ with vanishing d--curvatures. The geometric constructions are
stated by the Weitzenbock d--connection $^{[Wa]}\mathbf{\Gamma }_{\beta
\gamma }^{\alpha }=\mathbf{\Gamma }_{\bigtriangledown ~\beta \gamma
}^{\alpha }+\mathbf{\hat{Z}}_{~\beta \gamma }^{\alpha }+\mathbf{Z}_{~\beta
\gamma }^{\alpha }$ with distorsions without nonmetricity d--fields
preserving the condition of zero values for d--curvatures.

\item Teleparallel Berwald spaces (TB) are defined by distorsions of the
Weitzenbock connection on a manifold $V^{n+m}$ to any Berwald like
strucutre, $\ ^{[WB]}\mathbf{\Gamma }_{\beta \gamma }^{\alpha }=\mathbf{%
\Gamma }_{\bigtriangledown ~\beta \gamma }^{\alpha }+\mathbf{\hat{Z}}%
_{~\beta \gamma }^{\alpha }+\mathbf{Z}_{~\beta \gamma }^{\alpha }$
satisfying the condition that the curvature is zero. All constructions with
effective off--diagonal metrics can be adapted to the N--connection and
considered for d--objects. Such spaces satisfy the metricity conditions in
the h- and v--subspaces, $Q_{ijk}=0$ and $Q_{abc}=0,$ but, in general, there
are nontrivial nonmetricity d--fields, $Q_{iab}$ and $Q_{ajk}.$

\item Teleparallel Berwald spaces with prescribed torsion (TBT) are defined
by a more general class of distorsions resulting in the Weitzenbock
d--connection,
\begin{equation*}
\ ^{[WB\tau ]}\mathbf{\Gamma }_{\beta \gamma }^{\alpha }=\mathbf{\Gamma }%
_{\bigtriangledown ~\beta \gamma }^{\alpha }+\mathbf{\hat{Z}}_{~\beta \gamma
}^{\alpha }+\mathbf{Z}_{~\beta \gamma }^{\alpha },
\end{equation*}
having prescribed values $\tau _{\ jk}^{i}$ and $\tau _{\ bc}^{a}$ in
d--torsion\
\begin{equation*}
^{[WB]}\mathbf{T}_{\ \beta \gamma }^{\alpha }=[L_{\ [jk]}^{i},+\tau _{\
jk}^{i},0,\Omega _{ij}^{a},T_{\ bj}^{a},C_{\ [bc]}^{a}+\tau _{\ bc}^{a}]
\end{equation*}
and characterized by the condition $^{[WB\tau ]}\mathbf{R}_{\ \beta \gamma
\tau }^{\alpha }=0$ with certain nontrivial nonmetricity d--fields, $%
^{[WB\tau ]}\mathbf{Q}_{\alpha \beta \gamma }=\left( ^{[WB\tau ]}Q_{cij},\
^{[WB\tau ]}Q_{iab}\right) .$

\item Teleparallel generalized Lagrange spaces (TGL) are modeled as
Riemann--Cartan spaces of odd--dimension, $\mathbf{V}^{n+n},$ provided with
generalized Lagrange d--metric and associated N--connection inducing a
tangent bundle structure with zero d--curvature. The Weitzenblock--Lagrange
d--connection $\ ^{[Wa]}\mathbf{\Gamma }_{\ \alpha \beta }^{\gamma }$ $=( \
^{[Wa]}L_{jk}^{i},\ ^{[Wa]}C_{jc}^{i}) , $\ where $^{[Wa]}\mathbf{\Gamma }%
_{\beta \gamma }^{\alpha }=\mathbf{\Gamma }_{\bigtriangledown ~\beta \gamma
}^{\alpha }+\mathbf{\hat{Z}}_{~\beta \gamma }^{\alpha }+\mathbf{Z}_{~\beta
\gamma }^{\alpha },$ are defined by a d--metric $\mathbf{g}_{[a]}$ (\ref%
{dmglas}) with $\mathbf{Z}_{\ \ \beta }^{\alpha }$ inducing zero
nonmetricity d--fields, $^{[a]}\mathbf{Q}_{\alpha \beta \gamma }=0$ and zero
d--curvature, $^{[Wa]}\mathbf{R}_{\ \beta \gamma \tau }^{\alpha }=0.$

\item Teleparallel Lagrange spaces (TL, see section\ \ref{rlafs}) are
Riemann--Cartan spaces $\mathbf{V}^{n+n}$ provided with a Lagrange quadratic
form $g_{ij}^{[L]}(x,y)=\frac{1}{2}\frac{\partial ^{2}L^{2}}{\partial
y^{i}\partial y^{j}}$ (\ref{lagm}) inducing the canonical N--connection
structure $^{[cL]}\mathbf{N}=\{\ ^{[cL]}N_{j}^{i}\}$ (\ref{cncls})\ for a
Lagrange space $\mathbf{L}^{n}=\left( V^{n},g_{ij}(x,y)\right) $ but with
vanishing d--curvature. The d--connection structure $\ ^{[WL]}\mathbf{\Gamma
}_{\ \alpha \beta }^{\gamma }$ (of Weitzenblock--Lagrange type) is the
generated as a distortion by the Weitzenbock d--torsion, $^{[W]}\mathbf{T}%
_{\beta },$ but zero nonmetricity d--fields,$\ ^{[WL]}\mathbf{Q}_{\beta
\gamma \alpha }=0,$ when $^{[WL]}\mathbf{\Gamma }_{\ \alpha \beta }^{\gamma
}=\mathbf{\Gamma }_{\bigtriangledown ~\beta \gamma }^{\alpha }+\mathbf{\hat{Z%
}}_{~\beta \gamma }^{\alpha }+\mathbf{Z}_{~\beta \gamma }^{\alpha }.$

\item Teleparallel Finsler spaces (TF) are Riemann--Cartan manifolds $%
\mathbf{V}^{n+n}$ defined by a quadratic form $g_{ij}^{[F]}=\frac{1}{2}\frac{%
\partial ^{2}F^{2}}{\partial y^{i}\partial y^{j}}$ (\ref{finm2}) and a
Finsler metric $F\left( x^{i},y^{j}\right) .$ They are provided with a
canonical N--connection structure $\ ^{[F]}\mathbf{N}=\{\ ^{[F]}N_{j}^{i}\}$
(\ref{ncc})\ as in the Finsler space $\mathbf{F}^{n}=\left( V^{n},F\left(
x,y\right) \right) $ but with a Finsler--Weitzenbock d--connection structure
$\ ^{[WF]}\mathbf{\Gamma }_{\ \alpha \beta }^{\gamma },$ respective
d--torsion, $^{[WF]}\mathbf{T}_{\beta },$ and vanishing nonmetricity, $%
^{[WF]}\mathbf{Q}_{\beta \gamma \tau }=0,$ d--fields, $\ ^{[WF]}\mathbf{%
\Gamma }_{\ \alpha \beta }^{\gamma }=\mathbf{\Gamma }_{\bigtriangledown
~\beta \gamma }^{\alpha }+\mathbf{\hat{Z}}_{~\beta \gamma }^{\alpha }+%
\mathbf{Z}_{~\beta \gamma }^{\alpha },$where $\ \mathbf{\hat{Z}}_{~\beta
\gamma }^{\alpha }$ contains distorsions from the canonical Finsler
d--connection (\ref{dccfs}).
\end{enumerate}

\newpage

\begin{table}[h]
\begin{center}
%
\end{center}
\caption{Generalized Lagrange--affine spaces}
\label{tablegs}
\end{table}

\newpage

\begin{table}[h]
\begin{center}
%
%
\end{center}
\caption{Generalized Hamilton--affine spaces}
\label{tableghs}
\end{table}

\newpage

\begin{table}[h]
\begin{center}
%
%
\end{center}
\caption{Teleparallel Lagrange--affine spaces}
\label{tabletls}
\end{table}

\newpage

\begin{table}[h]
\begin{center}
%
%
\end{center}
\caption{Teleparallel Hamilton--affine spaces}
\label{tableths}
\end{table}

\newpage

\begin{table}[h]
\begin{center}
%
%
\end{center}
\caption{Generalized Finsler--Lagrange spaces}
\label{tablegfls}
\end{table}

\newpage

\begin{table}[h]
\begin{center}
%
%
\end{center}
\caption{Generalized Hamilton--Cartan spaces}
\label{tableghcs}
\end{table}

\newpage

\begin{table}[h]
\begin{center}
%
%
\end{center}
\caption{Teleparallel Finsler--Lagrange spaces}
\label{tabletfls}
\end{table}

\newpage

\begin{table}[h]
\begin{center}
%
%
\end{center}
\caption{Teleparallel Hamilton--Cartan spaces}
\label{tablehcs}
\end{table}

\newpage

\begin{table}[h]
\begin{center}
%
%
\end{center}
\caption{Distinguished Riemann--Cartan spaces}
\label{tabledrcs}
\end{table}
\

\newpage

\begin{table}[h]
\begin{center}
%
%
\end{center}
\caption{Distinguished (pseudo) Riemannian spaces}
\label{tableprs}
\end{table}

\newpage

\begin{table}[h]
\begin{center}
%
\\ \hline\hline
\end{tabular}%
\end{center}
\caption{Teleparallel spaces}
\label{tablets}
\end{table}

\end{document}